\documentclass[aps, prx, twocolumn, superscriptaddress, floatfix]{revtex4-2}\usepackage{mathrsfs} 
\usepackage{amsmath}
\usepackage[version=4]{mhchem}
\usepackage{amsfonts} 
\usepackage{amsmath,amssymb}
\usepackage{graphicx}
\usepackage{subfig}
\usepackage{float}
\usepackage[table]{xcolor}
\usepackage{colortbl}
\usepackage{tikz}
\usepackage{pgfplots}
\usepackage{bm}
\usepackage{multirow}
\usepackage[explicit]{titlesec}
\usepackage[most]{tcolorbox}
\usepackage{braket}
\usepackage{soul}
\usepackage{hyperref}
\hypersetup{
    colorlinks=true,
    citecolor=red,
    linkcolor=blue,
    filecolor=magenta,      
    urlcolor=red
    }
\pgfplotsset{compat=1.18}

\DeclareCaptionJustification{justified}{\leftskip=0pt \rightskip=0pt \parfillskip=0pt plus 1fil}

\definecolor{dy}{rgb}{0.9,0.9,0.4}
\definecolor{dr}{rgb}{0.95,0.65,0.55}
\definecolor{db}{rgb}{0.5,0.8,0.9}
\definecolor{dg}{rgb}{0.2,0.9,0.6}
\definecolor{BrickRed}{rgb}{0.8,0.3,0.3}
\definecolor{Navy}{rgb}{0.2,0.2,0.6}
\definecolor{DarkGreen}{rgb}{0.1,0.4,0.1}
\begin{document}

\title{A spin chain with non-Hermitian $\mathscr{PT}-$symmetric boundary couplings: exact solution, dissipative Kondo effect, and phase transitions on the edge}

\author{Pradip Kattel}
\email{pradip.kattel@rutgers.edu}
\affiliation{Department of Physics and Astronomy, Center for Material Theory, Rutgers University,
Piscataway, New Jersey 08854, United States of America}

\author{Parameshwar R. Pasnoori}
\affiliation{Department of Physics, University of Maryland, College Park, Maryland 20742, United
States of America}
\affiliation{Laboratory for Physical Sciences, 8050 Greenmead Dr, College Park, Maryland 20740,
United States of America}

\author{J. H. Pixley}
\affiliation{Department of Physics and Astronomy, Center for Material Theory, Rutgers University,
Piscataway, New Jersey 08854, United States of America}
\affiliation{Center for Computational Quantum Physics, Flatiron Institute, 162 5th Avenue, New York, NY 10010}

\author{Natan Andrei}
\affiliation{Department of Physics and Astronomy, Center for Material Theory, Rutgers University,
Piscataway, New Jersey 08854, United States of America}

\begin{abstract}
    We construct an exactly solvable $\mathscr{PT}-$symmetric non-Hermitian model where a spin$-\frac{1}{2}$ isotropic quantum Heisenberg spin chain is coupled to two spin$-\frac{1}{2}$ Kondo impurities at its boundaries with coupling strengths that are complex conjugates of each other. Solving the model by means of a combination of the Bethe Ansatz  and density matrix renormalization group (DMRG) techniques, we show  that the model exhibits three distinct boundary phases: a $\mathscr{PT}$ symmetric phase with a dissipative Kondo effect, a phase with bound modes and spontaneously broken $\mathscr{PT}$ symmetry,
    and a phase with an effectively unscreened spin (i.e. a free local moment). In the Kondo and the unscreened phases, the $\mathscr{PT}-$symmetry is unbroken, and hence all states have real energies, whereas in the bound mode phases,  in addition to the states with real energies, there exist states with complex energy eigenvalues that appear in complex conjugate pairs, signaling spontaneous breaking of the $\mathscr{PT}-$symmetry. The exact solution is used to provide an accessible benchmark for DMRG with a non-Hermitian matrix product operator representation that demonstrates an accuracy comparable to its Hermitian limit thus showing the power of DMRG to handle non-Hermitian many body calculations.
\end{abstract}

\maketitle
\section{Introduction}
   The fundamental axioms of quantum mechanics assert that the equilibrium behavior  of a quantum system as well as its dynamics are dictated by the Hamiltonian $\mathcal{H}$, a Hermitian operator acting  on the Hilbert space defined over $\mathbb{C}$. The eigenvalues of the Hamiltonian operator, $E_i$, are the system's energy levels and the  squared magnitude of the corresponding eigenvectors $|\psi_i(\vec r) \rangle$ give the probability density $P(\vec r)=||\braket{\psi_i(\vec r)|\psi_i(\vec r)}||$ associated with locating the system within the spatial position $\vec r$. However, this  description requires modification when describing the behavior of an open quantum system, namely, a system that interacts with its environment through the exchange of particles or energy. 
   The system then resides in non-equilibrium states with its surroundings, typically
expressed as density operators comprised of system and environment degrees of freedom. When the environment is weakly coupled and Markovian, one proceeds via Gorini-Kossakowski-Sudarshan-Lindblad (GKSL) formulation \cite{lindblad1976generators,gorini1976completely,manzano2020short}. This framework has now successfully described a myriad of natural and engineered condensed matter systems, optical systems, and  quantum information and quantum cosmology applications \cite{dattagupta2004dissipative,weiss2012quantum,koch2022quantum,cao2020reservoir,wouters2007excitations,fitzpatrick2017observation,botzung2021engineered,alberton2021entanglement,yu2008understanding,kaplanek2020hot,chakraborty2022thermal}.

    Effects of interactions with environment can sometimes be included directly into the Hamiltonian, resulting in a non-Hermitian Hamiltonian.  This may occur for example when the jump operators in the Lindbladian can be neglected and the remaining terms are lumped into the Hamiltonian. Other circumstances may lead to an effective non-Hermitian Hamiltonians. For example, non-Hermitian Hamiltonian  can be rigorously justified in cases where dynamics is confined only to the subspace of the system (for example, by employing projection operator methods projecting out the degrees of freedom involving the surrounding)\cite{rotter2009non,muller2009phase,savin2003concept}. Moreover, within the realm of quantum measurement theory, this justification is reinforced by conditioning quantum trajectories according to specific measurement outcomes \cite{ashida2020non,breuer2002theory}. Also various quantum master equations can be mapped to non-Hermitian Hamiltoians defined in  the doubled `Liouville-Fock' space \cite{prosen2008third,Essler-prozen,alba2023free}. These non-Hermitian effective Hamiltonians typically have  complex eigenvalues. The real part of the eigenvalues describes the energy of the system whereas the imaginary part gives an estimate of the lifetime of the dissipating states.

  Within the field of non-Hermitian Hamiltonians, those possessing both parity and time-reversal symmetry ($\mathscr{PT}-$symmetry), have found application across a broad spectrum of domains, notably including quantum optics\cite{castaldi2013p,yin2013unidirectional,zyablovsky2014pt,klauck2019observation}, and condensed matter systems \cite{kawabata2018parity,kornich2022signature,bagarello2015non,zhao2017p,turker2018pt}. In a seminal paper \cite{bender1998real}, Bender \textit{et. al} noticed that the non-Hermitian Hamiltonian of the form
  \begin{equation}
      \mathcal H=\hat{p}^2+\hat{x}^2+(i \hat{x})^N
      \label{expt},
  \end{equation}
where $N\geq 2$ have all real energies and they asserted that the reality of the Hamiltonian is due to $\mathscr{PT}-$symmetry. It is worth noting that $\mathscr{PT}$ is an anti-linear operator. The proof of the reality of eigenvalues for $\mathscr{PT}-$symmetric Hamiltonians traces back to an observation by Wigner \cite{wigner1960phenomenological}
\footnote{ Consider the Schrodinger equation
\begin{equation}
   \mathcal H \ket{\psi(t)} = E \ket{\psi(t)}.
\end{equation}
Acting with a generic anti-linear operator \(\hat{\mathcal{A}}\), we arrive at:
\begin{equation}
     \hat{\mathcal{A}} \mathcal H \hat{\mathcal{A}}^{-1} \hat{\mathcal{A}} \ket{\psi(t)} = E^* \hat{\mathcal{A}} \ket{\psi(t)}.
    \label{antlop}
\end{equation}
In Eq.\eqref{antlop}, we find that when the Hamiltonian \(\mathcal H\) obeys the antilinear symmetry condition \(\hat{\mathcal{A}} \mathcal H \hat{\mathcal{A}}^{-1} = \mathcal H\), two distinct scenarios emerge, as originally pointed out by Wigner:

    a) If the eigenfunctions satisfy \(\hat{\mathcal{A}}\ket{\psi(t)} = \ket{\psi(t)}\), then the corresponding eigenvalues are real.\\
   b)Alternatively, when the eigenfunctions exhibit the property \(\hat{\mathcal{A}}\ket{\psi(t)} \neq \ket{\psi(t)}\), the energies appear in conjugate pairs, and the conjugate eigenfunctions take the form $|\psi(t)\rangle \sim e^{-i E t}$ and $\hat{\mathcal{A}}|\psi(t)\rangle \sim e^{-i Et}$.
} that anti-unitary operators have either real eigenvalues or the complex ones that appear in complex conjugate pairs.

Some $\mathscr{PT}-$ symmetric Hamiltonians like Eq.\eqref{expt} have all the right eigenfunctions that are $\mathscr{PT}$ symmetric \textit{i.e.} $\mathscr{PT}\ket{\psi_i}=\ket{\psi_i}$ and  hence all the eigenvalues of such Hamiltonians are entirely real just like those of a Hermitian Hamiltinoian. In fact, it can be shown that such a Hamiltonian $\mathcal{H}$ is related to a Hermitian Hamiltonian $h$ by an invertible linear map $\eta$ \textit{i.e.}
    $\mathcal{H}=\eta h \eta^{-1}$. Such Hamiltonians might not yield any expansion to the Hermitian quantum framework \cite{mostafazadeh2003exact}. Nevertheless, when certain eigenvectors associated with $\mathscr{PT}-$symmetric Hamiltonians fail to exhibit $\mathscr{PT}$ symmetry (\textit{i.e.} there is spontaneous breaking of the $\mathscr{PT}$ symmetry), these Hamiltonians could characterize an open system with balanced loss and gain. 

In this paper, we  consider the latter form of the Hamiltonian by explicitly constructing an integrable Heisenberg  spin chain with non-Hermitian quantum impurities positioned at the edges. Choosing the two impurity coupling to be complex conjugate of each other, we impose the $\mathscr{PT}$  symmetry on the model. Because of the $\mathscr{PT}$ symmetry, most of the energy eigenvalues of the model are real and when they are complex, they come in complex conjugate pair. Thus, the model have a well defined ground state which is defined as the state with lowest real part of the energy eigenvalue.  We then proceed to solve the model exactly in the thermodynamic limit using Bethe Ansatz and identify various boundary phase transition in the model. By computing some physical quantities, we identify that the signature of the phase transition can be seen from the ground state observables. Moreover, we solve the model using DMRG implemented in ITensor library \cite{fishman2022itensor} by turning on the flag \textit{ishermitian=false} such that we can efficiently compute the ground state energy and obervables. We then compare the DMRG result with Bethe Ansatz to show that DMRG algorithm work very well for non-Hermitian problems. This bench-marking is essential such that we can trust the DMRG methods for the cases where exact solution is not possible, which is often the case. Analytically computing energy levels for $\mathscr{PT}$ -symmetric Hamiltonians is a challenge. Even basic single-particle quantum systems with simple potentials, like the $ix^3$ potential which is a special case of $N=3$ described in Eq.\eqref{expt} \cite{bender2004extension} or $\mathscr{PT}$-symmetric square well potential \cite{bender2006calculation}, can only be addressed through perturbative methods.

\section{The Model}
We shall construct a $\mathscr{PT}-$ symmetric spin chain Hamiltonian
\begin{equation}
   \mathcal H=J\sum_{j=1}^{\bar N-1}\bm\sigma_j\cdot\bm\sigma_{j+1}+J_{\mathrm{imp}}\bm\sigma_1\cdot\bm\sigma_L+J_{\mathrm{imp}}^*\bm\sigma_N\cdot\bm\sigma_R,
   \label{modelHam}
\end{equation}
where the bulk coupling $J$ is real but the boundary couplings $J_{\mathrm{imp}}$ and $J_{\mathrm{imp}}^*$ are complex conjugate of each other, $\bar N$ is the number of bulk sites, and $N=\bar N+2$ is the total number of sites including the two impurities at the two edges. The Heisenberg chain with periodic boundary conditions was first solved by Bethe \cite{1931_Bethe_ZP_71}. The model with open boundary conditions is integrable and was solved by boundary Bethe Ansatz method in \cite{alcaraz1987surface,sklyanin1988boundary,wang2015off,frahm2010functional}. The nature of the ground state and excited states depends on both the parity of the total number of sites and the ratio of the bulk and boundary couplings. Here, we will only focus on the case of an even number of total sites. We will often write the complex boundary coupling in terms of the bulk coupling as
\begin{equation}
    J_{\mathrm{imp}}=\frac{J}{1-(\beta+i\gamma)^2}
    \label{coeffrel},
\end{equation}
where $\beta$ and $\gamma$ are two real parameters. This unusual parameterization is important because, as we show below, only the parameter $\beta$ governs the phase transition of the model whereas both $\beta$ and $\gamma$ determine the energies of the states. 
\subsection{Limits and Symmetries}

In the limit when $\gamma=0$ and $\beta\neq 0$ or when $\gamma\neq 0$ and $\beta=0$, the model reduces to the Hermitian case considered in \cite{kattel2023kondo,wang1997exact} with equal impurity strength on two edges. Only when both $\beta$ and $\gamma$ are non-zero, the model describes the non-hermitian effects.  This model generalizes our recent study of  a many body $\mathscr{PT}-$symmetric system consisting of  Heisenberg spin chain with complex boundary fields at its edges for which we   provided an exact solution via Bethe Ansatz \cite{kattel2023exact}. The classical field at the edges are now replaced with quantum impurities such that the phase diagram is much richer in the current case.

Let's recall the action of time reversal and parity in a discrete system. Time reversal operator  ($\mathscr{T}$) flips the sign of the imaginary unit: $\mathscr{T} i \mathscr{T} = -i$ and parity operator ($\mathscr{P}$) flips spins across the midpoint: $\mathscr{P} \sigma_{l}^{\alpha} \mathscr{P} = \sigma_{N+1-l}^{\alpha}$ for a system of $N$ spins. Thus, it is evident that Hamiltonian Eq.\eqref{modelHam} is $\mathscr{PT}-$symmetric.

The Hermitian version of the problem with real boundary coupling has been studied by various authors \cite{wang1997exact, frahm1997open,kattel2023kondo,kattel2024kondo} where it was shown that depending on the ratio of the boundary and bulk couplings, the impurity could be screened or unscreened in the ground state.  Likewise, the nature of the excitations also strongly depends on the ratio of the couplings. For antiferromagnetic coupling, the impurity is
(Kondo) screened by a multiparticle ``cloud'' of spinon dynamically forming a manybody singlet is prevalent only in the small region in the phase space $\left( 0<{J_\mathrm{imp}}/{J}<{4}/{3}\right)$ and there exists a new phase where impurity is screened by impurity for $\left( {J_\mathrm{imp}}/{J}>{4}/{3}\right)$. For the ferromagnetic boundary coupling, the impurity is unscreened in the ground state. However, for certain ranges of the boundary coupling strengths $\left( \left\vert{J_\mathrm{imp}}/{J}\right\vert >{4}/{5}\right)$, a boundary bound modes exist in higher excited states.
Conversely, at  coupling strengths in the range $\left( 0<\left\vert\frac{J_\mathrm{imp}}{J}\right\vert <\frac{4}{5}\right)$ , the impurity remains unscreened at all energy scales \cite{kattel2023kondo}. Our model provides deep insight into the nature of Kondo physics in dissipative open quantum systems when the bulk is strongly interacting. 

\subsection{Application of the model}
Despite the model under consideration is far from the conventional Kondo model of quantum impurity in a Fermi sea, there are several ways to see that this model does contain similar phenomenology and is also interesting in its own right.
For example, one way to understand the model Hamiltonian in Eq.~\eqref{modelHam} is to consider it as a phenomenological model describing the two-body dissipation at the two boundaries, which could be experimentally realized in various engineered quantum systems ranging from ultra-cold atoms \cite{ren2022chiral,nakagawa2018non}, trapped ions \cite{cao2023probing,lu2024realizing}, and superconducting qubits \cite{dogra2021quantum,chen2021quantum}. To be more concrete, consider ultra cold spinor bosonic atoms in deep optical lattices  where the atoms become effectively localized on individual sites, forming a spinfull bosonic Mott insulator such that the dynamics of the remaining degrees of freedom is governed by effective spin–spin interactions realizing nearest-neighbour Heisenberg model \cite{zhang2020controlling,gorshkov2010two,jepsen2020spin}.
Using the anarhomicity of the lattice potential and applying the laser fields to the boundary spins, we can construct the lossy boundary terms in the Hamiltonian Eq.\eqref{modelHam} \cite{schwager2013dissipative}. In Alkaline-earth atoms system there exist durable metastable excited states that act as localized impurities and their ground states as conduction electrons. The interaction between ground and excited states results in two-body loss/gain through inelastic scattering, giving rise to the non-Hermitian Kondo effect \cite{nakagawa2018non, kattel2024dissipation}. Thus, the model considered here provides a natural description of the dissipative Kondo effect in interacting cold atom systems with balanced loss and gain.

\begin{figure}
    \centering
\includegraphics[width=1\linewidth]{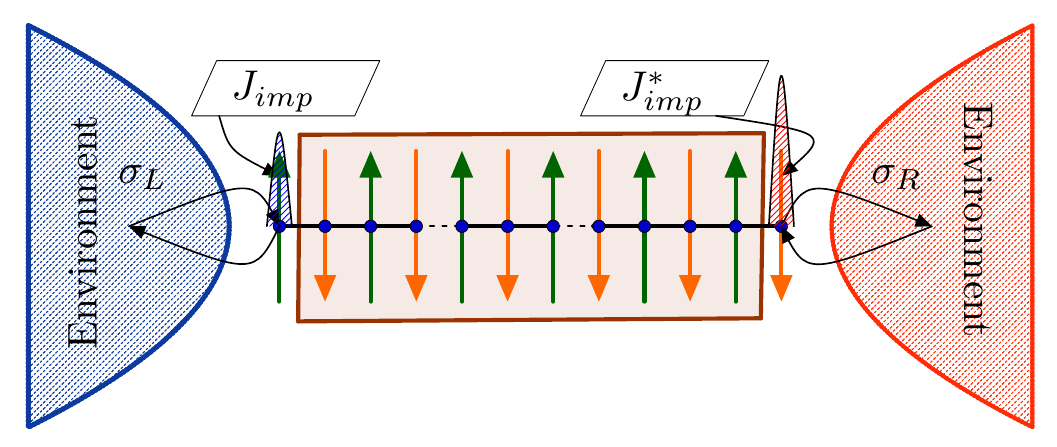}
    \caption{Cartoon realization of a spin chain interacting with magnetic bath via complex spin exchange interaction. The complex Kondo couplings at the two ends are complex conjugates of each other such that the model has $\mathscr{PT}-$symmetry.}
    \label{PTXXX-cartoon}
\end{figure}

As shown in the Fig.\ref{PTXXX-cartoon}, the model describes the spin chain interacting with magnetic bath with spin exchange interaction. To understand the application of the model in a broader context, let us recall the well known fact that the low energy effective field theoretic description of the Heisenberg chain is given by $SU(2)_1$ Weiss-Zumino-Witten theory perturbed by a marginal operator $\mathbf{J}_R
\cdot \mathbf{J}_L$ \cite{affleck1986exact}. In terms of the quadratic $SU(2)$ current, the low energy theory can be written as 
\begin{equation}
 \mathcal{H}_C=\frac{1}{k+2} \int d x\left(\mathbf{J}_R \cdot \mathbf{J}_R+\mathbf{J}_L \cdot \mathbf{J}_L\right) +g \int \mathrm{d}x  \mathbf{J}_R\cdot \mathbf{J}_L,
\end{equation}
where $g \propto J$ is the coupling constant for the marginally irrelevant Thirring term and the level $k=1$ in our case. Adding the boundary perturbation described in  Eq.\eqref{modelHam}, the low energy description of Eq.\eqref{modelHam} is given by the Hamiltonian \cite{laflorencie2008kondo}
\begin{equation}
    \mathscr{H}=  \mathcal{H}_C+J_{\mathrm{imp}}\mathbf{J}_L(-L)\cdot \vec S_L+J^*_{\mathrm{imp}}\mathbf{J}_L(L)\cdot \vec S_R,
    \label{fermionicHam}
\end{equation}
where $\vec S_L$ and $\vec S_R$ are the spin operators of the impurities at the edges of spin chain. 

From Eq.\eqref{fermionicHam}, it is evident that this model describes the spin part of an interacting system with two non-Hermitian Kondo-like impurities at the edges.  Recently, the non-Hermitian Kondo model in a non-interacting Fermi-liquid has been studied as an archetypal model for the Kondo effect in dissipative system\cite{nakagawa2018non,kattel2024dissipation}. 
Here we advance the study of quantum spin chains with boundary (Kondo) couplings to include interactions in the bulk part of the system thus fundamentally going beyond previous limits of the problem. At the same time, our results share many common features with the dissipative Kondo model including bound modes and first order like transitions that are fundamentally beyond what is possible in  the Hermitian model.

\section{Summary of the Results}
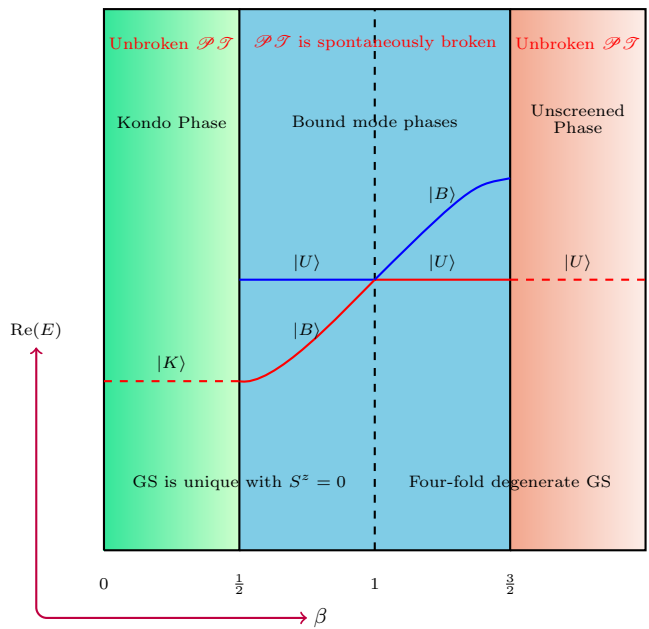
\begin{figure}
\definecolor{dy}{rgb}{0.9,0.9,0.4}
\definecolor{dr}{rgb}{0.95,0.65,0.55}
\definecolor{db}{rgb}{0.5,0.8,0.9}
\definecolor{dg}{rgb}{0.2,0.9,0.6}
\begin{center}
\begin{tikzpicture}[thick,scale=0.9, every node/.style={scale=0.9}]
\usepgflibrary {shadings}
\draw [<->, rounded corners, thick, purple] (-1,3) -- (-1,-1) -- (3,-1);

\node at (-1,3.25) {\scriptsize{$\operatorname{Re}(E)$}};
\node at (3.2,-1) {\small{$\beta$}};

\draw [left color=dr,right color=dr!20!white] (6,0)--(8,0)--(8,8)--(6,8);
\draw [fill=db] (2,0)--(6,0)--(6,8)--(2,8);
\draw [left color=dg,right color=green!20!white] (0,0)--(2,0)--(2,8)--(0,8)--(0,0);
\draw[dashed] (4,0)--(4,8);
\draw[blue] (2,4)--(4,4);
\draw[red] (4,4)--(6,4);
\draw[red,dashed](0,2.5)--(2,2.5);
\draw[red,dashed](6,4)--(8,4);
\draw[red] (2,2.5) .. controls (2.5,2.4) and (3.5,3.5) .. (4,4);
\draw[blue] (4,4) .. controls (5.5,5.5) and (5.5,5.4) .. (6,5.5);

\node at (0,-.5) {\scriptsize{0}};
\node at (2,-.5) {\scriptsize{$\frac12$}};
\node at (4,-.5) {\scriptsize{1}};
\node at (6,-.5) {\scriptsize{$\frac32$}};

\node at (1,6.325) {\scriptsize{Kondo Phase}};

\node at (1,7.5) {\scriptsize{\color{red}{Unbroken $\mathscr{PT}$}}};

\node at (2,1) {\scriptsize{GS is unique with $S^z=0$}};

\node at (7,6.5) {\scriptsize{Unscreened}};
\node at (7,6.25) {\scriptsize{Phase}};

\node at (7,7.5) {\scriptsize{\color{red}{Unbroken $\mathscr{PT}$}}};

\node at (4,7.5) {\scriptsize{{\color{red} $\mathscr{PT}$ is spontaneously broken}}};

\node at (6,1) {\scriptsize{Four-fold degenerate GS}};

\node at (4,6.325) {\scriptsize{Bound mode phases}};

\node at (1,2.75) {\scriptsize{$\ket{K}$}};

\node at (3,3.25) {\scriptsize{$\ket{B}$}};

\node at (3,4.25) {\scriptsize{$\ket{U}$}};

\node at (5,4.25) {\scriptsize{$\ket{U}$}};

\node at (5,5.25) {\scriptsize{$\ket{B}$}};

\node at (7,4.25) {\scriptsize{$\ket{U}$}};
\end{tikzpicture}
\end{center}
\caption{Phase diagram for $\gamma\neq 0$. The red line represent the energy of the ground state in various phases as a function of parameter $\beta$ where the complex impurity coupling is related to the bulk coupling $J$ and parameters $\beta$ and $\gamma$ by $J_{\mathrm{imp}}=\frac{J}{1-(\beta+i\gamma)^2}$. $\ket{K}$ is the unique ground state in Kondo phase where both impurities are screened by the Kondo clouds, $\ket{B}$ is a state where both impurities are screened by bound modes, and $\ket{U}$ is a four-fold degenerate state where both impurities are unscreened. The state  $\ket{B}$ is the ground state in the bound mode I phase and $\ket{U}$ is the ground state in the bound mode II and unscreened phases. }
\label{Fig:PD-PTK}
\end{figure}

Before describing our solution of the model, let us briefly summarize our main results. The model has three distinct phases characterized by $\beta$ (explicitly written in Eq.\eqref{betadef}) which is related to the bulk and boundary couplings via Eq.\eqref{coeffrel} as shown in the phase diagram Fig.\ref{Fig:PD-PTK}. In two of the three phases, the $\mathscr{PT}$ symmetry is unbroken, where as in one of the phases, the $\mathscr{PT}$ symmetry is spontaneously broken.
\begin{itemize}
    \item {\it Dissipative Kondo effect}: For even $N$, when $0 < \beta < \frac{1}{2}$, impurities are screened due to the multiparticle Kondo-like effect. The ground state is a unique singlet state, and no boundary excitations are possible at low energies. All states within this phase possess real energies in the thermodynamic limit, demonstrating unbroken $\mathscr{PT}-$symmetry. If the total number of sites $N$ is odd, then there is a spinon in the ground state state which could be oriented in the positive or negative $z-$direction thereby the ground state is two fold degenerate with $S^z=\pm \frac{1}{2}$ and both the impurities are screened by Kondo clouds at the respective edges. When a global magnetic field is applied by adding a Zeeman term to the Hamiltonian, the local impurity magnetization  changes smoothly from $0$ and $h=0$ to $0.5$ at $h=4J$ where the entire spin chain polarizes. 

    \item {\it Spontaneously broken $\mathscr{PT}$ symmetry}: When $\frac{1}{2}<\beta<1$, the ground state is a singlet where the bulk is a sea of singlets, and the boundary impurities are screened by bound modes formed at the impurity sites. Boundary excitations involve removal of the bound mode at one or both ends. The states containing an odd number of bound modes have complex energies whereas the ones containing an even number of bound modes have real energies.  When a global magnetic field is applied by adding a Zeeman term to the Hamiltonian, the local impurity magnetization no longer smoothly changes from $0$ to $0.5$ but rather has an abrupt jump at some value of magnetic field $h_c$. This abrupt jump in magnetic field shows that the bound mode is present in the ground state in this regime. \\
    
    Moreover, when $1<\beta<\frac{3}{2}$, the ground state is four-fold degenerate where both impurities are unscreened. Boundary excitations involve screening of the impurities because of the formation of bound modes at the impurity sites. The states containing either zero or two of bound modes have real energies, and those containing a single bound mode have complex energies. States featuring bound modes only at the left edge have corresponding partner states with bound modes only at the right edge, characterized by complex conjugate energies. The states with complex energies always appear in pair demonstrating that the $\mathscr{PT}-$symmetry is spontaneously broken in this phase.

\item {\it Effectively free local moment}: When $\beta>\frac{3}{2}$, the ground state is four-fold degenerate state where both impurities are unscreened. There are no excitations in this phase which can screen the impurities. In this phase, all states have real energies in the thermodynamic limit, illustrating the unbroken $\mathscr{PT}-$symmetry.
\end{itemize}

After the brief description of the result, we are now ready to present the Bethe Ansatz and DMRG results obtained for each phase.

\section{Bethe Ansatz and DMRG results}
We shall now proceed to discuss the Bethe Ansatz and DMRG results in much detail in this section. We shall describe the structure of the ground state and elementary excitations in all three boundary phases.
To make concrete quantitative comparison of the Bethe Ansatz and DMRG results, in the following we work at a fixed truncation error in the discarded singular values at $10^{-10}$ and we obtain the energy density from DMRG with non-Hermitian matrix product operators for various system sizes $(N=102,202,302,402,502)$. As there is always a ground state energy in this model, we fit the ground state energy density $E_{\mathrm{GS}}/N$ to determine the bulk and boundary contributions from the functional form
\begin{equation}
    E_{\mathrm{GS}}/N=E_\mathrm{B}+E_{\partial \mathrm{B}}/N,
    \label{formengdens}
\end{equation}
where the slope $E_{\partial \mathrm{B}}$ (as a function of $1/N$)  is the combined energy density contribution due to the open boundary and the interaction with the impurities. Henceforth, we refer to $E_{\partial \mathrm{B}}$ as the boundary part of energy density  and $E_\mathrm{B}$ is the energy density of the bulk. From the fit, we extract $E_{\partial \mathrm{B}}$ and $E_\mathrm{B}$ for various parameters and compare with the the exact result obtained for $N\to \infty$ from Bethe Ansatz (see Appendix \ref{DMRG} Fig.\ref{fig:b0p3g1} and Fig.\ref{extrapolation} for details).
In each boundary phase of the model, the exact value of $E_{\mathrm{B}}$ remains the same at $E_{\mathrm{B}}=1-4\ln(2)$, which is the energy density of the Hermitian XXX chain with periodic boundary conditions. On the other hand,  $E_{\partial \mathrm{B}}$ depends on each boundary phase and we present a detailed error estimate in the DMRG estimate of $E_{\partial \mathrm{B}}$ relative to its exact value we derive with Bethe ansatz in the following.
\subsection{The dissipative Kondo phase}
When the parameter $\beta$ takes the values between 0 and $\frac{1}{2}$, both impurities are screened by a dissipative multi particle Kondo clouds in the ground state.  As we shall see, all eigenstates in this phase have real eigenvalues. Thus, the ground state is defined in the usual sense as the state with lowest energy. The ground state is  a multiparticle singlet with real energy given by Eq.\eqref{gseng} which is calculated using both Bethe Ansatz and DMRG (see Appendix \ref{kphase} for details).        In the Kondo phase  $E_{\partial B}$ in the thermodynamic limit is
\begin{align}
  E_{\partial \mathrm{B}}=&2 \sum _{\alpha=\pm }\sum_{\delta=1}^2(-1)^{\delta+1} \Re\left(\psi ^{(0)}\left(\frac{\alpha  \beta }{2}+\frac{i \gamma }{2}+\frac{\delta}{2}\right)\right)\nonumber\\
  &+2 \Re\left(\frac{1}{1-(\beta +i \gamma )^2}\right)+\pi -3+3\log (4),
  \label{bexp}
\end{align}
where $\psi^0(z)$ is the logarithmic derivative of the gamma function also called as the digamma function. 

\begin{figure}
\includegraphics[width=1\linewidth]{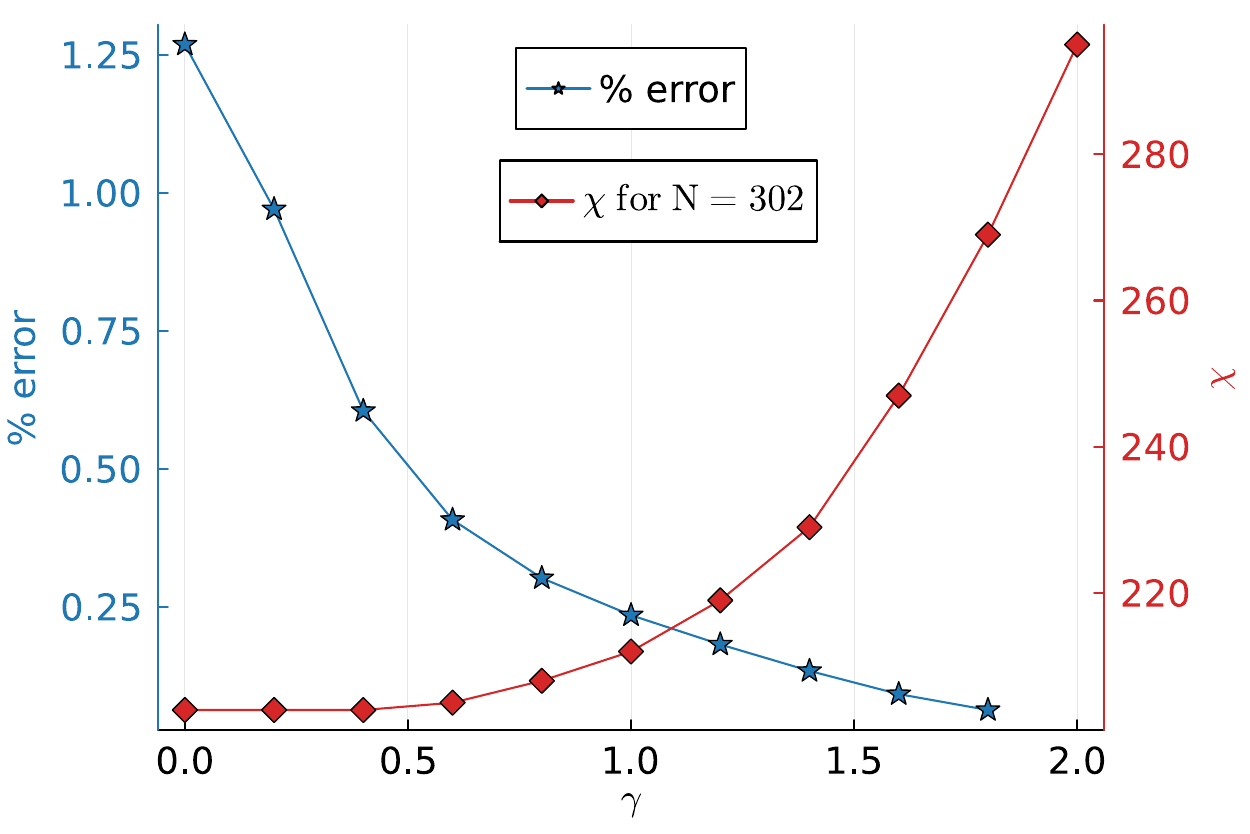}
    \caption{The relative error of the boundary part of the ground state energy density ($E_{\partial B}$) in the Kondo phase when $\beta=0.1$ and $\gamma$ is varied is extracted by fitting the energy density for $N=102,202,302,402$ and $502$ to Eq.~\eqref{formengdens}. The  relative error (left vertical axes in blue) between the DMRG and Bethe Ansatz result in Eq.~\eqref{bexp}. The bond dimension $\chi$ required for convergence (right vertical axes in red) with truncation cut-off set at $10^{-10}$ for $N=302$ for various values of $\gamma$ are shown in the right vertical axis in red. To ensure the convergence for each data points, 50 sweeps were performed. Notice that the bond dimension increases when $\gamma$ is increased. }
    \label{fig:diffplotKondo}
\end{figure}

We find that the bulk part of energy density $E_{ \mathrm{B}}$ obtained from the DRMG calculation and Bethe Ansatz result in thermodynamic limit has a relative percentage difference of about $5.0\times 10^{-4}\% $ for any value of $\gamma$ and $\beta$ within the Kondo phase. The boundary term $E_{\partial \mathrm{B}}$ suffers larger finite size effects and hence has a larger relative error compared to the bulk part.
The relative difference in the values of $E_{\partial \mathrm{B}}$ between the Bethe Ansatz result in Eq.~\eqref{bexp} and DMRG for a representative case of $\beta=0.1$ and various values of parameter $\gamma$ as shown in Fig.\ref{fig:diffplotKondo} along with the bond dimension required to attain convergence with the truncation cut off set $10^{-10}$ for $N=302$ sites.  We perform 50 sweeps to ensure convergence for every data points.
Importantly, as we show in Appendix \ref{herm-limit}, the quality of convergence in the boundary energy density is comparable to what we find in the same model in the limit of a Hermitian coupling showing that the DMRG does not suffer any loss of performance despite the non-Hermitian boundary couplings.

All the excited states are constructed by adding an even number of spinons, bulk strings, quartets etc \cite{destri1982analysis} all of which have real energies. Thus, all the excitations in this phase are completely real which shows that $\mathscr{PT}-$symmetry remains unbroken in this phase.

The non-Hermitian Kondo regime is characterized by a dynamically generated  complex scale $T_K$ \cite{kattel2024dissipation}. The real part of this temperature is the characteristic Kondo temperature, the dynamically generated energy scale below which  the impurities are screened by the Kondo effect and the imaginary part sets the scale for dynamically balanced loss and gain at the two edges of the chain.  The effect of this multiparticle screening can be seen in the enhancement of the local impurity density of states among other physical quantities. The ratio of the impurity and bulk contribution to the density of states is (explicitly computed in Appendix \ref{detsol})
\begin{equation}
     R(E)_\pm=\frac{N}{2}\frac{\rho_{\mathrm{imp}}(E)}{\rho_{\mathrm{bulk}}(E)}=
     \frac{4 \pi ^2 J^2 \cos (\pi  (\beta\pm i \gamma))}{E^2 \cos (2 \pi  (\beta\pm i \gamma))-E^2+8 \pi ^2 J^2},
     \label{redef}
\end{equation}
where the positive sign corresponds to left boundary impurity and the negative sign corresponds to the right boundary impurity. 

\begin{figure}
    \centering
    \includegraphics[width=\linewidth]{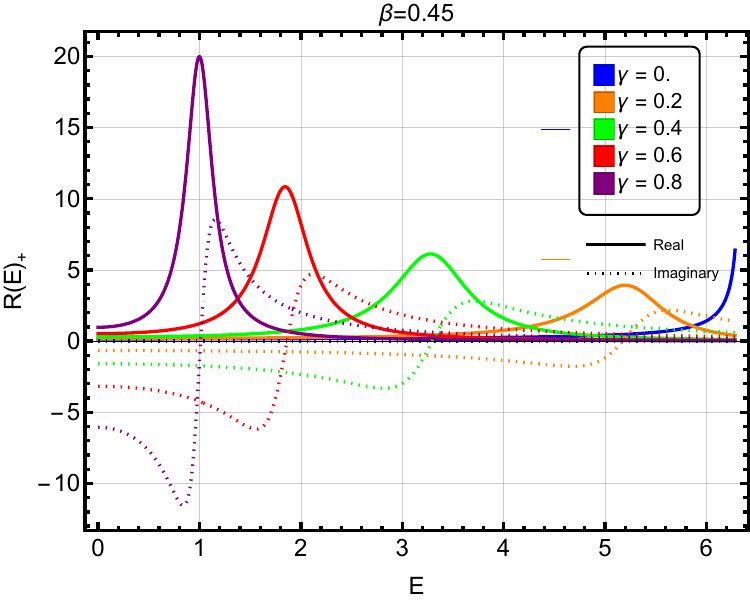}
    \caption{The plot shows that the spectral weight $R(E)_+$ given by Eq.\eqref{redef} of the spinons in the Kondo cloud is no longer situated at $E=0$ or $E=2\pi J$ as in Hermitian case \cite{kattel2023kondo}. The peak in the ratio of the impurity and bulk density of state shifts from $E=2\pi J$ at $\gamma=0$ and moves gradually to $E=0$ at large values of $\gamma$. Notice that the plot of $R(E)_-$ has the same real part as that of $R(E)_+$ but the sign of imaginary part is opposite.
    }
    \label{fig:REplus}
\end{figure}

Notice that even though the ground state is real because of the balanced gain and loss at the two boundaries, the physical quantities concerning only one boundary impurity is not real. Hence, the ratio of the density of states of the impurity to that of the bulk is not real for single impurity. This quantity is shown in Fig.\ref{fig:REplus} for various values of $\gamma$
and $\beta=0.45$. The shift in the density of states away from the Fermi level may be observed by STM measurements \cite{pcref}.

Defining the Kondo temperature as a characteristic energy scale below which the number of states is exactly half
of the total number of states associated with the impurity 
\textit{i.e} \cite{kattel2023kondo}
\begin{equation}
    \int_0^{T_K} d E \rho_{\mathrm{i m p}}(E)=\frac{1}{2} \int_0^{2\pi J} d E \rho_{\mathrm{i m p}}(E),
\end{equation}
where the impurity DOS is given by Eq.\eqref{impdos}
\begin{equation}
    { \rho_{\mathrm{imp}}}(E)_\pm= \frac{4 \pi  J^2 \cos (\pi  (\beta\pm i \gamma))/\sqrt{4 \pi ^2 J^2-E^2}}{ \left(E^2 \cos (2 \pi  (\beta\pm i \gamma))-E^2+8 \pi ^2 J^2\right)},
    \nonumber
\end{equation}
such that the Kondo temperature becomes
\begin{equation}
    {T_K}_\pm=\left(\frac{2\pi J}{\sqrt{1+\cos ^2(\pi (\beta\pm i\gamma)}}\right),
\end{equation}
where the $+$ve sign is for the impurity at the left end and the $-$ve sign is for the impurity situated at the right end of the chain. Due to the $\mathscr{PT}$ symmetry, the Kondo temperatures corresponding to the left and right impurities are complex conjugates of each other, reflecting the dynamically balanced loss and gain at the two edges of the chain.

\begin{figure}[h!]
    \begin{minipage}{0.47\textwidth}
        \includegraphics[width=\linewidth]{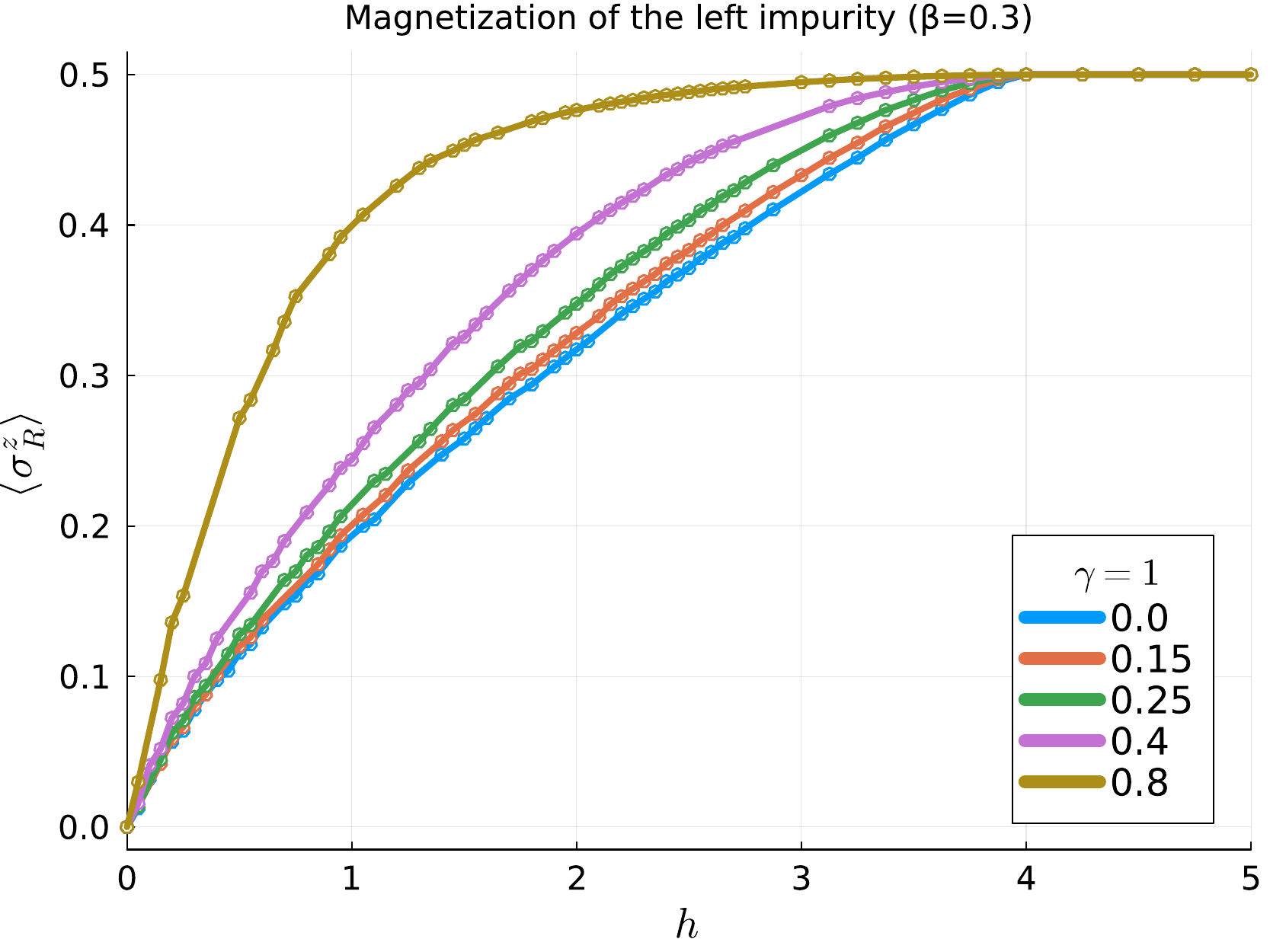}
    \end{minipage} \\
    \begin{minipage}{0.47\textwidth}
        \includegraphics[width=\linewidth]{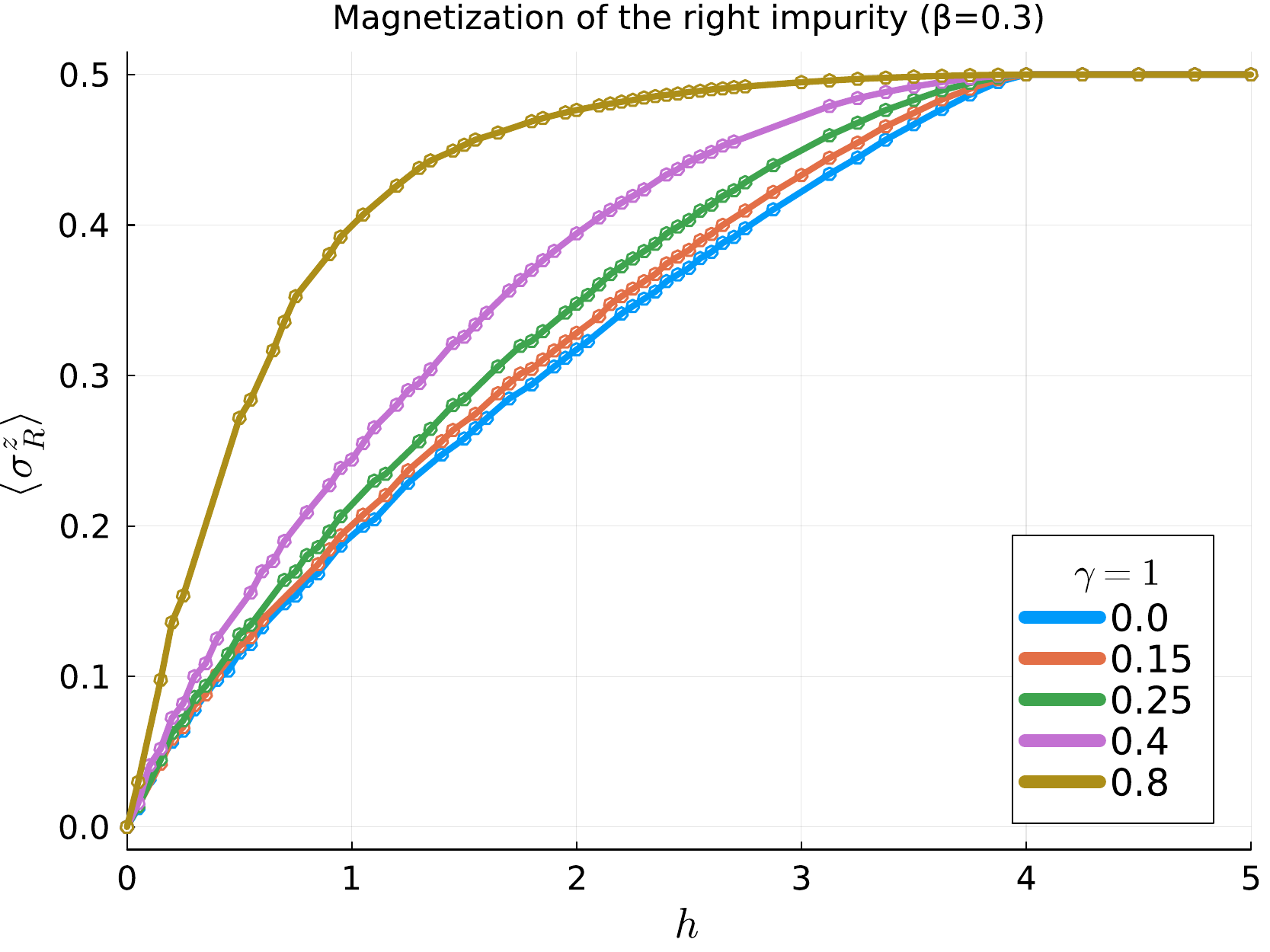}
    \end{minipage}%

\caption{Local impurity magnetization for the left and the right impurity when $\beta=0.3$ and $\gamma$ is varied is computed using DMRG. The local impurity magnetization is qualitatively the same for the left and right impurity and both of them show smooth crossover from 0 at $h=0$ to $\frac{1}{2}$ and $h=4J$ as which point the entire spin chain polarizes. All DMRG calculations are performed by turning on \textit{ishermitian=false} flag in the ITensor library \cite{fishman2022itensor} and setting the truncation cut-off at $10^{-10}$ and performing 80 sweeps to ensure convergence for all the data points. }
\label{fig:impmaglargeb}
\end{figure}

We shall now study the model in the presence of a global magnetic field by adding a Zeeman term
\begin{equation}
    \mathcal{H}_h=-h\sum_{j=1}^{N}\sigma_j^j -h\sigma^z_L-h\sigma^z_R
    \label{zeeman-term}
\end{equation}
to the Hamiltonian. In order to study the impurity physics in the presence of the magnetic field, we  shall use DMRG and study the local magnetization at the impurity site 
\begin{equation}
    M_{\mathrm{loc}}(h)=\langle \sigma_{R/L}^z\rangle
\label{eqn:Mloc}
\end{equation}
for a spin chain of size $\bar N=498$ bulk sites and 2 impurities (thus a spin chain of total $N=500$ sites) and perform DMRG using the ITensor library \cite{fishman2022itensor}. All DMRG calculations are performed with a truncation error cut-off of $10^{-10}$. Depending on the value of magnetic field and the boundary coupling, the calculation converges after different number of sweeps. Thus, to ensure the convergence for all ranges of parameter, we used 50 sweeps for all calculation.

In the Kondo phase, we fix $\beta$ between 0 and 0.5 and vary $\gamma$, and compute the local magnetization of the impurity. The plot of the impurity magnetization in Fig.\ref{fig:impmaglargeb} shows that it smoothly crosses over from 0 and $h=0$ to $\frac{1}{2}$ at finite field $h=4J$ exactly as in the Hermitian case \cite{kattel2023kondo} for all values of $\gamma$. Notice that both left and right impurity magnetization is qualitatively the same as shown in the representative figure below for $\beta=0.3$ which resembles the impurity magnetization behavior of the usual Kondo problem \cite{andrei1983solution}.

\subsection{Bound mode phases with spontaneously broken $\mathscr{PT}$ symmetry}
When $\frac{1}{2}<\beta<\frac{3}{2}$, a boundary bound mode appears in the spectrum of the model. The properties of the ground state in this phase depend on when $\beta$ is smaller or larger than $1$. Thus, we further divide the bound mode phase into two regimes: the bound mode phase I when $\frac{1}{2}<\beta<1$ where the bound mode has negative real part of the energy and hence it is contained in the ground state and the bound mode II phase when $1<\beta<\frac{3}{2}$ where the bound mode has positive real part of the energy and thus it  is not in the ground state. 
\subsubsection{Bound mode phase I}\label{bm1}

\begin{figure}
    \centering
    \includegraphics[width=\linewidth]{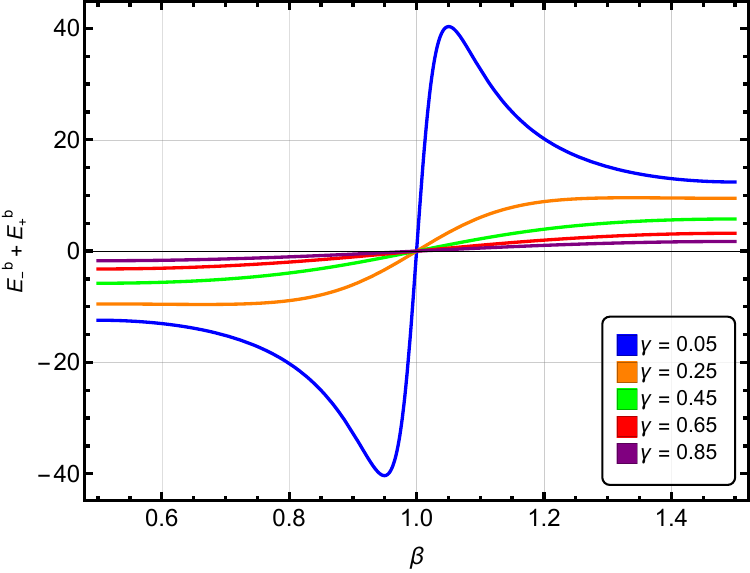}
    \caption{The sum of the energies of two bound modes with complex conjugate energies given by Eq.\eqref{bmengexp}. At the quantum critical point $\beta=1$, the energies of the bound mode diverge in the Hermitian limit $\gamma\to 0$ but any nonzero $\gamma$ leads to vanishing sum of the energies of the bound mode. As $\gamma$ increases, the energy of bound mode becomes smaller. Moreover, in the Hermitian limit, the bound mode energies are always greater than the maximum energy of a single spinon. However, in the presence of a large $\gamma$ parameter, the bound mode energy can be smaller than the energy of a single spinon.}
    \label{fig:reebsum}
\end{figure}

When $\frac{1}{2}<\beta<1$, both impurities are screened in the ground state by two bound modes formed at their respective edges. These bound modes are described by the two boundary string solutions of the Bethe Ansatz equation 
\begin{equation}
    \mu^b_\pm=\pm \gamma\pm i \left(\frac{1}{2}-\beta\right)
\end{equation}
and they have energy
\begin{equation}
    E^b_{\pm}=-2\pi \csc(\pi(\beta\pm i \gamma)).
    \label{bmengexp}
\end{equation}
Notice that in the Hermitian case, the boundary string solutions are purely imaginary solutions of the Bethe equation with real energy that describe the boundary excitations \cite{kapustin1996surface,pasnoori2020kondo,pasnoori2021boundary,pasnoori2022rise,kattel2023exact,rylands2020exact,wang1997exact,frahm1997open}. Here, the boundary string solutions are complex and they describe the dissipative boundary excitations in the model as the energy eigenvalues of these solutions are complex. The sum of the energies of two bound modes is real and negative as shown in Fig.\ref{fig:reebsum}.

\begin{figure}
    \centering
    \includegraphics[width=1\linewidth]{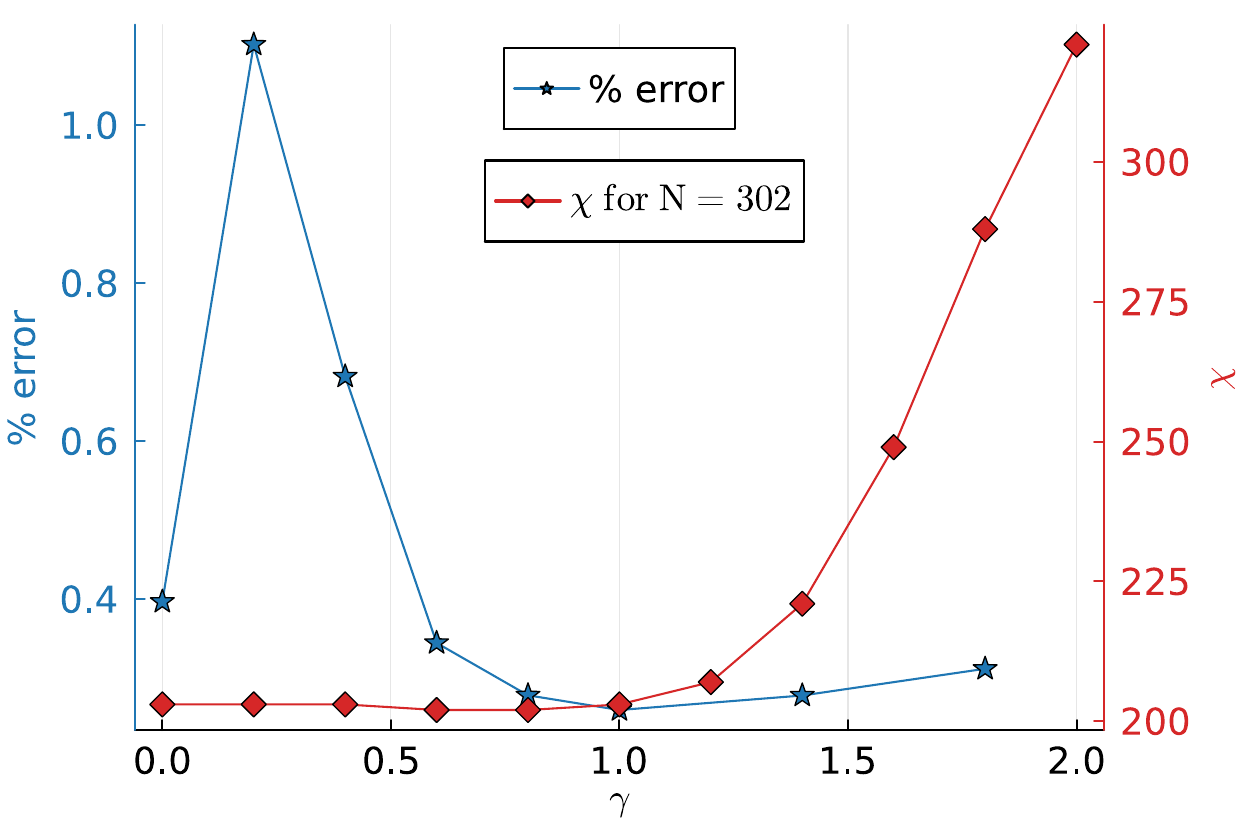} 
    \caption{The relative difference between the boundary part of the ground state energy density in the bound mode phase I when $\beta=0.6$ and $\gamma$ is varied is shown in the left vertical axis in blue. The boundary part of energy density is extracted via DMRG by fitting the energy density for $N=102,202,302,402$ and $502$ and the exact Bethe Ansatz result in thermodynamic limit is given by Eq.~\eqref{bexp}. Moreover, the bond dimension required for convergence with truncation cut-off set at $10^{-10}$ for $N=302$ for various values of $\gamma$ are shown in the right vertical axis in red. Just like in the Kondo phase, the bond dimension increases when $\gamma$ is increased.}
    \label{fig:GSbm1phaseDMRG}
\end{figure}

 Thus, the ground state contains both boundary strings such that the ground state energy is real as given by Eq.\eqref{gsengbm1phase}. The boundary part of the energy density (i.e. $a$ introduced in Eq.~\eqref{formengdens} and explicitly given by Eq.~\eqref{formengdens}) is computed from both Bethe Ansatz and DMRG and the relative difference between the two methods is shown in Fig.\ref{fig:GSbm1phaseDMRG} versus $\gamma$ for the representative case of $\beta=0.6$. The bond dimension required for the convergence when the truncation cut-off is set at $10^{-10}$ for $N=302$ total number of sites is also shown in Fig.\ref{fig:GSbm1phaseDMRG}.

In the Hermitian limit $\gamma\to 0$, the energies of the bound mode diverge when $\beta\to 1$. But any  non-zero $\gamma$, the real part of the energy of the bound mode vanishes at $\beta=1$. Recall that the bulk excitations (spinon) can have energies in the range $0<E_\theta<2\pi J$ and in the Hermitian limit, the energy of the bound mode is in the range $E^b\geq 2\pi J$ which shows the clear demarcation between the bulk and boundary excitations. However, as shown in Fig.\ref{fig:reebsum}, when $\gamma\neq 0$, the boundary excitation can also have any energy $E^b>0$, thereby making the demarcation of the bulk and boundary excitation flimsy. This behavior was also seen in the case of complex boundary field studied in \cite{kattel2023exact}. For any fixed $\beta$, $\Re(E_b)\to 0$, when $\gamma\to \infty$. Thus, for large $\gamma$, the bound mode ceases to exist and hence, the difference between the bound mode and unscreened phases becomes negligible.  

Since the sum of the energies of the two boundary strings are negative when $\frac{1}{2}<\beta<1$, the ground state contains these bound modes. The ratio of the density of states contribution of impurity to bulk $R_b(E)_\pm$ becomes
\begin{equation}
   R_b(E)_\pm= R(E)_\pm +\delta(E-E^b_-)+\delta(E-E^b_+),
\end{equation}
where $R(E)_\pm$ is given by Eq.\eqref{redef}. In this regime  real part of the $R(E)_\pm$ is negative and hence the only positive contribution to the screening of the impurity comes from the two bound modes each located at their respective edges of the spin chain. 

\begin{figure}
    \begin{tikzpicture}[scale=.95, every node/.style={scale=0.95}]
\node at (0,-0.5) {{1}};
\node at (2,-0.5) {{2}};
\node at (4,-0.5) {{3}};
\node at (6,-0.5) {{4}};

\node at (-1,8.5) {{$E$-$E_0$}};
\node at (6.75,0.0) {{$\mathrm{set}$}};

\draw [thick,DarkGreen, <->] (-1,8)--(-1,0)--(6.5,0);
\draw [dashed,DarkGreen] (-1,2.0)--(2,2.0);
\draw [dashed,DarkGreen] (-1,2.0)--(4,2.0);
\draw [dashed,DarkGreen] (-1,3.0)--(6,3.0);

\node at (-1.3,1) {\small{$E^b_+$}};
\node at (-1.5,2) {\small{$E^b_-,E^b_+$}};
\node at (-1.4,3.25) {\small{$2\Re E^b_+$}};

\draw [ultra thick,BrickRed] (0,0)--(0,5);
\draw [ultra thick,BrickRed] (2,2)--(2,7); 
\draw [ultra thick,BrickRed] (4,2)--(4,7); 
\draw [ultra thick,BrickRed] (6,3)--(6,8); 

\foreach \y in {0, 0.1, 0.2, 0.4, 0.8, 1.5, 2.9, 4.7}
    {
        \draw[thick,Navy] (-0.5,\y)--(0.5,\y);
    }
    
\foreach \y in {2.0, 2.1, 2.25, 2.5, 2.85, 3.65, 3.5, 4.5, 6.2, 6.8}
    {
        \draw[thick,Navy] (1.5,\y)--(2.5,\y);
    }

\foreach \y in {2.0, 2.1, 2.25, 2.5, 2.85, 3.65, 3.5, 4.5, 6.2, 6.8}
    {
        \draw[thick,Navy] (3.5,\y)--(4.5,\y);
    }

\foreach \y in {3.0, 3.1, 3.2, 3.4, 3.8, 4.5, 6.9, 6.7, 7.9, 6.0}
    {
        \draw[thick,Navy] (5.5,\y)--(6.5,\y);
    }
\end{tikzpicture}

    \caption{Schematic representation of four different sets of excited states. All the states in set 1 and set 4 have real energies whereas the states in set 2 and set 3 have complex energies that are conjugate to each other. Both impurities are screened by two bound modes formed at the two edges in the states lying in set 1 and both impurities are unscreened in the states in set 4. One impurity is screened by the bound mode whereas another is unscreened in the states contained in set 2 and set 3. Each sets of excited state is obtained by adding even number of spinons, strings, quartets etc to the four states described below in \ref{bothscreened},\ref{leftscreened}, \ref{rightscreened}, and \ref{bothunscreened} respectively. }
    \label{fig:ESBM-BM}
\end{figure}
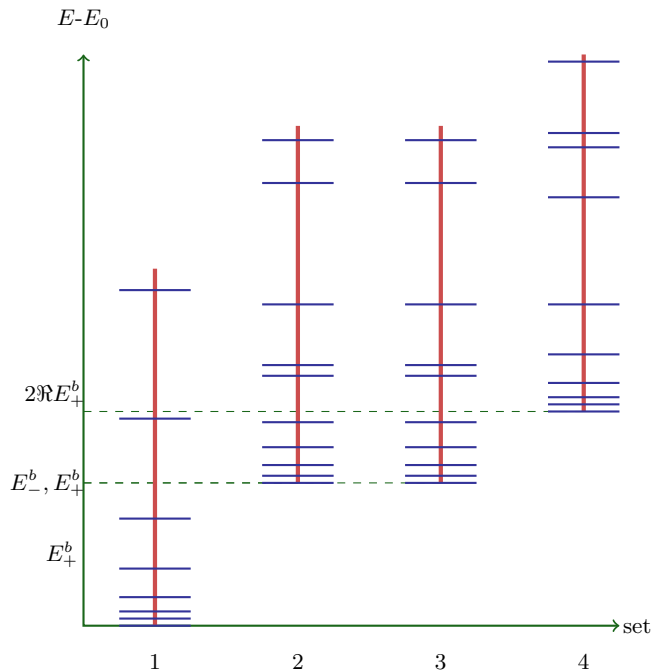

We can construct excited state solutions by removing one boundary string and adding a zero-energy spinon or by removing both boundary string solutions. Thus, there are four distinct kinds of elementary states that can be sorted into four distinct sets shown in the schematic Fig.\ref{fig:ESBM-BM} described below:
\begin{enumerate}
\item A unique state where each impurity is screened by a bound mode formed at the respective edge, which possesses real energy\label{bothscreened}.
\item A four-fold degenerate state with the left impurity screened by a bound mode, the right impurity unscreened, and a spinon propagating in the bulk. This state has complex energy.
\label{leftscreened}
\item A four-fold degenerate state with the right impurity screened by a bound mode, the left impurity unscreened, and a spinon propagating in the bulk. This state also has complex energy, with its energy being the complex conjugate of the state described in \ref{leftscreened}.\label{rightscreened}
\item A four-fold degenerate state with both impurities unscreened, which possesses real energy.\label{bothunscreened}
\end{enumerate}

  \begin{figure}[h!]
    \begin{minipage}{0.47\textwidth}
        \includegraphics[width=\linewidth]{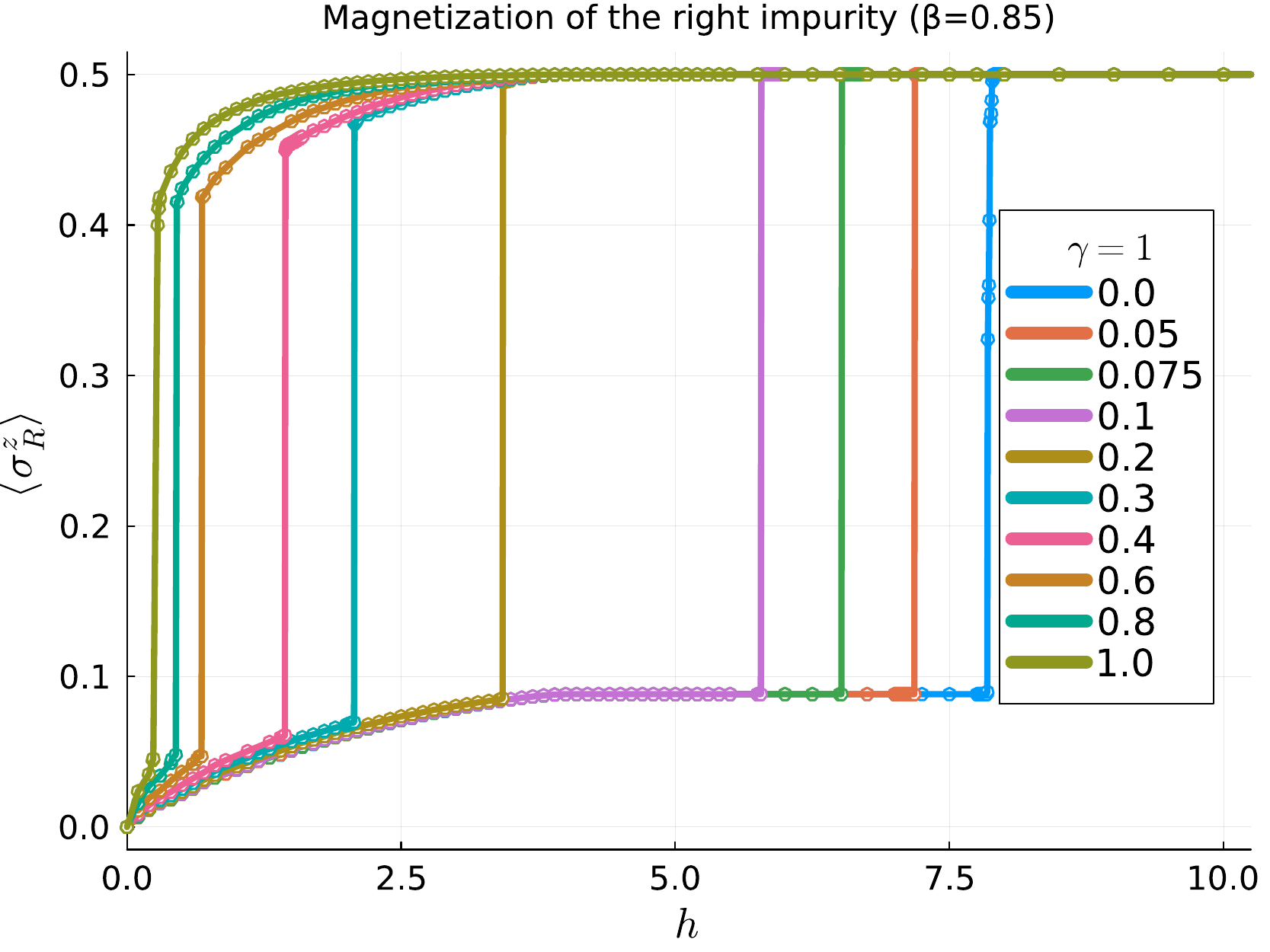}
    \end{minipage} \\
    \begin{minipage}{0.47\textwidth}
        \includegraphics[width=\linewidth]{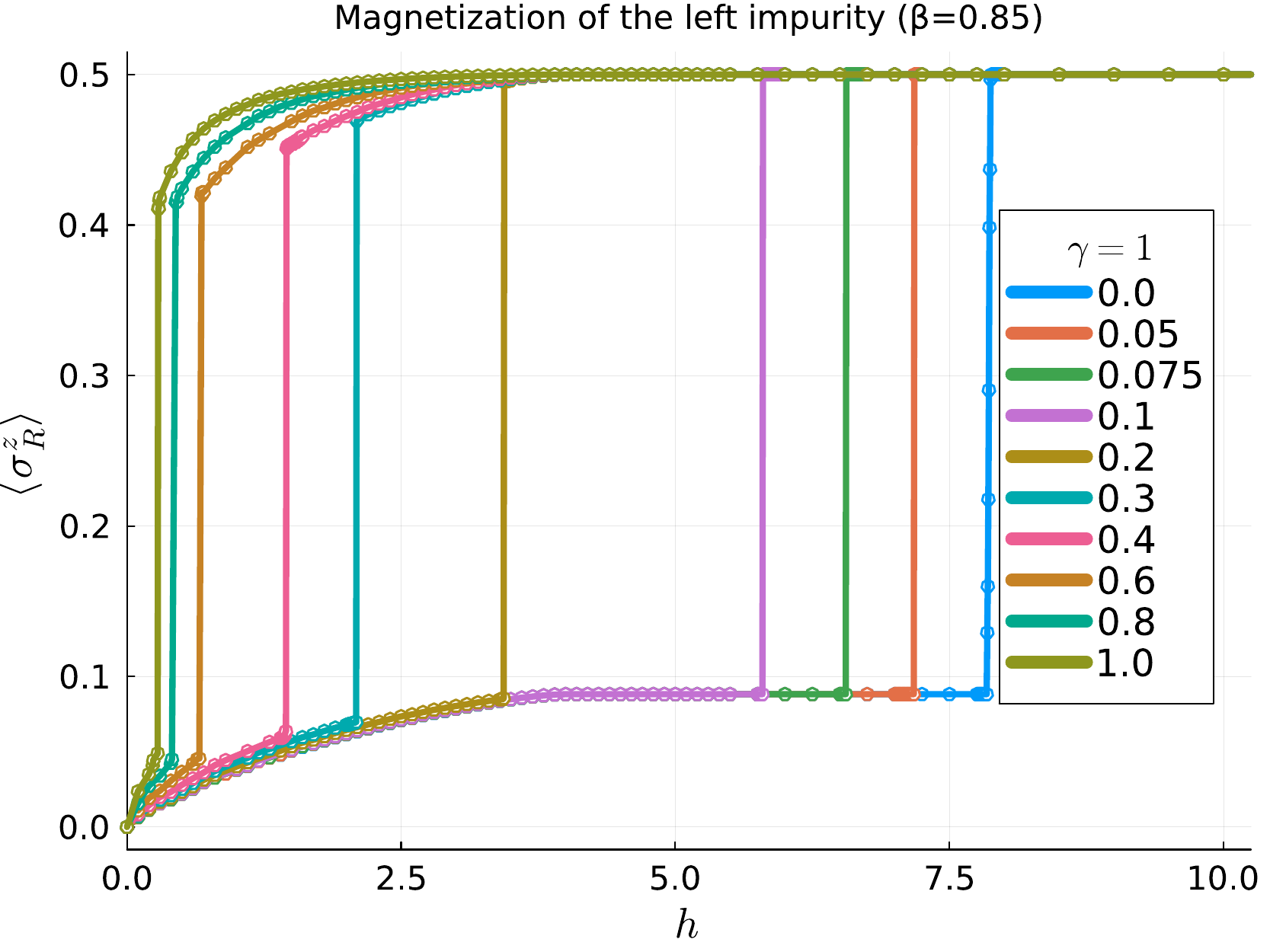}
    \end{minipage}%
\caption{Local impurity magnetization for the left and the right impurity when $\beta=0.85$ and $\gamma$ is varied. The local impurity magnetization is qualitatively the same for the left and right impurity. The impurity magnetization for smaller values of $\gamma$ grow till $h=4J$ and than saturate before finally jumping to 0.5 at some critical value of field whereas for larger value of $\gamma$, the jump occurs before $h=4J$ because the real part of the energy is small when $\gamma$ is increased. All DMRG calculations are performed by turning on \textit{ishermitian=false} flag in the ITensor library \cite{fishman2022itensor} and setting the truncation cut-off at $10^{-10}$ and performing 80 sweeps to ensure convergence for all the data points.}\label{fig:impmaglargeb2}
\end{figure}

\begin{figure*}[ht!]
    \begin{minipage}{0.33\textwidth}
        \includegraphics[height=\linewidth]{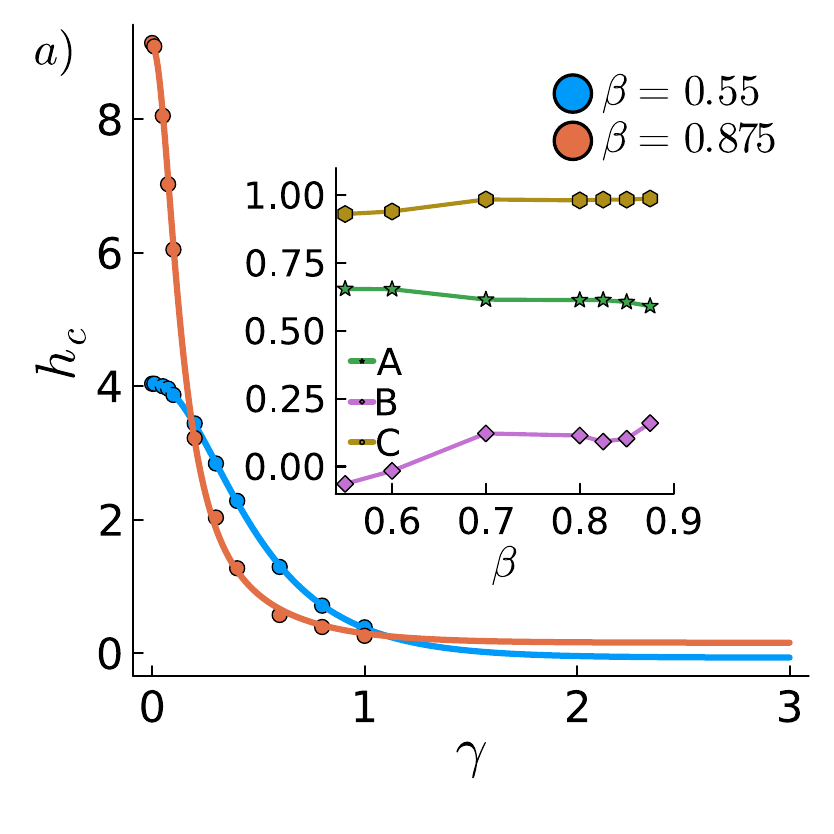}
    \end{minipage}%
    \begin{minipage}{0.33\textwidth}
        \includegraphics[height=\linewidth]{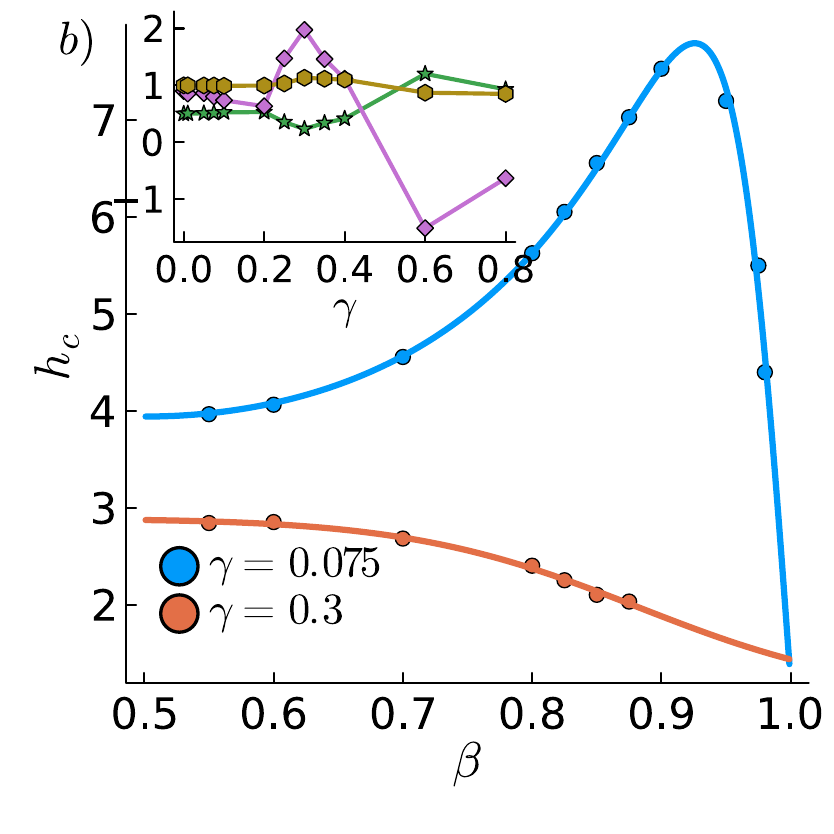}
    \end{minipage}%
    \begin{minipage}{0.33\textwidth}
        \includegraphics[height=\linewidth]{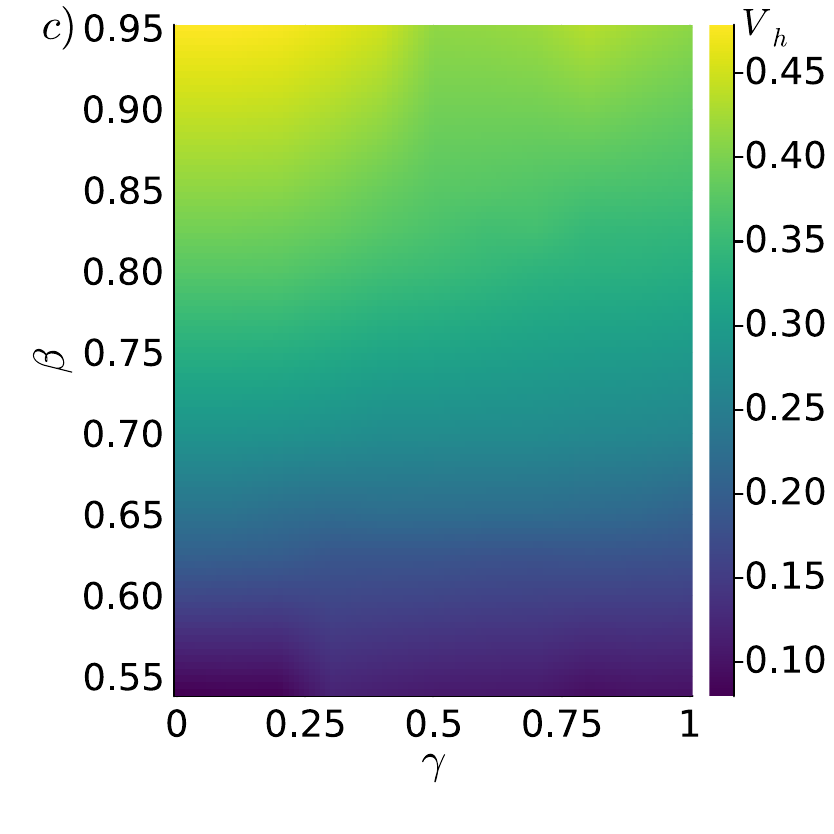}
    \end{minipage}
\caption{In the bound mode phase I, the local magnetization at the impurity sites jump a function of global magnetic field. The critical value of field $h_c$ at which magnetization jumps is expressed as Eq.\eqref{fitfunc}. a) Representative fitted curve showing the critical field for fixed values of $\beta=0.55$ and $\beta=0.875$ as $\gamma$ is varied. The inset shows the fitted parameters for any fixed value of $\gamma$ when $\beta$ is varied. b)Representative fitted curve showing the critical field for fixed values of $\gamma=0.075$ and $\gamma=0.3$ as $\beta$ is varied.  The inset shows the fitted parameters for any fixed value of $\beta$ when $\gamma$ is varied.  c) The vertical jump ($V_h=\lim_{\epsilon\to 0}M_{\mathrm{loc}}(h+\epsilon)-M_{\mathrm{loc}}(h-\epsilon)$) for for various values of $\gamma$ and $\beta$ is shown in the heatmap. Since the bound mode forms only when $\beta>\frac{1}{2}$, there is no jump when $\beta<\frac{1}{2}$ and as $\beta\to 1$ and $\gamma\to 0$, $V_h\to \frac{1}{2}$. 
}
    \label{fitfig}
\end{figure*}

To each of these states, we can add an even number of spinons, bulk strings, quartets and higher order boundary strings to form four distinct sets of excited states which are characterized by the number of the bound modes at two edges. All the states in the first set have two bound modes  whereas the ones in fourth set have no bound modes. All the states in both the first and fourth sets have real energies and an even number of spinons. The states in the second set have only one bound mode at the left end and all the states have complex energies. Moreover, all the states in the third set have only one bound mode at the right end and all the states have complex energies that are complex conjugate to the energies of the states in second set. The total number of spinons in each state in second and third set is odd. The presence of the states in pair with complex conjugate energy is the signature of the spontaneously broken $\mathscr{PT}-$symmetry. It is important to remember that only for smaller value of $\gamma$ where the real part of the boundary string given by Eq.\eqref{bmengexp} is sufficiently bigger than $0$, the distinction between the sets of eigenstates becomes apparent but when $\gamma\to 0$, the real part of the bound mode energy $\Re E_b$ approaches $0$ as shown in Fig.\ref{fig:reebsum} and hence experimentally detecting the effect of bound mode becomes difficult. Thus, for large $\gamma$, the demarcation between the different sets of excited states becomes rickety. In the Hermitian limit, since the boundary bound mode always have energy greater than the maximum energy of the spinon $2\pi J$, the four sets form a well defined towers of excited states \cite{kattel2023kondo,PPSPhysRevB.107.224412}.

We shall now study the model in the presence of global magnetic field by adding Zeeman term given by Eq.\eqref{zeeman-term}. In the bound mode phase I, by taking $\frac12<\beta<1$ and varying $\gamma$, we compute the local impurity magnetization.  Because of the presence of the boundary bound modes in the ground state, the local impurity magnetization abruptly jumps in in the bound mode phase just like in the Hermitian case \cite{kattel2023kondo}. Both left and right impurity magnetizations are qualitatively the same as shown in the representative figure Fig.\ref{fig:impmaglargeb2} for model parameter $\beta=0.85$ and various values of $\gamma$.

 It is well known that the periodic Heisenberg chain is fully polarized when $h=4J$. Here, we see an interesting trend that if the critical value of the field $h_c>4J$, then the local impurity magnetization smoothly increases from $0$ at $h=0$ and then attains a finite value at $h=4J$ up until which there are bulk degrees of freedom to polarize, then the magnetization becomes constant until $h_c$ where it suddenly jumps to $0.5$ as shown in Fig.\ref{fig:impmaglargeb2}. This sudden jump is because there are just two unpolarized bound modes left in the model above $h=4J$ and when the energies of these modes in the presence of the magnetic field are equal to the applied field, the modes flip thereby unscreening both impurities. However, if $h_c<4J$, then the magnetization increases smoothly till $h_c$, and makes a finite jump due to the contribution from the boundary string solution. This time, however, there are sill some bulk degrees of freedom to excite, thus the magnetization does not reach 0.5 after the jump but reaches some finite values. When $h$ is increased above $h_c$, the impurity magnetization further smoothly increases to $0.5$ at $h=4J$ at this point the impurity is completely unscreened.

For a fixed value of $\gamma$, if $\beta$ is varied or if $\beta$ is fixed and $\gamma$ is varied, then the value of the critical field $h_c$ at which the impurity magnetization jumps can be fitted to a function of the form
\begin{equation}
    h_c=\frac{2\pi A}{\sin(\pi C(\beta+i\gamma)}+B
    \label{fitfunc},
\end{equation}
where the parameters for various values of $\beta$ in the bound mode phase I are shown in Fig. \ref{fitfig}~a). Likewise, we fit the critical field $h_c$ for fixed $\beta$ when $\gamma$ is varied in Fig.\ref{fitfig}~b). Unlike in the Hermitian case, the relation between the critical field and the complex eigenvalue of the bound mode seems to be more complicated in this case. For example, the vertical jump $V_h$ i.e. the magnetization difference at the critical value of field  $\lim_{\epsilon\to 0}\braket{\sigma^z_L}(h_c+\epsilon)-\braket{\sigma^z_L}(h_c-\epsilon)$ for fixed $\beta$ as a function of parameter $\gamma$ is non-trivial as shown in Fig.\ref{fitfig}~c).

\subsubsection{Bound mode phase II}
When $1<\beta<\frac{3}{2}$, the sum of the energies of the two bound modes is positive as shown in Fig.\ref{fig:reebsum}. Hence the ground state is the state that does not contain any boundary strings. The ground state is thus constructed by adding all the real roots of Bethe Ansatz equations whose energy is explicitly given in Eq.~\eqref{GSengBM2phase}. Just as in the Kondo and bound mode phases, we write the energy as the bulk and boundary part given by Eq.\eqref{formengdens}. 
The boundary part of the energy density $E_{\partial \mathrm{B}}$ is now given by
\begin{align}
    E_{\partial \mathrm{B}}&=\pi -3+3 \log (4)+\Re\left(\frac{2}{1-(\beta +i \gamma )^2}-\frac{4}{\beta +i \gamma }\right)\nonumber\\
    &4 \Re\left(\psi ^{(0)}\left(\frac{1}{2} (\beta +i \gamma )+\frac{1}{2}\right)-\psi ^{(0)}\left(\frac{1}{2} (\beta +i \gamma )\right)\right).
    \label{bexpbm2usp}
\end{align}

\begin{figure}[H]
    \centering
    \includegraphics[width=1\linewidth]{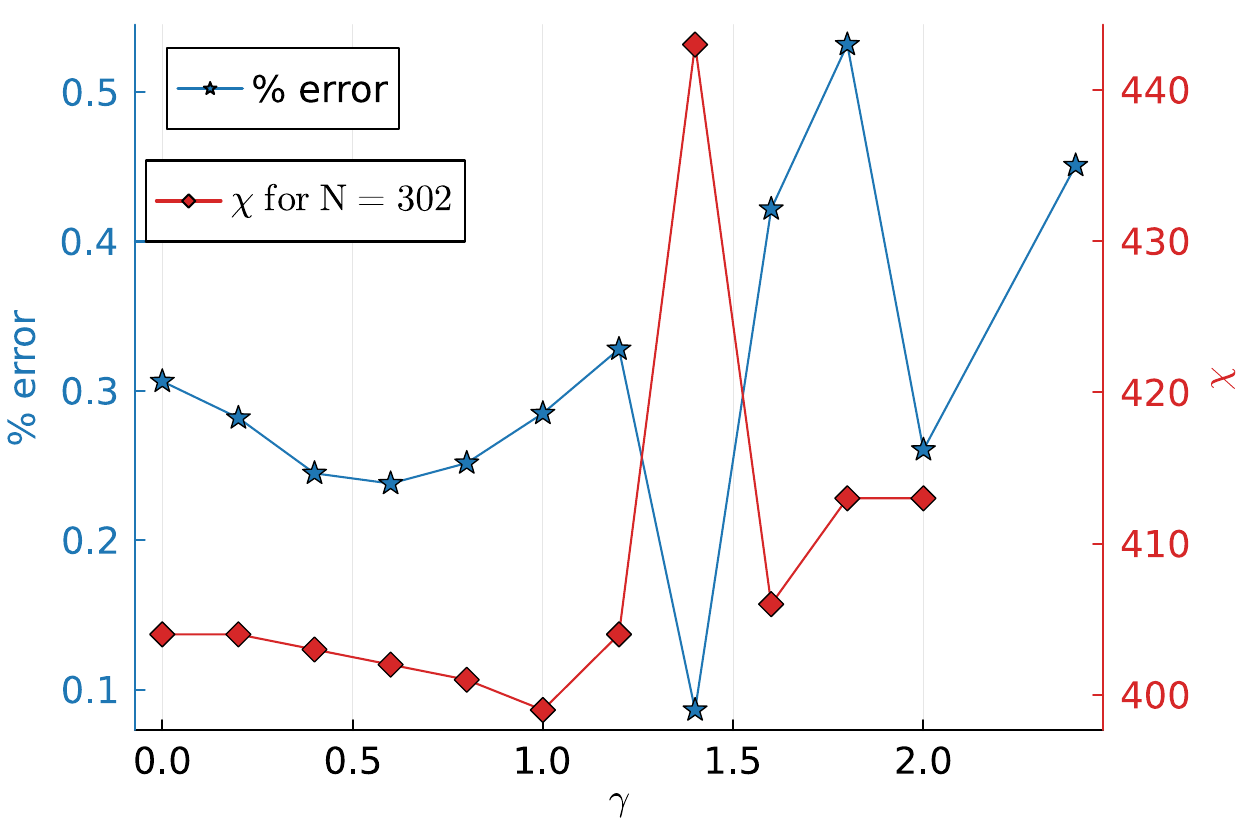} 
    \caption{The relative difference between the the boundary and impurity contribution to the ground state energy density in the bound mode II phase obtained from DMRG and Bethe Ansatz  when $\beta=1.3$ and $\gamma$ is varied is shown in the left vertical axis in blue. The DMRG result is obtained by fitting the energy density for $N=102, 202, 302$ and $402$ and the exact result in thermodynamic limit obtained via Bethe Ansatz is given in Eq.\eqref{bexpbm2usp}. Notice that the bond dimension required for convergence with truncation cut-off set at $10^{-10}$ for N=302 is shown in right vertical axis in red color. The bond dimension does not depend that strongly on the model parameter $\gamma$ in this phase.}
    \label{fig:GSbm2phaseDMRG}
\end{figure}

The boundary part of energy density $b$ extracted from DMRG by fitting data for various system size ($N=102,202,302$ and $402$) is plotted in Fig.\ref{fig:GSbm2phaseDMRG} for representative parameter $\beta=1.4$ as a function of the model parameter $\gamma$.

In the Hermitian limit where $\gamma\to 0$ and $\beta>1$, the sign of the impurity coupling changes from antiferromagnetic to ferromagnetic. However, when  $\gamma\neq 0$, the sign (of the real part) of the coupling varies based on the values of both $\beta$ and $\gamma$. This variation in sign is not indicative of a phase transition. The phase transition point is determined solely by the parameter $\beta$, regardless of where the coupling changes sign.

The same four distinct states listed in Bound mode phase I described in Sec.\ref{bm1} exist in this phase. But the sum of the energies of the two bound modes at the edges is positive as shown in Fig.\ref{fig:reebsum}. Thus energy of the state with no bound modes on both edges is smaller than the energy of the state with contain both bound modes. Thus, both impurities are unscreened in the ground state but there are excited state where one or both boundary impurities are screened. Just as in Bound mode regime I, there are four distinct sets of excited states 
\begin{enumerate}
    \item The set of excited states that contain states with no bound modes at both ends have all the states with real energies.  Both impurities are unscreened in each state in this set and each state have an even number of spinons.

    \item The set of excited states where all states have left impurity screened by bound mode but right impurity is unscreened. There are odd number of spinons in each state in this set and all states have complex energies.\label{tower2}

    \item The set of excited states where all states have right impurity screened by bound mode but left impurity is unscreened. There are odd number of spinons in in each state in this set and all states have complex energies which are complex conjugates of the energies of states described in \ref{tower2}.

    \item The set of excited states that contain states with two bound modes at each end have all the states with real energies.  Both impurities are screened in each state in this set and each state have an even number of spinons.
\end{enumerate}

\subsection{Unscreened local moment phase}
\begin{figure}
    \centering
    \includegraphics[width=1\linewidth]{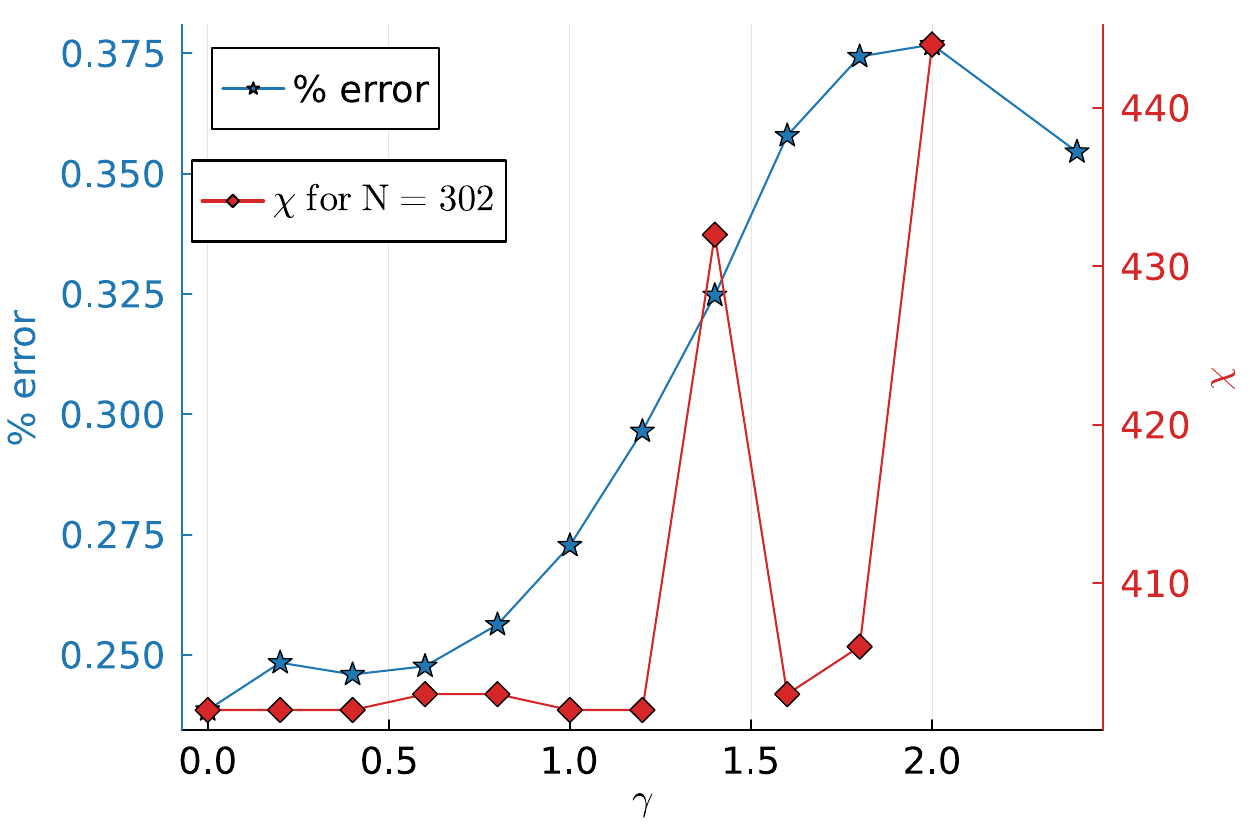} 
    \caption{The relative difference between the the boundary and impurity contribution to the ground state energy density in the bound mode II phase obtained from DMRG and Bethe Ansatz  when $\beta=1.8$ and $\gamma$ is varied is shown in the blue in the left vertical axis. The DMRG result is obtained by fitting the energy density for $N=102, 202, 302$ and $402$ and the exact result in thermodynamic limit obtained via Bethe Ansatz is given in Eq.\eqref{bexpbm2usp}. The bond dimension required for convergence with truncation cut-off set at $10^{-10}$ for N=302 is shown in the red color in right vertical axis. The bond dimension does not depend that strongly on the model parameter $\gamma$ in this phase.}
    \label{fig:GSUSphaseDMRG}
\end{figure}

When $\beta>\frac{3}{2}$, the impurities can not be screened in any eigenstates in this phase. Both impurities are unscreened in the ground state and hence the ground state is four-fold degenerate in the thermodynamic limit with real energy given by Eq.\eqref{GSengBM2phase}. The boundary part of the energy density $E_{\partial\mathrm{B}}$ is given by Eq.\eqref{bexpbm2usp}. We extract $E_{\partial\mathrm{B}}$ from DMRG by fitting the energy for various system size to Eq.\eqref{formengdens} and plot the relative difference between the result obtained from Bethe Ansatz in thermodynamic limit to the result obtained from DMRG in Fig.\ref{fig:GSUSphaseDMRG} for $\beta=1.8$ as a function of $\gamma$. 
All the excited states can be constructed by adding an even number of spinons, bulk strings, quartets etc. all of which have real energies. Thus, all the excited states in this phase have real energies and an even number of spinons. 


\section{Conclusion and Outlook}
We considered the integrable Heisenberg spin chain with complex boundary impurity and studied it using a combination of Bethe Ansatz and DMRG techniques. We showed that $\beta$, a complicated function of the bulk coupling $J$, the absolute value of the boundary coupling $|J_{\mathrm{imp}}|$ and the real part of the boundary coupling $\mathrm{Re}(J_{\mathrm{imp}})$ given in Eq.\eqref{betadef}, governs the phase transitions. There are four distinct phases in the model. When $0<\beta<\frac{1}{2}$, the model is in the Kondo phase where both impurities are screened by multiparticle Kondo cloud characterized by a peak in local density of states. When $\frac{1}{2}<\beta<1$, the model is in the bound mode phase I where both impurities are screened by bound modes formed at each impurity site in the ground state. There are, however, excited states where impurities can be unscreened by removing the bound mode solution.  When $\beta$ increases to the range $1<\beta<\frac{3}{2}$, the model is in the bound mode II phase, where the impurities are unscreened in the ground state, but they can be screened by bound modes formed at the impurity sites in the excited states. Finally, when the parameter $\beta>\frac{3}{2}$, the impurities cannot be screened in any eigenstates, and therefore the model is in the unscreened phase. 

In the Kondo and unscreened phases, all the eigenstates have real energies, which shows that the $\mathscr{PT}-$symmetry is unbroken in these phases. However, in the two bound mode phases, there are states with real energies and complex energies. For each state with complex energy, there is a state with complex conjugate energy which shows that the $\mathscr{PT}-$symmetry is spontaneously broken in these phases. 

As one moves across the phase transition point between the Kondo and bound mode phase I, the ground state undergoes a change where the two multiparticle Kondo clouds which screen the two impurities in the Kondo phase turn into two bound modes which screen the respective impurities. The eigenstates also undergo drastic changes reorganizing the entire Hilbert space as states with all real energies divide into four distinct sets of excited states: two with real energies and two with complex energies. 

We showed that the ITensor library's DMRG algorithm works well for an fully interacting many-body non-Hermitian model by comparing the ground state energy with exact Bethe Ansatz result and then used DMRG to compute local impurity magnetization. A through study of benchmarking the non-Hermitian DMRG will be a subject of upcoming publication.

\section{Acknowledgement}
We thank Miles Stoudenmire for helpful suggestions and discussions. J.H.P. is partially supported by the Army Research Office Grant No.~W911NF-23-1-0144 and NSF Career Grant No.~DMR- 1941569.

\bibliography{ref}

\begin{thebibliography}{72}%
\makeatletter
\providecommand \@ifxundefined [1]{%
 \@ifx{#1\undefined}
}%
\providecommand \@ifnum [1]{%
 \ifnum #1\expandafter \@firstoftwo
 \else \expandafter \@secondoftwo
 \fi
}%
\providecommand \@ifx [1]{%
 \ifx #1\expandafter \@firstoftwo
 \else \expandafter \@secondoftwo
 \fi
}%
\providecommand \natexlab [1]{#1}%
\providecommand \enquote  [1]{``#1''}%
\providecommand \bibnamefont  [1]{#1}%
\providecommand \bibfnamefont [1]{#1}%
\providecommand \citenamefont [1]{#1}%
\providecommand \href@noop [0]{\@secondoftwo}%
\providecommand \href [0]{\begingroup \@sanitize@url \@href}%
\providecommand \@href[1]{\@@startlink{#1}\@@href}%
\providecommand \@@href[1]{\endgroup#1\@@endlink}%
\providecommand \@sanitize@url [0]{\catcode `\\12\catcode `\$12\catcode
  `\&12\catcode `\#12\catcode `\^12\catcode `\_12\catcode `\%12\relax}%
\providecommand \@@startlink[1]{}%
\providecommand \@@endlink[0]{}%
\providecommand \url  [0]{\begingroup\@sanitize@url \@url }%
\providecommand \@url [1]{\endgroup\@href {#1}{\urlprefix }}%
\providecommand \urlprefix  [0]{URL }%
\providecommand \Eprint [0]{\href }%
\providecommand \doibase [0]{https://doi.org/}%
\providecommand \selectlanguage [0]{\@gobble}%
\providecommand \bibinfo  [0]{\@secondoftwo}%
\providecommand \bibfield  [0]{\@secondoftwo}%
\providecommand \translation [1]{[#1]}%
\providecommand \BibitemOpen [0]{}%
\providecommand \bibitemStop [0]{}%
\providecommand \bibitemNoStop [0]{.\EOS\space}%
\providecommand \EOS [0]{\spacefactor3000\relax}%
\providecommand \BibitemShut  [1]{\csname bibitem#1\endcsname}%
\let\auto@bib@innerbib\@empty
\bibitem [{\citenamefont {Lindblad}(1976)}]{lindblad1976generators}%
  \BibitemOpen
  \bibfield  {author} {\bibinfo {author} {\bibfnamefont {G.}~\bibnamefont
  {Lindblad}},\ }\bibfield  {title} {\bibinfo {title} {On the generators of
  quantum dynamical semigroups},\ }\href@noop {} {\bibfield  {journal}
  {\bibinfo  {journal} {Communications in Mathematical Physics}\ }\textbf
  {\bibinfo {volume} {48}},\ \bibinfo {pages} {119} (\bibinfo {year}
  {1976})}\BibitemShut {NoStop}%
\bibitem [{\citenamefont {Gorini}\ \emph {et~al.}(1976)\citenamefont {Gorini},
  \citenamefont {Kossakowski},\ and\ \citenamefont
  {Sudarshan}}]{gorini1976completely}%
  \BibitemOpen
  \bibfield  {author} {\bibinfo {author} {\bibfnamefont {V.}~\bibnamefont
  {Gorini}}, \bibinfo {author} {\bibfnamefont {A.}~\bibnamefont
  {Kossakowski}},\ and\ \bibinfo {author} {\bibfnamefont {E.~C.~G.}\
  \bibnamefont {Sudarshan}},\ }\bibfield  {title} {\bibinfo {title} {Completely
  positive dynamical semigroups of n-level systems},\ }\href@noop {} {\bibfield
   {journal} {\bibinfo  {journal} {Journal of Mathematical Physics}\ }\textbf
  {\bibinfo {volume} {17}},\ \bibinfo {pages} {821} (\bibinfo {year}
  {1976})}\BibitemShut {NoStop}%
\bibitem [{\citenamefont {Manzano}(2020)}]{manzano2020short}%
  \BibitemOpen
  \bibfield  {author} {\bibinfo {author} {\bibfnamefont {D.}~\bibnamefont
  {Manzano}},\ }\bibfield  {title} {\bibinfo {title} {A short introduction to
  the lindblad master equation},\ }\href@noop {} {\bibfield  {journal}
  {\bibinfo  {journal} {Aip Advances}\ }\textbf {\bibinfo {volume} {10}}
  (\bibinfo {year} {2020})}\BibitemShut {NoStop}%
\bibitem [{\citenamefont {Dattagupta}\ and\ \citenamefont
  {Puri}(2004)}]{dattagupta2004dissipative}%
  \BibitemOpen
  \bibfield  {author} {\bibinfo {author} {\bibfnamefont {S.}~\bibnamefont
  {Dattagupta}}\ and\ \bibinfo {author} {\bibfnamefont {S.}~\bibnamefont
  {Puri}},\ }\href@noop {} {\emph {\bibinfo {title} {Dissipative phenomena in
  condensed matter: some applications}}},\ Vol.~\bibinfo {volume} {71}\
  (\bibinfo  {publisher} {Springer Science \& Business Media},\ \bibinfo {year}
  {2004})\BibitemShut {NoStop}%
\bibitem [{\citenamefont {Weiss}(2012)}]{weiss2012quantum}%
  \BibitemOpen
  \bibfield  {author} {\bibinfo {author} {\bibfnamefont {U.}~\bibnamefont
  {Weiss}},\ }\href@noop {} {\emph {\bibinfo {title} {Quantum dissipative
  systems}}}\ (\bibinfo  {publisher} {World Scientific},\ \bibinfo {year}
  {2012})\BibitemShut {NoStop}%
\bibitem [{\citenamefont {Koch}\ and\ \citenamefont
  {Budich}(2022)}]{koch2022quantum}%
  \BibitemOpen
  \bibfield  {author} {\bibinfo {author} {\bibfnamefont {F.}~\bibnamefont
  {Koch}}\ and\ \bibinfo {author} {\bibfnamefont {J.~C.}\ \bibnamefont
  {Budich}},\ }\bibfield  {title} {\bibinfo {title} {Quantum non-hermitian
  topological sensors},\ }\href@noop {} {\bibfield  {journal} {\bibinfo
  {journal} {Physical Review Research}\ }\textbf {\bibinfo {volume} {4}},\
  \bibinfo {pages} {013113} (\bibinfo {year} {2022})}\BibitemShut {NoStop}%
\bibitem [{\citenamefont {Cao}\ \emph {et~al.}(2020)\citenamefont {Cao},
  \citenamefont {Lu}, \citenamefont {Meng}, \citenamefont {Sun}, \citenamefont
  {Shen},\ and\ \citenamefont {Xiao}}]{cao2020reservoir}%
  \BibitemOpen
  \bibfield  {author} {\bibinfo {author} {\bibfnamefont {W.}~\bibnamefont
  {Cao}}, \bibinfo {author} {\bibfnamefont {X.}~\bibnamefont {Lu}}, \bibinfo
  {author} {\bibfnamefont {X.}~\bibnamefont {Meng}}, \bibinfo {author}
  {\bibfnamefont {J.}~\bibnamefont {Sun}}, \bibinfo {author} {\bibfnamefont
  {H.}~\bibnamefont {Shen}},\ and\ \bibinfo {author} {\bibfnamefont
  {Y.}~\bibnamefont {Xiao}},\ }\bibfield  {title} {\bibinfo {title}
  {Reservoir-mediated quantum correlations in non-hermitian optical system},\
  }\href@noop {} {\bibfield  {journal} {\bibinfo  {journal} {Physical Review
  Letters}\ }\textbf {\bibinfo {volume} {124}},\ \bibinfo {pages} {030401}
  (\bibinfo {year} {2020})}\BibitemShut {NoStop}%
\bibitem [{\citenamefont {Wouters}\ and\ \citenamefont
  {Carusotto}(2007)}]{wouters2007excitations}%
  \BibitemOpen
  \bibfield  {author} {\bibinfo {author} {\bibfnamefont {M.}~\bibnamefont
  {Wouters}}\ and\ \bibinfo {author} {\bibfnamefont {I.}~\bibnamefont
  {Carusotto}},\ }\bibfield  {title} {\bibinfo {title} {Excitations in a
  nonequilibrium bose-einstein condensate of exciton polaritons},\ }\href@noop
  {} {\bibfield  {journal} {\bibinfo  {journal} {Physical review letters}\
  }\textbf {\bibinfo {volume} {99}},\ \bibinfo {pages} {140402} (\bibinfo
  {year} {2007})}\BibitemShut {NoStop}%
\bibitem [{\citenamefont {Fitzpatrick}\ \emph {et~al.}(2017)\citenamefont
  {Fitzpatrick}, \citenamefont {Sundaresan}, \citenamefont {Li}, \citenamefont
  {Koch},\ and\ \citenamefont {Houck}}]{fitzpatrick2017observation}%
  \BibitemOpen
  \bibfield  {author} {\bibinfo {author} {\bibfnamefont {M.}~\bibnamefont
  {Fitzpatrick}}, \bibinfo {author} {\bibfnamefont {N.~M.}\ \bibnamefont
  {Sundaresan}}, \bibinfo {author} {\bibfnamefont {A.~C.}\ \bibnamefont {Li}},
  \bibinfo {author} {\bibfnamefont {J.}~\bibnamefont {Koch}},\ and\ \bibinfo
  {author} {\bibfnamefont {A.~A.}\ \bibnamefont {Houck}},\ }\bibfield  {title}
  {\bibinfo {title} {Observation of a dissipative phase transition in a
  one-dimensional circuit qed lattice},\ }\href@noop {} {\bibfield  {journal}
  {\bibinfo  {journal} {Physical Review X}\ }\textbf {\bibinfo {volume} {7}},\
  \bibinfo {pages} {011016} (\bibinfo {year} {2017})}\BibitemShut {NoStop}%
\bibitem [{\citenamefont {Botzung}\ \emph {et~al.}(2021)\citenamefont
  {Botzung}, \citenamefont {Diehl},\ and\ \citenamefont
  {M{\"u}ller}}]{botzung2021engineered}%
  \BibitemOpen
  \bibfield  {author} {\bibinfo {author} {\bibfnamefont {T.}~\bibnamefont
  {Botzung}}, \bibinfo {author} {\bibfnamefont {S.}~\bibnamefont {Diehl}},\
  and\ \bibinfo {author} {\bibfnamefont {M.}~\bibnamefont {M{\"u}ller}},\
  }\bibfield  {title} {\bibinfo {title} {Engineered dissipation induced
  entanglement transition in quantum spin chains: From logarithmic growth to
  area law},\ }\href@noop {} {\bibfield  {journal} {\bibinfo  {journal}
  {Physical Review B}\ }\textbf {\bibinfo {volume} {104}},\ \bibinfo {pages}
  {184422} (\bibinfo {year} {2021})}\BibitemShut {NoStop}%
\bibitem [{\citenamefont {Alberton}\ \emph {et~al.}(2021)\citenamefont
  {Alberton}, \citenamefont {Buchhold},\ and\ \citenamefont
  {Diehl}}]{alberton2021entanglement}%
  \BibitemOpen
  \bibfield  {author} {\bibinfo {author} {\bibfnamefont {O.}~\bibnamefont
  {Alberton}}, \bibinfo {author} {\bibfnamefont {M.}~\bibnamefont {Buchhold}},\
  and\ \bibinfo {author} {\bibfnamefont {S.}~\bibnamefont {Diehl}},\ }\bibfield
   {title} {\bibinfo {title} {Entanglement transition in a monitored
  free-fermion chain: From extended criticality to area law},\ }\href@noop {}
  {\bibfield  {journal} {\bibinfo  {journal} {Physical Review Letters}\
  }\textbf {\bibinfo {volume} {126}},\ \bibinfo {pages} {170602} (\bibinfo
  {year} {2021})}\BibitemShut {NoStop}%
\bibitem [{\citenamefont {Yu}\ and\ \citenamefont
  {Zhang}(2008)}]{yu2008understanding}%
  \BibitemOpen
  \bibfield  {author} {\bibinfo {author} {\bibfnamefont {H.}~\bibnamefont
  {Yu}}\ and\ \bibinfo {author} {\bibfnamefont {J.}~\bibnamefont {Zhang}},\
  }\bibfield  {title} {\bibinfo {title} {Understanding hawking radiation in the
  framework of open quantum systems},\ }\href@noop {} {\bibfield  {journal}
  {\bibinfo  {journal} {Physical Review D}\ }\textbf {\bibinfo {volume} {77}},\
  \bibinfo {pages} {024031} (\bibinfo {year} {2008})}\BibitemShut {NoStop}%
\bibitem [{\citenamefont {Kaplanek}\ and\ \citenamefont
  {Burgess}(2020)}]{kaplanek2020hot}%
  \BibitemOpen
  \bibfield  {author} {\bibinfo {author} {\bibfnamefont {G.}~\bibnamefont
  {Kaplanek}}\ and\ \bibinfo {author} {\bibfnamefont {C.}~\bibnamefont
  {Burgess}},\ }\bibfield  {title} {\bibinfo {title} {Hot cosmic qubits:
  late-time de sitter evolution and critical slowing down},\ }\href@noop {}
  {\bibfield  {journal} {\bibinfo  {journal} {Journal of High Energy Physics}\
  }\textbf {\bibinfo {volume} {2020}},\ \bibinfo {pages} {1} (\bibinfo {year}
  {2020})}\BibitemShut {NoStop}%
\bibitem [{\citenamefont {Chakraborty}\ \emph {et~al.}(2022)\citenamefont
  {Chakraborty}, \citenamefont {Camblong},\ and\ \citenamefont
  {Ord{\'o}{\~n}ez}}]{chakraborty2022thermal}%
  \BibitemOpen
  \bibfield  {author} {\bibinfo {author} {\bibfnamefont {A.}~\bibnamefont
  {Chakraborty}}, \bibinfo {author} {\bibfnamefont {H.}~\bibnamefont
  {Camblong}},\ and\ \bibinfo {author} {\bibfnamefont {C.}~\bibnamefont
  {Ord{\'o}{\~n}ez}},\ }\bibfield  {title} {\bibinfo {title} {Thermal effect in
  a causal diamond: Open quantum systems approach},\ }\href@noop {} {\bibfield
  {journal} {\bibinfo  {journal} {Physical Review D}\ }\textbf {\bibinfo
  {volume} {106}},\ \bibinfo {pages} {045027} (\bibinfo {year}
  {2022})}\BibitemShut {NoStop}%
\bibitem [{\citenamefont {Rotter}(2009)}]{rotter2009non}%
  \BibitemOpen
  \bibfield  {author} {\bibinfo {author} {\bibfnamefont {I.}~\bibnamefont
  {Rotter}},\ }\bibfield  {title} {\bibinfo {title} {A non-hermitian hamilton
  operator and the physics of open quantum systems},\ }\href@noop {} {\bibfield
   {journal} {\bibinfo  {journal} {Journal of Physics A: Mathematical and
  Theoretical}\ }\textbf {\bibinfo {volume} {42}},\ \bibinfo {pages} {153001}
  (\bibinfo {year} {2009})}\BibitemShut {NoStop}%
\bibitem [{\citenamefont {M{\"u}ller}\ and\ \citenamefont
  {Rotter}(2009)}]{muller2009phase}%
  \BibitemOpen
  \bibfield  {author} {\bibinfo {author} {\bibfnamefont {M.}~\bibnamefont
  {M{\"u}ller}}\ and\ \bibinfo {author} {\bibfnamefont {I.}~\bibnamefont
  {Rotter}},\ }\bibfield  {title} {\bibinfo {title} {Phase lapses in open
  quantum systems and the non-hermitian hamilton operator},\ }\href@noop {}
  {\bibfield  {journal} {\bibinfo  {journal} {Physical Review A}\ }\textbf
  {\bibinfo {volume} {80}},\ \bibinfo {pages} {042705} (\bibinfo {year}
  {2009})}\BibitemShut {NoStop}%
\bibitem [{\citenamefont {Savin}\ \emph {et~al.}(2003)\citenamefont {Savin},
  \citenamefont {Sokolov},\ and\ \citenamefont {Sommers}}]{savin2003concept}%
  \BibitemOpen
  \bibfield  {author} {\bibinfo {author} {\bibfnamefont {D.~V.}\ \bibnamefont
  {Savin}}, \bibinfo {author} {\bibfnamefont {V.~V.}\ \bibnamefont {Sokolov}},\
  and\ \bibinfo {author} {\bibfnamefont {H.-J.}\ \bibnamefont {Sommers}},\
  }\bibfield  {title} {\bibinfo {title} {Is the concept of the non-hermitian
  effective hamiltonian relevant in the case of potential scattering?},\
  }\href@noop {} {\bibfield  {journal} {\bibinfo  {journal} {Physical Review
  E}\ }\textbf {\bibinfo {volume} {67}},\ \bibinfo {pages} {026215} (\bibinfo
  {year} {2003})}\BibitemShut {NoStop}%
\bibitem [{\citenamefont {Ashida}\ \emph {et~al.}(2020)\citenamefont {Ashida},
  \citenamefont {Gong},\ and\ \citenamefont {Ueda}}]{ashida2020non}%
  \BibitemOpen
  \bibfield  {author} {\bibinfo {author} {\bibfnamefont {Y.}~\bibnamefont
  {Ashida}}, \bibinfo {author} {\bibfnamefont {Z.}~\bibnamefont {Gong}},\ and\
  \bibinfo {author} {\bibfnamefont {M.}~\bibnamefont {Ueda}},\ }\bibfield
  {title} {\bibinfo {title} {Non-hermitian physics},\ }\href@noop {} {\bibfield
   {journal} {\bibinfo  {journal} {Advances in Physics}\ }\textbf {\bibinfo
  {volume} {69}},\ \bibinfo {pages} {249} (\bibinfo {year} {2020})}\BibitemShut
  {NoStop}%
\bibitem [{\citenamefont {Breuer}\ and\ \citenamefont
  {Petruccione}(2002)}]{breuer2002theory}%
  \BibitemOpen
  \bibfield  {author} {\bibinfo {author} {\bibfnamefont {H.-P.}\ \bibnamefont
  {Breuer}}\ and\ \bibinfo {author} {\bibfnamefont {F.}~\bibnamefont
  {Petruccione}},\ }\href@noop {} {\emph {\bibinfo {title} {The theory of open
  quantum systems}}}\ (\bibinfo  {publisher} {Oxford University Press, USA},\
  \bibinfo {year} {2002})\BibitemShut {NoStop}%
\bibitem [{\citenamefont {Prosen}(2008)}]{prosen2008third}%
  \BibitemOpen
  \bibfield  {author} {\bibinfo {author} {\bibfnamefont {T.}~\bibnamefont
  {Prosen}},\ }\bibfield  {title} {\bibinfo {title} {Third quantization: a
  general method to solve master equations for quadratic open fermi systems},\
  }\href@noop {} {\bibfield  {journal} {\bibinfo  {journal} {New Journal of
  Physics}\ }\textbf {\bibinfo {volume} {10}},\ \bibinfo {pages} {043026}
  (\bibinfo {year} {2008})}\BibitemShut {NoStop}%
\bibitem [{\citenamefont {Medvedyeva}\ \emph {et~al.}(2016)\citenamefont
  {Medvedyeva}, \citenamefont {Essler},\ and\ \citenamefont
  {Prosen}}]{Essler-prozen}%
  \BibitemOpen
  \bibfield  {author} {\bibinfo {author} {\bibfnamefont {M.~V.}\ \bibnamefont
  {Medvedyeva}}, \bibinfo {author} {\bibfnamefont {F.~H.~L.}\ \bibnamefont
  {Essler}},\ and\ \bibinfo {author} {\bibfnamefont {T.~c.~v.}\ \bibnamefont
  {Prosen}},\ }\bibfield  {title} {\bibinfo {title} {Exact bethe ansatz
  spectrum of a tight-binding chain with dephasing noise},\ }\href
  {https://doi.org/10.1103/PhysRevLett.117.137202} {\bibfield  {journal}
  {\bibinfo  {journal} {Phys. Rev. Lett.}\ }\textbf {\bibinfo {volume} {117}},\
  \bibinfo {pages} {137202} (\bibinfo {year} {2016})}\BibitemShut {NoStop}%
\bibitem [{\citenamefont {Alba}(2023)}]{alba2023free}%
  \BibitemOpen
  \bibfield  {author} {\bibinfo {author} {\bibfnamefont {V.}~\bibnamefont
  {Alba}},\ }\bibfield  {title} {\bibinfo {title} {Free fermions with dephasing
  and boundary driving: Bethe ansatz results},\ }\href@noop {} {\bibfield
  {journal} {\bibinfo  {journal} {arXiv preprint arXiv:2309.12978}\ } (\bibinfo
  {year} {2023})}\BibitemShut {NoStop}%
\bibitem [{\citenamefont {Castaldi}\ \emph {et~al.}(2013)\citenamefont
  {Castaldi}, \citenamefont {Savoia}, \citenamefont {Galdi}, \citenamefont
  {Alu},\ and\ \citenamefont {Engheta}}]{castaldi2013p}%
  \BibitemOpen
  \bibfield  {author} {\bibinfo {author} {\bibfnamefont {G.}~\bibnamefont
  {Castaldi}}, \bibinfo {author} {\bibfnamefont {S.}~\bibnamefont {Savoia}},
  \bibinfo {author} {\bibfnamefont {V.}~\bibnamefont {Galdi}}, \bibinfo
  {author} {\bibfnamefont {A.}~\bibnamefont {Alu}},\ and\ \bibinfo {author}
  {\bibfnamefont {N.}~\bibnamefont {Engheta}},\ }\bibfield  {title} {\bibinfo
  {title} {P t metamaterials via complex-coordinate transformation optics},\
  }\href@noop {} {\bibfield  {journal} {\bibinfo  {journal} {Physical review
  letters}\ }\textbf {\bibinfo {volume} {110}},\ \bibinfo {pages} {173901}
  (\bibinfo {year} {2013})}\BibitemShut {NoStop}%
\bibitem [{\citenamefont {Yin}\ and\ \citenamefont
  {Zhang}(2013)}]{yin2013unidirectional}%
  \BibitemOpen
  \bibfield  {author} {\bibinfo {author} {\bibfnamefont {X.}~\bibnamefont
  {Yin}}\ and\ \bibinfo {author} {\bibfnamefont {X.}~\bibnamefont {Zhang}},\
  }\bibfield  {title} {\bibinfo {title} {Unidirectional light propagation at
  exceptional points},\ }\href@noop {} {\bibfield  {journal} {\bibinfo
  {journal} {Nature materials}\ }\textbf {\bibinfo {volume} {12}},\ \bibinfo
  {pages} {175} (\bibinfo {year} {2013})}\BibitemShut {NoStop}%
\bibitem [{\citenamefont {Zyablovsky}\ \emph {et~al.}(2014)\citenamefont
  {Zyablovsky}, \citenamefont {Vinogradov}, \citenamefont {Pukhov},
  \citenamefont {Dorofeenko},\ and\ \citenamefont
  {Lisyansky}}]{zyablovsky2014pt}%
  \BibitemOpen
  \bibfield  {author} {\bibinfo {author} {\bibfnamefont {A.~A.}\ \bibnamefont
  {Zyablovsky}}, \bibinfo {author} {\bibfnamefont {A.~P.}\ \bibnamefont
  {Vinogradov}}, \bibinfo {author} {\bibfnamefont {A.~A.}\ \bibnamefont
  {Pukhov}}, \bibinfo {author} {\bibfnamefont {A.~V.}\ \bibnamefont
  {Dorofeenko}},\ and\ \bibinfo {author} {\bibfnamefont {A.~A.}\ \bibnamefont
  {Lisyansky}},\ }\bibfield  {title} {\bibinfo {title} {Pt-symmetry in
  optics},\ }\href@noop {} {\bibfield  {journal} {\bibinfo  {journal}
  {Physics-Uspekhi}\ }\textbf {\bibinfo {volume} {57}},\ \bibinfo {pages}
  {1063} (\bibinfo {year} {2014})}\BibitemShut {NoStop}%
\bibitem [{\citenamefont {Klauck}\ \emph {et~al.}(2019)\citenamefont {Klauck},
  \citenamefont {Teuber}, \citenamefont {Ornigotti}, \citenamefont {Heinrich},
  \citenamefont {Scheel},\ and\ \citenamefont
  {Szameit}}]{klauck2019observation}%
  \BibitemOpen
  \bibfield  {author} {\bibinfo {author} {\bibfnamefont {F.}~\bibnamefont
  {Klauck}}, \bibinfo {author} {\bibfnamefont {L.}~\bibnamefont {Teuber}},
  \bibinfo {author} {\bibfnamefont {M.}~\bibnamefont {Ornigotti}}, \bibinfo
  {author} {\bibfnamefont {M.}~\bibnamefont {Heinrich}}, \bibinfo {author}
  {\bibfnamefont {S.}~\bibnamefont {Scheel}},\ and\ \bibinfo {author}
  {\bibfnamefont {A.}~\bibnamefont {Szameit}},\ }\bibfield  {title} {\bibinfo
  {title} {Observation of pt-symmetric quantum interference},\ }\href@noop {}
  {\bibfield  {journal} {\bibinfo  {journal} {Nature Photonics}\ }\textbf
  {\bibinfo {volume} {13}},\ \bibinfo {pages} {883} (\bibinfo {year}
  {2019})}\BibitemShut {NoStop}%
\bibitem [{\citenamefont {Kawabata}\ \emph {et~al.}(2018)\citenamefont
  {Kawabata}, \citenamefont {Ashida}, \citenamefont {Katsura},\ and\
  \citenamefont {Ueda}}]{kawabata2018parity}%
  \BibitemOpen
  \bibfield  {author} {\bibinfo {author} {\bibfnamefont {K.}~\bibnamefont
  {Kawabata}}, \bibinfo {author} {\bibfnamefont {Y.}~\bibnamefont {Ashida}},
  \bibinfo {author} {\bibfnamefont {H.}~\bibnamefont {Katsura}},\ and\ \bibinfo
  {author} {\bibfnamefont {M.}~\bibnamefont {Ueda}},\ }\bibfield  {title}
  {\bibinfo {title} {Parity-time-symmetric topological superconductor},\
  }\href@noop {} {\bibfield  {journal} {\bibinfo  {journal} {Physical Review
  B}\ }\textbf {\bibinfo {volume} {98}},\ \bibinfo {pages} {085116} (\bibinfo
  {year} {2018})}\BibitemShut {NoStop}%
\bibitem [{\citenamefont {Kornich}\ and\ \citenamefont
  {Trauzettel}(2022)}]{kornich2022signature}%
  \BibitemOpen
  \bibfield  {author} {\bibinfo {author} {\bibfnamefont {V.}~\bibnamefont
  {Kornich}}\ and\ \bibinfo {author} {\bibfnamefont {B.}~\bibnamefont
  {Trauzettel}},\ }\bibfield  {title} {\bibinfo {title} {Signature of p
  t-symmetric non-hermitian superconductivity in angle-resolved photoelectron
  fluctuation spectroscopy},\ }\href@noop {} {\bibfield  {journal} {\bibinfo
  {journal} {Physical Review Research}\ }\textbf {\bibinfo {volume} {4}},\
  \bibinfo {pages} {L022018} (\bibinfo {year} {2022})}\BibitemShut {NoStop}%
\bibitem [{\citenamefont {Bagarello}\ \emph {et~al.}(2015)\citenamefont
  {Bagarello}, \citenamefont {Lattuca}, \citenamefont {Passante}, \citenamefont
  {Rizzuto},\ and\ \citenamefont {Spagnolo}}]{bagarello2015non}%
  \BibitemOpen
  \bibfield  {author} {\bibinfo {author} {\bibfnamefont {F.}~\bibnamefont
  {Bagarello}}, \bibinfo {author} {\bibfnamefont {M.}~\bibnamefont {Lattuca}},
  \bibinfo {author} {\bibfnamefont {R.}~\bibnamefont {Passante}}, \bibinfo
  {author} {\bibfnamefont {L.}~\bibnamefont {Rizzuto}},\ and\ \bibinfo {author}
  {\bibfnamefont {S.}~\bibnamefont {Spagnolo}},\ }\bibfield  {title} {\bibinfo
  {title} {Non-hermitian hamiltonian for a modulated jaynes-cummings model with
  pt symmetry},\ }\href@noop {} {\bibfield  {journal} {\bibinfo  {journal}
  {Physical Review A}\ }\textbf {\bibinfo {volume} {91}},\ \bibinfo {pages}
  {042134} (\bibinfo {year} {2015})}\BibitemShut {NoStop}%
\bibitem [{\citenamefont {Zhao}\ and\ \citenamefont {Lu}(2017)}]{zhao2017p}%
  \BibitemOpen
  \bibfield  {author} {\bibinfo {author} {\bibfnamefont {Y.}~\bibnamefont
  {Zhao}}\ and\ \bibinfo {author} {\bibfnamefont {Y.}~\bibnamefont {Lu}},\
  }\bibfield  {title} {\bibinfo {title} {P t-symmetric real dirac fermions and
  semimetals},\ }\href@noop {} {\bibfield  {journal} {\bibinfo  {journal}
  {Physical review letters}\ }\textbf {\bibinfo {volume} {118}},\ \bibinfo
  {pages} {056401} (\bibinfo {year} {2017})}\BibitemShut {NoStop}%
\bibitem [{\citenamefont {Turker}\ \emph {et~al.}(2018)\citenamefont {Turker},
  \citenamefont {Tombuloglu},\ and\ \citenamefont {Yuce}}]{turker2018pt}%
  \BibitemOpen
  \bibfield  {author} {\bibinfo {author} {\bibfnamefont {Z.}~\bibnamefont
  {Turker}}, \bibinfo {author} {\bibfnamefont {S.}~\bibnamefont {Tombuloglu}},\
  and\ \bibinfo {author} {\bibfnamefont {C.}~\bibnamefont {Yuce}},\ }\bibfield
  {title} {\bibinfo {title} {Pt symmetric floquet topological phase in ssh
  model},\ }\href@noop {} {\bibfield  {journal} {\bibinfo  {journal} {Physics
  Letters A}\ }\textbf {\bibinfo {volume} {382}},\ \bibinfo {pages} {2013}
  (\bibinfo {year} {2018})}\BibitemShut {NoStop}%
\bibitem [{\citenamefont {Bender}\ and\ \citenamefont
  {Boettcher}(1998)}]{bender1998real}%
  \BibitemOpen
  \bibfield  {author} {\bibinfo {author} {\bibfnamefont {C.~M.}\ \bibnamefont
  {Bender}}\ and\ \bibinfo {author} {\bibfnamefont {S.}~\bibnamefont
  {Boettcher}},\ }\bibfield  {title} {\bibinfo {title} {Real spectra in
  non-hermitian hamiltonians having p t symmetry},\ }\href@noop {} {\bibfield
  {journal} {\bibinfo  {journal} {Physical review letters}\ }\textbf {\bibinfo
  {volume} {80}},\ \bibinfo {pages} {5243} (\bibinfo {year}
  {1998})}\BibitemShut {NoStop}%
\bibitem [{\citenamefont {Wigner}(1960)}]{wigner1960phenomenological}%
  \BibitemOpen
  \bibfield  {author} {\bibinfo {author} {\bibfnamefont {E.~P.}\ \bibnamefont
  {Wigner}},\ }\bibfield  {title} {\bibinfo {title} {Phenomenological
  distinction between unitary and antiunitary symmetry operators},\ }\href@noop
  {} {\bibfield  {journal} {\bibinfo  {journal} {Journal of Mathematical
  Physics}\ }\textbf {\bibinfo {volume} {1}},\ \bibinfo {pages} {414} (\bibinfo
  {year} {1960})}\BibitemShut {NoStop}%
\bibitem [{Note1()}]{Note1}%
  \BibitemOpen
  \bibinfo {note} {Consider the Schrodinger equation \begin {equation} \protect
  \mathcal H \mathinner {|{\psi (t)}\rangle } = E \mathinner {|{\psi
  (t)}\rangle }. \end {equation} Acting with a generic anti-linear operator
  \(\protect \hat {\protect \mathcal {A}}\), we arrive at: \begin {equation}
  \protect \hat {\protect \mathcal {A}} \protect \mathcal H \protect \hat
  {\protect \mathcal {A}}^{-1} \protect \hat {\protect \mathcal {A}} \mathinner
  {|{\psi (t)}\rangle } = E^* \protect \hat {\protect \mathcal {A}} \mathinner
  {|{\psi (t)}\rangle }. \label {antlop} \end {equation} In Eq.\protect \eqref
  {antlop}, we find that when the Hamiltonian \(\protect \mathcal H\) obeys the
  antilinear symmetry condition \(\protect \hat {\protect \mathcal {A}}
  \protect \mathcal H \protect \hat {\protect \mathcal {A}}^{-1} = \protect
  \mathcal H\), two distinct scenarios emerge, as originally pointed out by
  Wigner: \par a) If the eigenfunctions satisfy \(\protect \hat {\protect
  \mathcal {A}}\mathinner {|{\psi (t)}\rangle } = \mathinner {|{\psi
  (t)}\rangle }\), then the corresponding eigenvalues are real.\\
  b)Alternatively, when the eigenfunctions exhibit the property \(\protect \hat
  {\protect \mathcal {A}}\mathinner {|{\psi (t)}\rangle } \protect \neq
  \mathinner {|{\psi (t)}\rangle }\), the energies appear in conjugate pairs,
  and the conjugate eigenfunctions take the form $|\psi (t)\rangle \sim e^{-i E
  t}$ and $\protect \hat {\protect \mathcal {A}}|\psi (t)\rangle \sim e^{-i
  Et}$.}\BibitemShut {Stop}%
\bibitem [{\citenamefont {Mostafazadeh}(2003)}]{mostafazadeh2003exact}%
  \BibitemOpen
  \bibfield  {author} {\bibinfo {author} {\bibfnamefont {A.}~\bibnamefont
  {Mostafazadeh}},\ }\bibfield  {title} {\bibinfo {title} {Exact pt-symmetry is
  equivalent to hermiticity},\ }\href@noop {} {\bibfield  {journal} {\bibinfo
  {journal} {Journal of Physics A: Mathematical and General}\ }\textbf
  {\bibinfo {volume} {36}},\ \bibinfo {pages} {7081} (\bibinfo {year}
  {2003})}\BibitemShut {NoStop}%
\bibitem [{\citenamefont {Fishman}\ \emph {et~al.}(2022)\citenamefont
  {Fishman}, \citenamefont {White},\ and\ \citenamefont
  {Stoudenmire}}]{fishman2022itensor}%
  \BibitemOpen
  \bibfield  {author} {\bibinfo {author} {\bibfnamefont {M.}~\bibnamefont
  {Fishman}}, \bibinfo {author} {\bibfnamefont {S.}~\bibnamefont {White}},\
  and\ \bibinfo {author} {\bibfnamefont {E.}~\bibnamefont {Stoudenmire}},\
  }\bibfield  {title} {\bibinfo {title} {The itensor software library for
  tensor network calculations},\ }\href@noop {} {\bibfield  {journal} {\bibinfo
   {journal} {SciPost Physics Codebases}\ ,\ \bibinfo {pages} {004}} (\bibinfo
  {year} {2022})}\BibitemShut {NoStop}%
\bibitem [{\citenamefont {Bender}\ \emph {et~al.}(2004)\citenamefont {Bender},
  \citenamefont {Brody},\ and\ \citenamefont {Jones}}]{bender2004extension}%
  \BibitemOpen
  \bibfield  {author} {\bibinfo {author} {\bibfnamefont {C.~M.}\ \bibnamefont
  {Bender}}, \bibinfo {author} {\bibfnamefont {D.~C.}\ \bibnamefont {Brody}},\
  and\ \bibinfo {author} {\bibfnamefont {H.~F.}\ \bibnamefont {Jones}},\
  }\bibfield  {title} {\bibinfo {title} {Extension of pt-symmetric quantum
  mechanics to quantum field theory with cubic interaction},\ }\href@noop {}
  {\bibfield  {journal} {\bibinfo  {journal} {Physical Review D}\ }\textbf
  {\bibinfo {volume} {70}},\ \bibinfo {pages} {025001} (\bibinfo {year}
  {2004})}\BibitemShut {NoStop}%
\bibitem [{\citenamefont {Bender}\ and\ \citenamefont
  {Tan}(2006)}]{bender2006calculation}%
  \BibitemOpen
  \bibfield  {author} {\bibinfo {author} {\bibfnamefont {C.~M.}\ \bibnamefont
  {Bender}}\ and\ \bibinfo {author} {\bibfnamefont {B.}~\bibnamefont {Tan}},\
  }\bibfield  {title} {\bibinfo {title} {Calculation of the hidden symmetry
  operator for a-symmetric square well},\ }\href@noop {} {\bibfield  {journal}
  {\bibinfo  {journal} {Journal of Physics A: Mathematical and General}\
  }\textbf {\bibinfo {volume} {39}},\ \bibinfo {pages} {1945} (\bibinfo {year}
  {2006})}\BibitemShut {NoStop}%
\bibitem [{\citenamefont {Bethe}(1931)}]{1931_Bethe_ZP_71}%
  \BibitemOpen
  \bibfield  {author} {\bibinfo {author} {\bibfnamefont {H.~A.}\ \bibnamefont
  {Bethe}},\ }\bibfield  {title} {\bibinfo {title} {Zur {T}heorie der
  {M}etalle. i. {E}igenwerte und {E}igenfunktionen der linearen {A}tomkette},\
  }\href {https://doi.org/10.1007/BF01341708} {\bibfield  {journal} {\bibinfo
  {journal} {Zeit. f\"ur Physik}\ }\textbf {\bibinfo {volume} {71}},\ \bibinfo
  {pages} {205} (\bibinfo {year} {1931})}\BibitemShut {NoStop}%
\bibitem [{\citenamefont {Alcaraz}\ \emph {et~al.}(1987)\citenamefont
  {Alcaraz}, \citenamefont {Barber}, \citenamefont {Batchelor}, \citenamefont
  {Baxter},\ and\ \citenamefont {Quispel}}]{alcaraz1987surface}%
  \BibitemOpen
  \bibfield  {author} {\bibinfo {author} {\bibfnamefont {F.~C.}\ \bibnamefont
  {Alcaraz}}, \bibinfo {author} {\bibfnamefont {M.~N.}\ \bibnamefont {Barber}},
  \bibinfo {author} {\bibfnamefont {M.~T.}\ \bibnamefont {Batchelor}}, \bibinfo
  {author} {\bibfnamefont {R.}~\bibnamefont {Baxter}},\ and\ \bibinfo {author}
  {\bibfnamefont {G.}~\bibnamefont {Quispel}},\ }\bibfield  {title} {\bibinfo
  {title} {Surface exponents of the quantum xxz, ashkin-teller and potts
  models},\ }\href@noop {} {\bibfield  {journal} {\bibinfo  {journal} {Journal
  of Physics A: mathematical and general}\ }\textbf {\bibinfo {volume} {20}},\
  \bibinfo {pages} {6397} (\bibinfo {year} {1987})}\BibitemShut {NoStop}%
\bibitem [{\citenamefont {Sklyanin}(1988)}]{sklyanin1988boundary}%
  \BibitemOpen
  \bibfield  {author} {\bibinfo {author} {\bibfnamefont {E.~K.}\ \bibnamefont
  {Sklyanin}},\ }\bibfield  {title} {\bibinfo {title} {Boundary conditions for
  integrable quantum systems},\ }\href@noop {} {\bibfield  {journal} {\bibinfo
  {journal} {Journal of Physics A: Mathematical and General}\ }\textbf
  {\bibinfo {volume} {21}},\ \bibinfo {pages} {2375} (\bibinfo {year}
  {1988})}\BibitemShut {NoStop}%
\bibitem [{\citenamefont {Wang}\ \emph {et~al.}(2015)\citenamefont {Wang},
  \citenamefont {Yang}, \citenamefont {Cao},\ and\ \citenamefont
  {Shi}}]{wang2015off}%
  \BibitemOpen
  \bibfield  {author} {\bibinfo {author} {\bibfnamefont {Y.}~\bibnamefont
  {Wang}}, \bibinfo {author} {\bibfnamefont {W.-L.}\ \bibnamefont {Yang}},
  \bibinfo {author} {\bibfnamefont {J.}~\bibnamefont {Cao}},\ and\ \bibinfo
  {author} {\bibfnamefont {K.}~\bibnamefont {Shi}},\ }\href@noop {} {\emph
  {\bibinfo {title} {Off-diagonal Bethe ansatz for exactly solvable models}}}\
  (\bibinfo  {publisher} {Springer},\ \bibinfo {year} {2015})\BibitemShut
  {NoStop}%
\bibitem [{\citenamefont {Frahm}\ \emph {et~al.}(2010)\citenamefont {Frahm},
  \citenamefont {Grelik}, \citenamefont {Seel},\ and\ \citenamefont
  {Wirth}}]{frahm2010functional}%
  \BibitemOpen
  \bibfield  {author} {\bibinfo {author} {\bibfnamefont {H.}~\bibnamefont
  {Frahm}}, \bibinfo {author} {\bibfnamefont {J.~H.}\ \bibnamefont {Grelik}},
  \bibinfo {author} {\bibfnamefont {A.}~\bibnamefont {Seel}},\ and\ \bibinfo
  {author} {\bibfnamefont {T.}~\bibnamefont {Wirth}},\ }\bibfield  {title}
  {\bibinfo {title} {Functional bethe ansatz methods for the open xxx chain},\
  }\href@noop {} {\bibfield  {journal} {\bibinfo  {journal} {Journal of Physics
  A: Mathematical and Theoretical}\ }\textbf {\bibinfo {volume} {44}},\
  \bibinfo {pages} {015001} (\bibinfo {year} {2010})}\BibitemShut {NoStop}%
\bibitem [{\citenamefont {Kattel}\ \emph
  {et~al.}(2024{\natexlab{a}})\citenamefont {Kattel}, \citenamefont {Pasnoori},
  \citenamefont {Pixley}, \citenamefont {Azaria},\ and\ \citenamefont
  {Andrei}}]{kattel2023kondo}%
  \BibitemOpen
  \bibfield  {author} {\bibinfo {author} {\bibfnamefont {P.}~\bibnamefont
  {Kattel}}, \bibinfo {author} {\bibfnamefont {P.~R.}\ \bibnamefont
  {Pasnoori}}, \bibinfo {author} {\bibfnamefont {J.~H.}\ \bibnamefont
  {Pixley}}, \bibinfo {author} {\bibfnamefont {P.}~\bibnamefont {Azaria}},\
  and\ \bibinfo {author} {\bibfnamefont {N.}~\bibnamefont {Andrei}},\
  }\bibfield  {title} {\bibinfo {title} {Kondo effect in the isotropic
  heisenberg spin chain},\ }\href {https://doi.org/10.1103/PhysRevB.109.174416}
  {\bibfield  {journal} {\bibinfo  {journal} {Phys. Rev. B}\ }\textbf {\bibinfo
  {volume} {109}},\ \bibinfo {pages} {174416} (\bibinfo {year}
  {2024}{\natexlab{a}})}\BibitemShut {NoStop}%
\bibitem [{\citenamefont {Wang}(1997)}]{wang1997exact}%
  \BibitemOpen
  \bibfield  {author} {\bibinfo {author} {\bibfnamefont {Y.}~\bibnamefont
  {Wang}},\ }\bibfield  {title} {\bibinfo {title} {Exact solution of the open
  heisenberg chain with two impurities},\ }\href@noop {} {\bibfield  {journal}
  {\bibinfo  {journal} {Physical Review B}\ }\textbf {\bibinfo {volume} {56}},\
  \bibinfo {pages} {14045} (\bibinfo {year} {1997})}\BibitemShut {NoStop}%
\bibitem [{\citenamefont {Kattel}\ \emph {et~al.}(2023)\citenamefont {Kattel},
  \citenamefont {Pasnoori},\ and\ \citenamefont {Andrei}}]{kattel2023exact}%
  \BibitemOpen
  \bibfield  {author} {\bibinfo {author} {\bibfnamefont {P.}~\bibnamefont
  {Kattel}}, \bibinfo {author} {\bibfnamefont {P.~R.}\ \bibnamefont
  {Pasnoori}},\ and\ \bibinfo {author} {\bibfnamefont {N.}~\bibnamefont
  {Andrei}},\ }\bibfield  {title} {\bibinfo {title} {Exact solution of a
  non-hermitian-symmetric spin chain},\ }\href@noop {} {\bibfield  {journal}
  {\bibinfo  {journal} {Journal of Physics A: Mathematical and Theoretical}\
  }\textbf {\bibinfo {volume} {56}},\ \bibinfo {pages} {325001} (\bibinfo
  {year} {2023})}\BibitemShut {NoStop}%
\bibitem [{\citenamefont {Frahm}\ and\ \citenamefont
  {Zvyagin}(1997)}]{frahm1997open}%
  \BibitemOpen
  \bibfield  {author} {\bibinfo {author} {\bibfnamefont {H.}~\bibnamefont
  {Frahm}}\ and\ \bibinfo {author} {\bibfnamefont {A.~A.}\ \bibnamefont
  {Zvyagin}},\ }\bibfield  {title} {\bibinfo {title} {The open spin chain with
  impurity: an exact solution},\ }\href@noop {} {\bibfield  {journal} {\bibinfo
   {journal} {Journal of Physics: Condensed Matter}\ }\textbf {\bibinfo
  {volume} {9}},\ \bibinfo {pages} {9939} (\bibinfo {year} {1997})}\BibitemShut
  {NoStop}%
\bibitem [{\citenamefont {Kattel}\ \emph
  {et~al.}(2024{\natexlab{b}})\citenamefont {Kattel}, \citenamefont {Tang},
  \citenamefont {Pixley},\ and\ \citenamefont {Andrei}}]{kattel2024kondo}%
  \BibitemOpen
  \bibfield  {author} {\bibinfo {author} {\bibfnamefont {P.}~\bibnamefont
  {Kattel}}, \bibinfo {author} {\bibfnamefont {Y.}~\bibnamefont {Tang}},
  \bibinfo {author} {\bibfnamefont {J.~H.}\ \bibnamefont {Pixley}},\ and\
  \bibinfo {author} {\bibfnamefont {N.}~\bibnamefont {Andrei}},\ }\bibfield
  {title} {\bibinfo {title} {The kondo effect in the quantum xx spin chain},\
  }\href@noop {} {\bibfield  {journal} {\bibinfo  {journal} {Journal of Physics
  A: Mathematical and Theoretical}\ } (\bibinfo {year}
  {2024}{\natexlab{b}})}\BibitemShut {NoStop}%
\bibitem [{\citenamefont {Ren}\ \emph {et~al.}(2022)\citenamefont {Ren},
  \citenamefont {Liu}, \citenamefont {Zhao}, \citenamefont {He}, \citenamefont
  {Pak}, \citenamefont {Li},\ and\ \citenamefont {Jo}}]{ren2022chiral}%
  \BibitemOpen
  \bibfield  {author} {\bibinfo {author} {\bibfnamefont {Z.}~\bibnamefont
  {Ren}}, \bibinfo {author} {\bibfnamefont {D.}~\bibnamefont {Liu}}, \bibinfo
  {author} {\bibfnamefont {E.}~\bibnamefont {Zhao}}, \bibinfo {author}
  {\bibfnamefont {C.}~\bibnamefont {He}}, \bibinfo {author} {\bibfnamefont
  {K.~K.}\ \bibnamefont {Pak}}, \bibinfo {author} {\bibfnamefont
  {J.}~\bibnamefont {Li}},\ and\ \bibinfo {author} {\bibfnamefont {G.-B.}\
  \bibnamefont {Jo}},\ }\bibfield  {title} {\bibinfo {title} {Chiral control of
  quantum states in non-hermitian spin--orbit-coupled fermions},\ }\href@noop
  {} {\bibfield  {journal} {\bibinfo  {journal} {Nature Physics}\ }\textbf
  {\bibinfo {volume} {18}},\ \bibinfo {pages} {385} (\bibinfo {year}
  {2022})}\BibitemShut {NoStop}%
\bibitem [{\citenamefont {Nakagawa}\ \emph {et~al.}(2018)\citenamefont
  {Nakagawa}, \citenamefont {Kawakami},\ and\ \citenamefont
  {Ueda}}]{nakagawa2018non}%
  \BibitemOpen
  \bibfield  {author} {\bibinfo {author} {\bibfnamefont {M.}~\bibnamefont
  {Nakagawa}}, \bibinfo {author} {\bibfnamefont {N.}~\bibnamefont {Kawakami}},\
  and\ \bibinfo {author} {\bibfnamefont {M.}~\bibnamefont {Ueda}},\ }\bibfield
  {title} {\bibinfo {title} {Non-hermitian kondo effect in ultracold
  alkaline-earth atoms},\ }\href@noop {} {\bibfield  {journal} {\bibinfo
  {journal} {Physical review letters}\ }\textbf {\bibinfo {volume} {121}},\
  \bibinfo {pages} {203001} (\bibinfo {year} {2018})}\BibitemShut {NoStop}%
\bibitem [{\citenamefont {Cao}\ \emph {et~al.}(2023)\citenamefont {Cao},
  \citenamefont {Li}, \citenamefont {Zhao}, \citenamefont {Guo}, \citenamefont
  {Qi}, \citenamefont {Chang}, \citenamefont {Zhou}, \citenamefont {Xu},\ and\
  \citenamefont {Duan}}]{cao2023probing}%
  \BibitemOpen
  \bibfield  {author} {\bibinfo {author} {\bibfnamefont {M.-M.}\ \bibnamefont
  {Cao}}, \bibinfo {author} {\bibfnamefont {K.}~\bibnamefont {Li}}, \bibinfo
  {author} {\bibfnamefont {W.-D.}\ \bibnamefont {Zhao}}, \bibinfo {author}
  {\bibfnamefont {W.-X.}\ \bibnamefont {Guo}}, \bibinfo {author} {\bibfnamefont
  {B.-X.}\ \bibnamefont {Qi}}, \bibinfo {author} {\bibfnamefont {X.-Y.}\
  \bibnamefont {Chang}}, \bibinfo {author} {\bibfnamefont {Z.-C.}\ \bibnamefont
  {Zhou}}, \bibinfo {author} {\bibfnamefont {Y.}~\bibnamefont {Xu}},\ and\
  \bibinfo {author} {\bibfnamefont {L.-M.}\ \bibnamefont {Duan}},\ }\bibfield
  {title} {\bibinfo {title} {Probing complex-energy topology via non-hermitian
  absorption spectroscopy in a trapped ion simulator},\ }\href@noop {}
  {\bibfield  {journal} {\bibinfo  {journal} {Physical Review Letters}\
  }\textbf {\bibinfo {volume} {130}},\ \bibinfo {pages} {163001} (\bibinfo
  {year} {2023})}\BibitemShut {NoStop}%
\bibitem [{\citenamefont {Lu}\ \emph {et~al.}(2024)\citenamefont {Lu},
  \citenamefont {Liu}, \citenamefont {Liu}, \citenamefont {Rao}, \citenamefont
  {Lao}, \citenamefont {Wu}, \citenamefont {Zhu},\ and\ \citenamefont
  {Luo}}]{lu2024realizing}%
  \BibitemOpen
  \bibfield  {author} {\bibinfo {author} {\bibfnamefont {P.}~\bibnamefont
  {Lu}}, \bibinfo {author} {\bibfnamefont {T.}~\bibnamefont {Liu}}, \bibinfo
  {author} {\bibfnamefont {Y.}~\bibnamefont {Liu}}, \bibinfo {author}
  {\bibfnamefont {X.}~\bibnamefont {Rao}}, \bibinfo {author} {\bibfnamefont
  {Q.}~\bibnamefont {Lao}}, \bibinfo {author} {\bibfnamefont {H.}~\bibnamefont
  {Wu}}, \bibinfo {author} {\bibfnamefont {F.}~\bibnamefont {Zhu}},\ and\
  \bibinfo {author} {\bibfnamefont {L.}~\bibnamefont {Luo}},\ }\bibfield
  {title} {\bibinfo {title} {Realizing quantum speed limit in open system with
  a-symmetric trapped-ion qubit},\ }\href@noop {} {\bibfield  {journal}
  {\bibinfo  {journal} {New Journal of Physics}\ }\textbf {\bibinfo {volume}
  {26}},\ \bibinfo {pages} {013043} (\bibinfo {year} {2024})}\BibitemShut
  {NoStop}%
\bibitem [{\citenamefont {Dogra}\ \emph {et~al.}(2021)\citenamefont {Dogra},
  \citenamefont {Melnikov},\ and\ \citenamefont {Paraoanu}}]{dogra2021quantum}%
  \BibitemOpen
  \bibfield  {author} {\bibinfo {author} {\bibfnamefont {S.}~\bibnamefont
  {Dogra}}, \bibinfo {author} {\bibfnamefont {A.~A.}\ \bibnamefont
  {Melnikov}},\ and\ \bibinfo {author} {\bibfnamefont {G.~S.}\ \bibnamefont
  {Paraoanu}},\ }\bibfield  {title} {\bibinfo {title} {Quantum simulation of
  parity--time symmetry breaking with a superconducting quantum processor},\
  }\href@noop {} {\bibfield  {journal} {\bibinfo  {journal} {Communications
  Physics}\ }\textbf {\bibinfo {volume} {4}},\ \bibinfo {pages} {26} (\bibinfo
  {year} {2021})}\BibitemShut {NoStop}%
\bibitem [{\citenamefont {Chen}\ \emph {et~al.}(2021)\citenamefont {Chen},
  \citenamefont {Abbasi}, \citenamefont {Joglekar},\ and\ \citenamefont
  {Murch}}]{chen2021quantum}%
  \BibitemOpen
  \bibfield  {author} {\bibinfo {author} {\bibfnamefont {W.}~\bibnamefont
  {Chen}}, \bibinfo {author} {\bibfnamefont {M.}~\bibnamefont {Abbasi}},
  \bibinfo {author} {\bibfnamefont {Y.~N.}\ \bibnamefont {Joglekar}},\ and\
  \bibinfo {author} {\bibfnamefont {K.~W.}\ \bibnamefont {Murch}},\ }\bibfield
  {title} {\bibinfo {title} {Quantum jumps in the non-hermitian dynamics of a
  superconducting qubit},\ }\href@noop {} {\bibfield  {journal} {\bibinfo
  {journal} {Physical Review Letters}\ }\textbf {\bibinfo {volume} {127}},\
  \bibinfo {pages} {140504} (\bibinfo {year} {2021})}\BibitemShut {NoStop}%
\bibitem [{\citenamefont {Zhang}\ \emph {et~al.}(2020)\citenamefont {Zhang},
  \citenamefont {Cheng}, \citenamefont {Zhang},\ and\ \citenamefont
  {Zhai}}]{zhang2020controlling}%
  \BibitemOpen
  \bibfield  {author} {\bibinfo {author} {\bibfnamefont {R.}~\bibnamefont
  {Zhang}}, \bibinfo {author} {\bibfnamefont {Y.}~\bibnamefont {Cheng}},
  \bibinfo {author} {\bibfnamefont {P.}~\bibnamefont {Zhang}},\ and\ \bibinfo
  {author} {\bibfnamefont {H.}~\bibnamefont {Zhai}},\ }\bibfield  {title}
  {\bibinfo {title} {Controlling the interaction of ultracold alkaline-earth
  atoms},\ }\href@noop {} {\bibfield  {journal} {\bibinfo  {journal} {Nature
  Reviews Physics}\ }\textbf {\bibinfo {volume} {2}},\ \bibinfo {pages} {213}
  (\bibinfo {year} {2020})}\BibitemShut {NoStop}%
\bibitem [{\citenamefont {Gorshkov}\ \emph {et~al.}(2010)\citenamefont
  {Gorshkov}, \citenamefont {Hermele}, \citenamefont {Gurarie}, \citenamefont
  {Xu}, \citenamefont {Julienne}, \citenamefont {Ye}, \citenamefont {Zoller},
  \citenamefont {Demler}, \citenamefont {Lukin},\ and\ \citenamefont
  {Rey}}]{gorshkov2010two}%
  \BibitemOpen
  \bibfield  {author} {\bibinfo {author} {\bibfnamefont {A.~V.}\ \bibnamefont
  {Gorshkov}}, \bibinfo {author} {\bibfnamefont {M.}~\bibnamefont {Hermele}},
  \bibinfo {author} {\bibfnamefont {V.}~\bibnamefont {Gurarie}}, \bibinfo
  {author} {\bibfnamefont {C.}~\bibnamefont {Xu}}, \bibinfo {author}
  {\bibfnamefont {P.~S.}\ \bibnamefont {Julienne}}, \bibinfo {author}
  {\bibfnamefont {J.}~\bibnamefont {Ye}}, \bibinfo {author} {\bibfnamefont
  {P.}~\bibnamefont {Zoller}}, \bibinfo {author} {\bibfnamefont
  {E.}~\bibnamefont {Demler}}, \bibinfo {author} {\bibfnamefont {M.~D.}\
  \bibnamefont {Lukin}},\ and\ \bibinfo {author} {\bibfnamefont
  {A.}~\bibnamefont {Rey}},\ }\bibfield  {title} {\bibinfo {title} {Two-orbital
  su (n) magnetism with ultracold alkaline-earth atoms},\ }\href@noop {}
  {\bibfield  {journal} {\bibinfo  {journal} {Nature physics}\ }\textbf
  {\bibinfo {volume} {6}},\ \bibinfo {pages} {289} (\bibinfo {year}
  {2010})}\BibitemShut {NoStop}%
\bibitem [{\citenamefont {Jepsen}\ \emph {et~al.}(2020)\citenamefont {Jepsen},
  \citenamefont {Amato-Grill}, \citenamefont {Dimitrova}, \citenamefont {Ho},
  \citenamefont {Demler},\ and\ \citenamefont {Ketterle}}]{jepsen2020spin}%
  \BibitemOpen
  \bibfield  {author} {\bibinfo {author} {\bibfnamefont {P.~N.}\ \bibnamefont
  {Jepsen}}, \bibinfo {author} {\bibfnamefont {J.}~\bibnamefont {Amato-Grill}},
  \bibinfo {author} {\bibfnamefont {I.}~\bibnamefont {Dimitrova}}, \bibinfo
  {author} {\bibfnamefont {W.~W.}\ \bibnamefont {Ho}}, \bibinfo {author}
  {\bibfnamefont {E.}~\bibnamefont {Demler}},\ and\ \bibinfo {author}
  {\bibfnamefont {W.}~\bibnamefont {Ketterle}},\ }\bibfield  {title} {\bibinfo
  {title} {Spin transport in a tunable heisenberg model realized with ultracold
  atoms},\ }\href@noop {} {\bibfield  {journal} {\bibinfo  {journal} {Nature}\
  }\textbf {\bibinfo {volume} {588}},\ \bibinfo {pages} {403} (\bibinfo {year}
  {2020})}\BibitemShut {NoStop}%
\bibitem [{\citenamefont {Schwager}\ \emph {et~al.}(2013)\citenamefont
  {Schwager}, \citenamefont {Cirac},\ and\ \citenamefont
  {Giedke}}]{schwager2013dissipative}%
  \BibitemOpen
  \bibfield  {author} {\bibinfo {author} {\bibfnamefont {H.}~\bibnamefont
  {Schwager}}, \bibinfo {author} {\bibfnamefont {J.~I.}\ \bibnamefont
  {Cirac}},\ and\ \bibinfo {author} {\bibfnamefont {G.}~\bibnamefont
  {Giedke}},\ }\bibfield  {title} {\bibinfo {title} {Dissipative spin chains:
  Implementation with cold atoms and steady-state properties},\ }\href@noop {}
  {\bibfield  {journal} {\bibinfo  {journal} {Physical Review A}\ }\textbf
  {\bibinfo {volume} {87}},\ \bibinfo {pages} {022110} (\bibinfo {year}
  {2013})}\BibitemShut {NoStop}%
\bibitem [{\citenamefont {Kattel}\ \emph
  {et~al.}(2024{\natexlab{c}})\citenamefont {Kattel}, \citenamefont {Zhakenov},
  \citenamefont {Pasnoori}, \citenamefont {Azaria},\ and\ \citenamefont
  {Andrei}}]{kattel2024dissipation}%
  \BibitemOpen
  \bibfield  {author} {\bibinfo {author} {\bibfnamefont {P.}~\bibnamefont
  {Kattel}}, \bibinfo {author} {\bibfnamefont {A.}~\bibnamefont {Zhakenov}},
  \bibinfo {author} {\bibfnamefont {P.~R.}\ \bibnamefont {Pasnoori}}, \bibinfo
  {author} {\bibfnamefont {P.}~\bibnamefont {Azaria}},\ and\ \bibinfo {author}
  {\bibfnamefont {N.}~\bibnamefont {Andrei}},\ }\bibfield  {title} {\bibinfo
  {title} {Dissipation driven phase transition in the non-hermitian kondo
  model},\ }\href@noop {} {\bibfield  {journal} {\bibinfo  {journal} {arXiv
  preprint arXiv:2402.09510}\ } (\bibinfo {year}
  {2024}{\natexlab{c}})}\BibitemShut {NoStop}%
\bibitem [{\citenamefont {Affleck}(1986)}]{affleck1986exact}%
  \BibitemOpen
  \bibfield  {author} {\bibinfo {author} {\bibfnamefont {I.}~\bibnamefont
  {Affleck}},\ }\bibfield  {title} {\bibinfo {title} {Exact critical exponents
  for quantum spin chains, non-linear $\sigma$-models at $\theta$= $\pi$ and
  the quantum hall effect},\ }\href@noop {} {\bibfield  {journal} {\bibinfo
  {journal} {Nuclear Physics B}\ }\textbf {\bibinfo {volume} {265}},\ \bibinfo
  {pages} {409} (\bibinfo {year} {1986})}\BibitemShut {NoStop}%
\bibitem [{\citenamefont {Laflorencie}\ \emph {et~al.}(2008)\citenamefont
  {Laflorencie}, \citenamefont {S{\o}rensen},\ and\ \citenamefont
  {Affleck}}]{laflorencie2008kondo}%
  \BibitemOpen
  \bibfield  {author} {\bibinfo {author} {\bibfnamefont {N.}~\bibnamefont
  {Laflorencie}}, \bibinfo {author} {\bibfnamefont {E.~S.}\ \bibnamefont
  {S{\o}rensen}},\ and\ \bibinfo {author} {\bibfnamefont {I.}~\bibnamefont
  {Affleck}},\ }\bibfield  {title} {\bibinfo {title} {The kondo effect in spin
  chains},\ }\href@noop {} {\bibfield  {journal} {\bibinfo  {journal} {Journal
  of Statistical Mechanics: Theory and Experiment}\ }\textbf {\bibinfo {volume}
  {2008}},\ \bibinfo {pages} {P02007} (\bibinfo {year} {2008})}\BibitemShut
  {NoStop}%
\bibitem [{\citenamefont {Destri}\ and\ \citenamefont
  {Lowenstein}(1982)}]{destri1982analysis}%
  \BibitemOpen
  \bibfield  {author} {\bibinfo {author} {\bibfnamefont {C.}~\bibnamefont
  {Destri}}\ and\ \bibinfo {author} {\bibfnamefont {J.}~\bibnamefont
  {Lowenstein}},\ }\bibfield  {title} {\bibinfo {title} {Analysis of the
  bethe-ansatz equations of the chiral-invariant gross-neveu model},\
  }\href@noop {} {\bibfield  {journal} {\bibinfo  {journal} {Nuclear Physics
  B}\ }\textbf {\bibinfo {volume} {205}},\ \bibinfo {pages} {369} (\bibinfo
  {year} {1982})}\BibitemShut {NoStop}%
\bibitem [{\citenamefont {Avraham}(2024)}]{pcref}%
  \BibitemOpen
  \bibfield  {author} {\bibinfo {author} {\bibfnamefont {N.}~\bibnamefont
  {Avraham}},\ }\href@noop {} {\bibinfo {title} {private communication}}
  (\bibinfo {year} {2024})\BibitemShut {NoStop}%
\bibitem [{\citenamefont {Andrei}\ \emph {et~al.}(1983)\citenamefont {Andrei},
  \citenamefont {Furuya},\ and\ \citenamefont
  {Lowenstein}}]{andrei1983solution}%
  \BibitemOpen
  \bibfield  {author} {\bibinfo {author} {\bibfnamefont {N.}~\bibnamefont
  {Andrei}}, \bibinfo {author} {\bibfnamefont {K.}~\bibnamefont {Furuya}},\
  and\ \bibinfo {author} {\bibfnamefont {J.}~\bibnamefont {Lowenstein}},\
  }\bibfield  {title} {\bibinfo {title} {Solution of the kondo problem},\
  }\href@noop {} {\bibfield  {journal} {\bibinfo  {journal} {Reviews of modern
  physics}\ }\textbf {\bibinfo {volume} {55}},\ \bibinfo {pages} {331}
  (\bibinfo {year} {1983})}\BibitemShut {NoStop}%
\bibitem [{\citenamefont {Kapustin}\ and\ \citenamefont
  {Skorik}(1996)}]{kapustin1996surface}%
  \BibitemOpen
  \bibfield  {author} {\bibinfo {author} {\bibfnamefont {A.}~\bibnamefont
  {Kapustin}}\ and\ \bibinfo {author} {\bibfnamefont {S.}~\bibnamefont
  {Skorik}},\ }\bibfield  {title} {\bibinfo {title} {Surface excitations and
  surface energy of the antiferromagnetic xxz chain by the bethe ansatz
  approach},\ }\href@noop {} {\bibfield  {journal} {\bibinfo  {journal}
  {Journal of Physics A: Mathematical and General}\ }\textbf {\bibinfo {volume}
  {29}},\ \bibinfo {pages} {1629} (\bibinfo {year} {1996})}\BibitemShut
  {NoStop}%
\bibitem [{\citenamefont {Pasnoori}\ \emph {et~al.}(2020)\citenamefont
  {Pasnoori}, \citenamefont {Rylands},\ and\ \citenamefont
  {Andrei}}]{pasnoori2020kondo}%
  \BibitemOpen
  \bibfield  {author} {\bibinfo {author} {\bibfnamefont {P.~R.}\ \bibnamefont
  {Pasnoori}}, \bibinfo {author} {\bibfnamefont {C.}~\bibnamefont {Rylands}},\
  and\ \bibinfo {author} {\bibfnamefont {N.}~\bibnamefont {Andrei}},\
  }\bibfield  {title} {\bibinfo {title} {Kondo impurity at the edge of a
  superconducting wire},\ }\href@noop {} {\bibfield  {journal} {\bibinfo
  {journal} {Physical Review Research}\ }\textbf {\bibinfo {volume} {2}},\
  \bibinfo {pages} {013006} (\bibinfo {year} {2020})}\BibitemShut {NoStop}%
\bibitem [{\citenamefont {Pasnoori}\ \emph {et~al.}(2021)\citenamefont
  {Pasnoori}, \citenamefont {Andrei},\ and\ \citenamefont
  {Azaria}}]{pasnoori2021boundary}%
  \BibitemOpen
  \bibfield  {author} {\bibinfo {author} {\bibfnamefont {P.~R.}\ \bibnamefont
  {Pasnoori}}, \bibinfo {author} {\bibfnamefont {N.}~\bibnamefont {Andrei}},\
  and\ \bibinfo {author} {\bibfnamefont {P.}~\bibnamefont {Azaria}},\
  }\bibfield  {title} {\bibinfo {title} {Boundary-induced topological and
  mid-gap states in charge conserving one-dimensional superconductors:
  Fractionalization transition},\ }\href@noop {} {\bibfield  {journal}
  {\bibinfo  {journal} {Physical Review B}\ }\textbf {\bibinfo {volume}
  {104}},\ \bibinfo {pages} {134519} (\bibinfo {year} {2021})}\BibitemShut
  {NoStop}%
\bibitem [{\citenamefont {Pasnoori}\ \emph {et~al.}(2022)\citenamefont
  {Pasnoori}, \citenamefont {Andrei}, \citenamefont {Rylands},\ and\
  \citenamefont {Azaria}}]{pasnoori2022rise}%
  \BibitemOpen
  \bibfield  {author} {\bibinfo {author} {\bibfnamefont {P.~R.}\ \bibnamefont
  {Pasnoori}}, \bibinfo {author} {\bibfnamefont {N.}~\bibnamefont {Andrei}},
  \bibinfo {author} {\bibfnamefont {C.}~\bibnamefont {Rylands}},\ and\ \bibinfo
  {author} {\bibfnamefont {P.}~\bibnamefont {Azaria}},\ }\bibfield  {title}
  {\bibinfo {title} {Rise and fall of yu-shiba-rusinov bound states in
  charge-conserving s-wave one-dimensional superconductors},\ }\href@noop {}
  {\bibfield  {journal} {\bibinfo  {journal} {Physical Review B}\ }\textbf
  {\bibinfo {volume} {105}},\ \bibinfo {pages} {174517} (\bibinfo {year}
  {2022})}\BibitemShut {NoStop}%
\bibitem [{\citenamefont {Rylands}(2020)}]{rylands2020exact}%
  \BibitemOpen
  \bibfield  {author} {\bibinfo {author} {\bibfnamefont {C.}~\bibnamefont
  {Rylands}},\ }\bibfield  {title} {\bibinfo {title} {Exact boundary modes in
  an interacting quantum wire},\ }\href@noop {} {\bibfield  {journal} {\bibinfo
   {journal} {Physical Review B}\ }\textbf {\bibinfo {volume} {101}},\ \bibinfo
  {pages} {085133} (\bibinfo {year} {2020})}\BibitemShut {NoStop}%
\bibitem [{\citenamefont {Pasnoori}\ \emph {et~al.}(2023)\citenamefont
  {Pasnoori}, \citenamefont {Lee}, \citenamefont {Pixley}, \citenamefont
  {Andrei},\ and\ \citenamefont {Azaria}}]{PPSPhysRevB.107.224412}%
  \BibitemOpen
  \bibfield  {author} {\bibinfo {author} {\bibfnamefont {P.~R.}\ \bibnamefont
  {Pasnoori}}, \bibinfo {author} {\bibfnamefont {J.}~\bibnamefont {Lee}},
  \bibinfo {author} {\bibfnamefont {J.~H.}\ \bibnamefont {Pixley}}, \bibinfo
  {author} {\bibfnamefont {N.}~\bibnamefont {Andrei}},\ and\ \bibinfo {author}
  {\bibfnamefont {P.}~\bibnamefont {Azaria}},\ }\bibfield  {title} {\bibinfo
  {title} {Boundary quantum phase transitions in the spin-$\frac{1}{2}$
  heisenberg chain with boundary magnetic fields},\ }\href
  {https://doi.org/10.1103/PhysRevB.107.224412} {\bibfield  {journal} {\bibinfo
   {journal} {Phys. Rev. B}\ }\textbf {\bibinfo {volume} {107}},\ \bibinfo
  {pages} {224412} (\bibinfo {year} {2023})}\BibitemShut {NoStop}%
\bibitem [{Note2()}]{Note2}%
  \BibitemOpen
  \bibinfo {note} {The energy density becomes independent of the system size
  for some values of $\gamma $ for $\beta $ approximately from 0.35 to 1.17.
  However, since the energy function is written in terms of digamma functions
  which are difficult to invert, the analytic expression for the relation
  between $\beta $ and $\gamma $ where the energy density becomes independent
  of system size is difficult to obtain.}\BibitemShut {Stop}%
\bibitem [{Note3()}]{Note3}%
  \BibitemOpen
  \bibinfo {note} {In the Hermitian limit $\gamma \to 0$, notice that the
  expression in Equation \protect \eqref {engallreal} exhibits superfluous
  divergence due to $E_b=\csc (\pi (\beta \pm i\gamma ))$ blowing up for $\beta
  \in \protect \mathbb {Z^+}$. However, there is also a divergent factor in
  $E_{\mathinner {|{0}\rangle }}$ in the digamma functions that cancels out the
  divergences in the energy of the boundary string. Thus, one can explicitly
  write $E_{ar}$ as shown in Eq.\protect \eqref {engallreal} which is
  divergence free.}\BibitemShut {Stop}%
\end{thebibliography}%

\begin{appendix}
\begin{widetext}
\section{Integrability of the Hamiltonian}
The rational 6-vertex $R$ matrix 
\begin{equation}
    R(\lambda)=\lambda \mathbb{I}+P=\left(
\begin{array}{cccc}
 \lambda +1 & 0 & 0 & 0 \\
 0 & \lambda  & 1 & 0 \\
 0 & 1 & \lambda  & 0 \\
 0 & 0 & 0 & \lambda +1 \\
\end{array}
\right),
\end{equation}
where $\mathbb{I}_{ab,cd}=\delta_{ab}\delta_{cd}$ is the identity matrix and $P_{ab,cd}=\delta_{a d}\delta_{c b} $ is the permutation operator $P(x\otimes y)=y \otimes x$, acts non trivially on the product space $\mathbb{C}^2\otimes \mathbb{C}^2$ such that in $\operatorname{End}(\mathbb{C}^2\otimes \mathbb{C}^2 \otimes \mathbb{C}^2$), it satisfies the a non-linear matrix equation
\begin{equation}
    R_{ij}(\lambda-\mu)R_{ik}(\lambda-\nu)R_{jk}(\mu-\nu)=R_{jk}(\mu-\nu)R_{ik}(\lambda-\nu)R_{ij}(\lambda-\mu)
    \label{qYBE}
\end{equation}
called quantum Yang-Baxter equation (qYBE).
We introduce two transfer matrices $T_0(\lambda)$ and $\hat T_0(\lambda)$ 
\begin{align}
T_0(\lambda)&=R_{0,R}(\lambda-b)R_{0,\bar N}(\lambda)R_{0,\bar N-1}(\lambda)\cdots R_{0,2}(\lambda)R_{0,1}(\lambda)R_{0,L}(\lambda-b^*)\\
\hat T_0(\lambda)&=R_{0,L}(\lambda+b^*)R_{0,1}(\lambda)R_{0,2}(\lambda)\cdots R_{0,\bar N-1}(\lambda)L_{0,\bar N}(\lambda)R_{0,R}(\lambda+b)
\end{align}
Here each $R$ matrix  $R_{0j}(\lambda)$ with the spectral parameter $\lambda \in \mathbb{C}$  is a linear operator acting on the product space $\mathfrak h_0\otimes \mathfrak h_k$ where $\mathfrak h_k \cong \mathbb{C}^2$ is the local Hilbert space of either the bulk spin (with index 1 through $\bar N$) or two boundary impurity spins (with index L and R) and $\mathfrak h_0 \cong \mathbb{C}^2$ is called auxiliary space. The spectral parameter is shifted by $b\in \mathbb{C}$ and its complex conjugate $b^*$ for the right and left impurities, respectively. The trace of the product of two transfer matrices over the auxiliary space is defined as the double row transfer matrix
\begin{equation}
    t(\lambda)=\operatorname{tr}_0\left( T_0(\lambda)\hat T_0(\lambda)\right)
    \label{tmat}
\end{equation}

It follows from the qYBE Eq.\eqref{qYBE} that the transfer matrix is a one-parameter family of commuting operator \textit{i.e.}
\begin{equation}
    [t(\lambda),t(\rho)]=0 \quad\quad \text{where } \lambda,\rho \in \mathbb{C}
\end{equation}
such that it can act as a generator of infinite number of commuting operators which can be diagonalized simultaneously. One of such operators is the Hamiltonian Eq.\eqref{modelHam} which is obtained from the transfer matrix as
\begin{equation}
\mathcal{H} = J \frac{\mathrm{d}}{\mathrm{d}\lambda} \log(\lambda)\bigg\vert_{\lambda\to 0} -\bar NJ - \frac{J}{1 - b^2} - \frac{J}{1 - (b^*)^2}
    \label{ham}
\end{equation}
upon identifying the complex boundary coupling constant as
\begin{equation}
    J_\mathrm{imp}=\frac{J}{1-b^2} .
    \label{parrel}
\end{equation}

Each eigenvalues $\Lambda(\lambda)$ of the transfer matrix Eq.\eqref{tmat} is given by the Baxter's T-Q relation
\begin{equation}
\Lambda(\lambda) = \frac{(\lambda+1)^{2\bar N+1}((\lambda+1)^2 - b^2)((\lambda+1)^2 - (b^*)^2)}{2\lambda+1} \frac{Q(\lambda-1)}{Q(\lambda)} \\+ \frac{(\lambda^2 - b^2)(\lambda^2 - (b^*)^2) \lambda^{2\bar N+1}}{2\lambda+1} \frac{Q(\lambda+1)}{Q(\lambda)}
        \label{eval}
\end{equation}

where the \(Q\)-function is given by
\begin{equation}
Q(\lambda)=\prod_{\ell=1}^M (\lambda-\lambda_\ell)(\lambda+\lambda_\ell+1)
\end{equation}

Regularity of the \(T-Q\) equation gives the BAEs
\begin{equation}
\left(\frac{\lambda_j +1}{\lambda_j}\right)^{2 \bar N+1}\frac{\lambda_j +b+1}{\lambda_j +b}\frac{\lambda_j -b+1}{\lambda_j -b}\frac{\lambda_j +b^*+1}{\lambda_j +b^*}\frac{\lambda_j -b^*+1}{\lambda_j -b^*}\\=-\prod_{\ell=1}^M\frac{(\lambda_j-\lambda_\ell+1)(\lambda_j+\lambda_\ell+2)}{(\lambda_j-\lambda_\ell-1)(\lambda_j+\lambda_\ell)}
\end{equation}

Changing the variable \(\lambda_j=i\mu_j-\frac{1}{2}\) and introducing $b=\beta+i\gamma$ where $\beta,\gamma\in \mathbb{R}$, we rewrite the above equation as
\begin{multline}
   \left(\frac{\mu _j-\frac{i}{2}}{\mu _j+\frac{i}{2}}\right)^{2 \bar N} \frac{\mu_j-\gamma-i\left(\frac{1}{2}-\beta\right)}{\mu_j-\gamma+i\left(\frac{1}{2}-\beta\right)} 
   \frac{\mu_j+\gamma-i\left(\frac{1}{2}-\beta\right)}{\mu_j+\gamma+i\left(\frac{1}{2}-\beta\right)}
   \frac{\mu_j+\gamma-i\left(\frac{1}{2}+\beta\right)}{\mu_j+\gamma+i\left(\frac{1}{2}+\beta\right)}
   \frac{\mu_j-\gamma-i\left(\frac{1}{2}+\beta\right)}{\mu_j-\gamma+i\left(\frac{1}{2}+\beta\right)}\\=\prod_{j \neq \ell=1}^{M}\left(\frac{\mu_{j}-\mu_{\ell}-i}{\mu_{j}-\mu_{\ell}+i}\right)\left(\frac{\mu_{j}+\mu_{\ell}-i}{\mu_{j}+\mu_{\ell}+i}\right)
\label{bae}
\end{multline}

From Eq.\eqref{ham} and Eq.\eqref{eval}, the energy eigenvalues are obtained
\begin{equation}
E=-\sum _{j=1}^M \frac{2J}{\mu _j^2+\frac{1}{4}}+J (\bar N-1)+\frac{2 J \left(\gamma ^2-\beta ^2+1\right)}{\beta ^4+2 \beta ^2 \left(\gamma ^2-1\right)+\left(\gamma ^2+1\right)^2}
\label{cengeng}
\end{equation}

In terms of the bulk coupling $J$ and boundary coupling $J_\mathrm{imp}$, the parameters $\beta$ and $\gamma$ can be written as
\begin{equation}
    \beta = \frac{\sqrt{| J_{\mathrm{imp}} | ^2 \left(2 | J_{\mathrm{imp}} | ^2 - 2 J \operatorname{Re}(J_{\mathrm{imp}})\right) + 2 | J_{\mathrm{imp}} | ^3 | J - J_{\mathrm{imp}} | }}{2 | J_{\mathrm{imp}} | ^2}
    \label{betadef}
\end{equation}
and
\begin{equation}
    \gamma = \frac{\sqrt{| (J - J_{\mathrm{imp}}) J_{\mathrm{imp}} | + | J_{\mathrm{imp}} | ^2 - J \operatorname{Re}(J_{\mathrm{imp}})} \left(| (J - J_{\mathrm{imp}}) J_{\mathrm{imp}} | - J_{\mathrm{imp}} J_{\mathrm{imp}}^* + J \operatorname{Re}(J_{\mathrm{imp}})\right)}{\sqrt{2} J | J_{\mathrm{imp}} | \operatorname{Im}(J_{\mathrm{imp}})}
    \label{gammadef}
\end{equation}
In the logarithmic form, the Bethe equations Eq.\eqref{bae} becomes
\begin{multline}
(2\bar N+1)\arctan(2\mu_j)+\arctan\left(\frac{\mu_j-\gamma}{\frac{1}{2}-\beta} \right)+\arctan\left(\frac{\mu_j+\gamma}{\frac{1}{2}-\beta} \right)+\arctan\left(\frac{\mu_j-\gamma}{\frac{1}{2}+\beta} \right)\\+\arctan\left(\frac{\mu_j+\gamma}{\frac{1}{2}+\beta} \right)=\sum_{\ell=1}^M\left[\tan ^{-1}\left(\mu_j-\mu_{\ell}\right)+\tan ^{-1}\left(\mu_j+\mu_{\ell}\right)\right]+\pi I_j .
    \label{cFBAE}
\end{multline}

Quantum numbers $I_j$ occupy a symmetric interval around the origin. Moreover, thanks to the presence of $\mathscr{PT-}$symmetry in the Hamiltonian, the rapidities $\mu_j$ in the ground state are real. This, to analyze Eq.\eqref{cFBAE} in thermodynamics limit, we define the density of Bethe roots as
\begin{equation}
\rho(\mu_j)=\frac{1}{\mu_{j+1}-\mu_j}
\end{equation}
Such that we convert the sums over $j$ in Eq.\eqref{cengeng} and Eq.\eqref{cFBAE} into integral over $\mu$ as
\begin{equation}
   E=-\int_{-\infty}^\infty  \mathrm{d}\mu \rho(\mu) \frac{2J}{\mu _j^2+\frac{1}{4}}+J (\bar N-1)2+\frac{2 J \left(\gamma ^2-\beta ^2+1\right)}{\beta ^4+2 \beta ^2 \left(\gamma ^2-1\right)+\left(\gamma ^2+1\right)^2}
   \label{engcont} 
3w\end{equation}
and
\begin{multline}
(2 \bar N+1)\arctan(2\mu_j)+\arctan\left(\frac{\mu_j-\gamma}{\frac{1}{2}-\beta} \right)+\arctan\left(\frac{\mu_j+\gamma}{\frac{1}{2}-\beta} \right)+\arctan\left(\frac{\mu_j-\gamma}{\frac{1}{2}+\beta} \right)\\+\arctan\left(\frac{\mu_j+\gamma}{\frac{1}{2}+\beta} \right)=\int_{-\infty}^\infty \mathrm{d}\mu' \rho(\mu')\left[\tan^{-1}(\mu_j-\mu')+\tan^{-1}(\mu_j+\mu') \right]+\pi I_j
\label{BAEcont}
\end{multline}

We extract the density of roots in the ground state $\rho_0(\mu)$ by subtracting Eq.\eqref{BAEcont} written for $\mu_j$ from the same equation written for $\mu_{j+1}$ and expanding in the difference $\Delta\mu=\mu_{j+1}-\mu_j$. This gives
\begin{equation}
2\rho_0(\mu)=f(\mu)-\int_{-\infty}^\infty\mathrm{d}\mu'~ K(\mu-\mu')\rho_0(\mu')-\int_{-\infty}^\infty\mathrm{d}\mu'~ K(\mu+\mu')\rho_0(\mu')+\mathcal{O}\left(\frac{1}{\bar N}\right)
\label{SolDensity}
\end{equation}
Since, we are interested in thermodynamics limit $\bar N\to \infty$, the higher order terms are negligible.

Here,
\begin{align}
f(\mu)&=(2\bar N+1)a_{\frac12}(\mu)+a_{\frac{1}{2}-\beta}(\mu-\gamma)+a_{\frac{1}{2}-\beta}(\mu+\gamma) +a_{\frac{1}{2}+\beta}(\mu-\gamma)+a_{\frac{1}{2}+\beta}(\mu+\gamma) \\
a_\gamma(\mu)&=\frac{1}{\pi}\frac{\gamma}{\mu^2+\gamma^2}\\
K(\mu)&=\frac{1}{\pi  \left(\mu ^2+1\right)}=a_1(\mu)\label{kernel}
\end{align}

We will now solve the Bethe equation in various phases which are determined by values of $\beta$ and $\gamma$. It is well known that the solution depends on whether the total number of particles is even or odd. We will focus on the case of an even number of particles. The case for an odd number can be easily generalized as shown in detail in Hermitian case in \cite{kattel2023kondo}.

\section{Solution of Bethe Ansatz equations \texorpdfstring{Eq.~\eqref{bae}}{Eq. (1)}}\label{detsol}

We will now solve the Bethe Ansatz equations when both $\beta$ and $\gamma$ are nonzero. In this case the model is non-Hermitian but $\mathscr{PT}-$symmetric as mentioned earlier. 

\subsection{Kondo phase \texorpdfstring{$(0<\beta<\frac{1}{2})$ and $\gamma\neq 0$}{(0 < beta < 1/2) and gamma != 0}}\label{kphase}

The density of the solutions of Bethe equations can be obtained from an integral equation 
\begin{align}
2\rho_{\ket{0}}(\mu)&=(2\bar N+1)a_{\frac12}(\mu)+a_{\frac{1}{2}-\beta}(\mu-\gamma)+a_{\frac{1}{2}-\beta}(\mu+\gamma) +a_{\frac{1}{2}+\beta}(\mu-\gamma)+a_{\frac{1}{2}+\beta}(\mu+\gamma)\nonumber\\
&-\sum_{\upsilon=\pm}\int_{-\infty}^\infty\mathrm{d}\mu~ a_1(\mu+\upsilon\mu')\rho_{\left|0\right\rangle}(\mu')-\delta(\mu)
\label{densimp}
\end{align}

where the delta function is added at the origin because the eigenvector corresponding to $\mu_j=0$ vanishes.

The solution of Eq.\eqref{densimp} is immediate in Fourier space

\begin{align}
\tilde\rho_{\ket{0}}(\omega)=\frac{1}{4} \text{sech}\left(\frac{| \omega | }{2}\right) \left(4 \cos (\gamma  \omega ) \cosh (\beta  | \omega | )-e^{\frac{| \omega | }{2}}+2\bar N+1\right)
\label{soldenskk}
\end{align}
such that the total number of Bethe roots is
\begin{equation}
    M_{\ket{0}}=\int_{-\infty}^\infty \rho_{\ket{0}}(\mu)\mathrm{d}\mu =\tilde \rho_{\ket{0}}(0)=\frac{\bar N+2}{2}
\end{equation}
and hence the total spin in the ground state is 
\begin{equation}
    S^z_{\ket{0}}=\frac{\bar N+2}{2}-M_{\ket{0}}=0
\end{equation}
which shows that the impurity is screened by multiparticle Kondo cloud in the ground state.

Using Eq.\eqref{cengeng}, the energy of this state is 
\begin{multline}
   E_{\ket{0}} = \frac{2 J \left(-\beta ^2+\gamma ^2+1\right)}{\beta ^4+2 \beta ^2 \left(\gamma ^2-1\right)+\left(\gamma ^2+1\right)^2}+J (\bar N-1)-J (2\bar N+1) \log (4)+\pi  J \\
   +2 \operatorname{Re}\Biggl(-J \psi ^{(0)}\left(\frac{\beta }{2}+\frac{i \gamma }{2}+1\right)+J \psi ^{(0)}\left(\frac{\beta }{2}+\frac{i \gamma }{2}+\frac{1}{2}\right)
   -J \psi ^{(0)}\left(-\frac{\beta }{2}+\frac{i \gamma }{2}+1\right)+J \psi ^{(0)}\left(-\frac{\beta }{2}+\frac{i \gamma }{2}+\frac{1}{2}\right)\Biggr)
   \label{gseng}
\end{multline}

We energy density
\begin{equation}
    E_{\mathrm{dens}}=\frac{1}{\bar N+2}E_{\ket{0}}
\end{equation}
for various values of $\beta$ and $\gamma$ in the Kondo phase below using the result from both Bethe Ansatz and DMRG are shown in Fig.\ref{ris}. Notice that since the system has a boundary and impurity, the scattering phase shift from the boundary contributes to the energy and hence the energy density is not independent of the system size.
\begin{figure}[H]
\begin{minipage}[h]{0.47\linewidth}
\begin{center}
\includegraphics[width=1\linewidth]{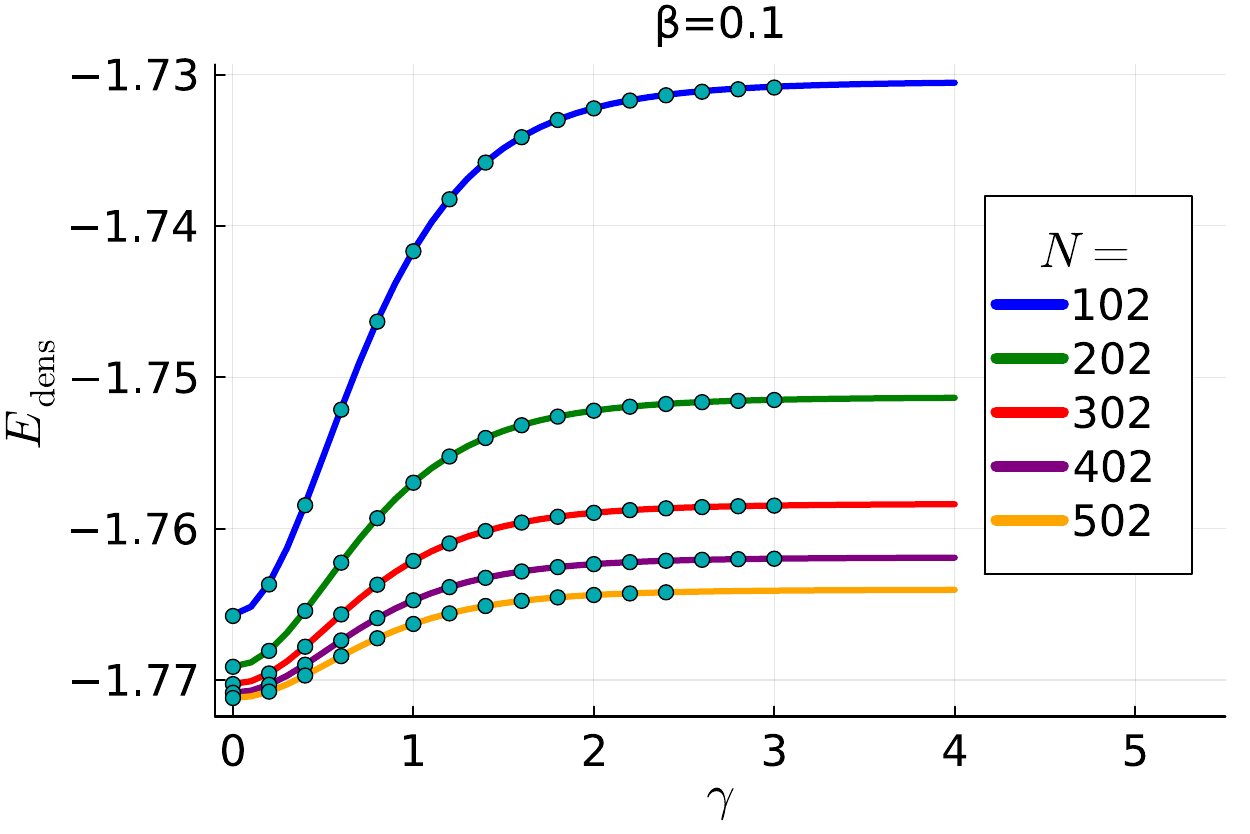} 
\label{qwe1}
\end{center} 
\end{minipage}
\hfill
\vspace{0.2 cm}
\begin{minipage}[h]{0.47\linewidth}
\begin{center}
\includegraphics[width=1\linewidth]{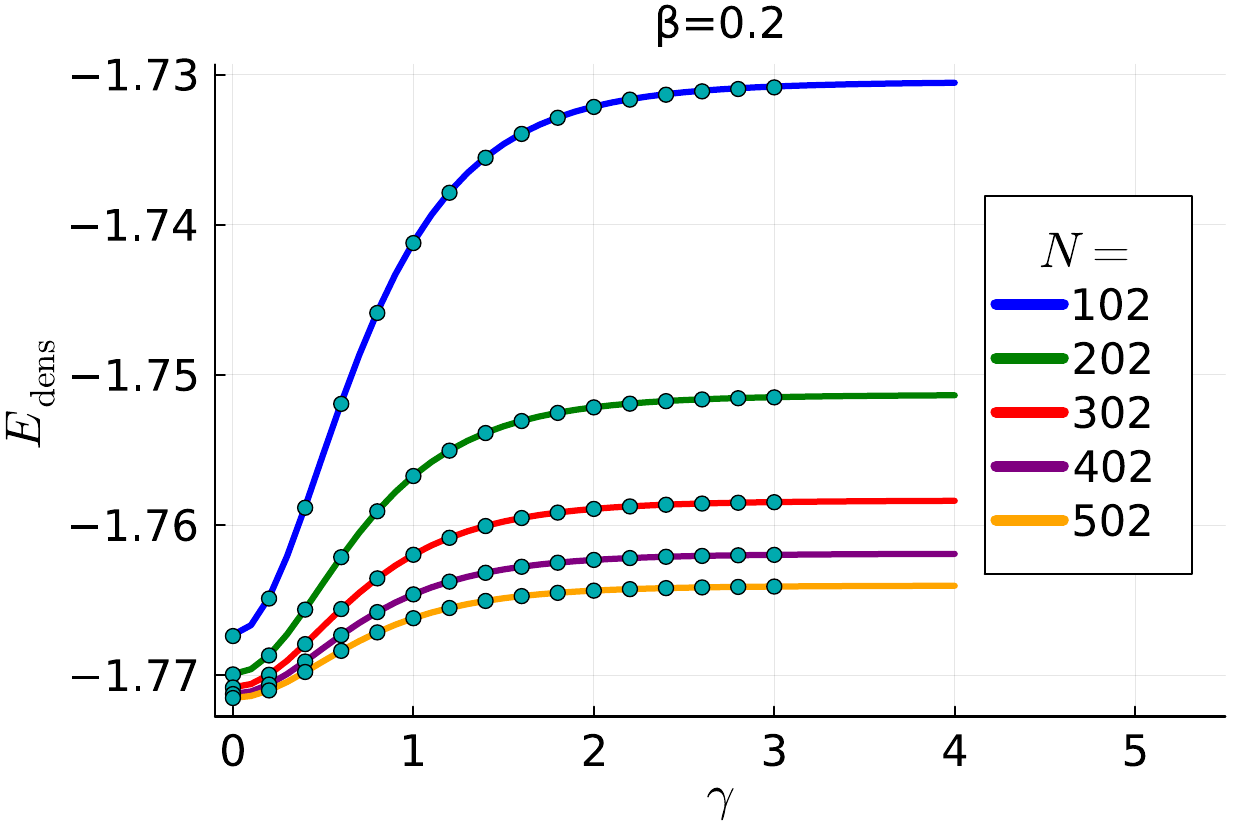} 
\label{qwe2}
\end{center}
\end{minipage}
\vfill
\vspace{0.2 cm}
\begin{minipage}[h]{0.47\linewidth}
\begin{center}
\includegraphics[width=1\linewidth]{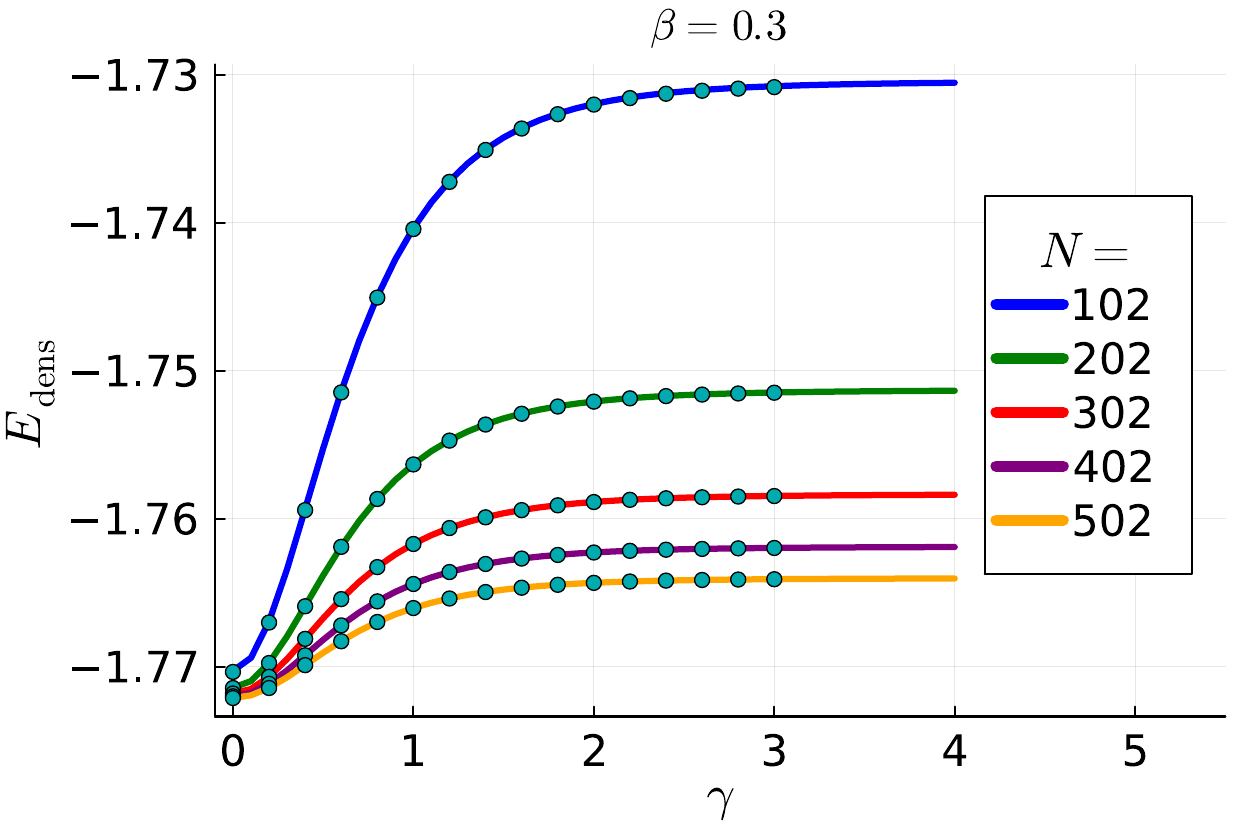} 
\label{qwe3}
\end{center}
\end{minipage}
\hfill
\begin{minipage}[h]{0.47\linewidth}
\begin{center}
\includegraphics[width=1\linewidth]{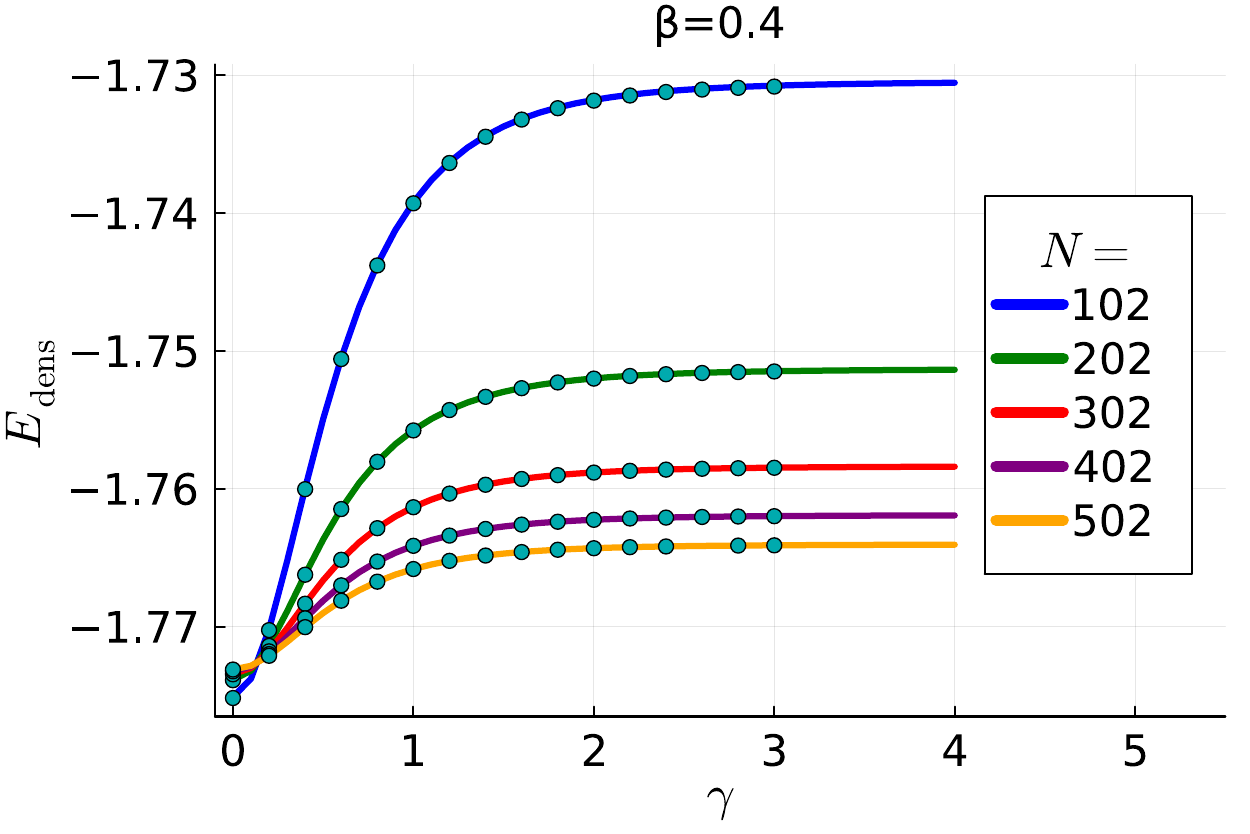} 
\label{qwe4}
\end{center}
\end{minipage}
\caption{Bethe Ansatz and DMRG calculation of ground state energies for various values of complex impurity couplings. Here $N=\bar N+2$ is the total number of spins including the two impurities. Note that the energy density obtained from Bethe Ansatz in the thermodynamic limit and the ones obtained from DMRG using finite size algorithm have a relative difference of about $10^{-5}\%$. }
\label{ris}
\end{figure}

It is important to note that the ground state has real energy in this phase. Also notice that Kondo phase lacks boundary excitations entirely. Moreover, any other excitations in this phase can be constructed by adding an even number of spinons, bulk strings, and quartets, all of which possess real energies. Consequently, all states in the Kondo phase have real energies. Thus, the $\mathscr{PT-}$symmetry is unbroken in this phase.  
\subsubsection{Spinon density of states}
\label{DOS}
From Eq.\eqref{soldenskk}, we compute the 
density of state contribution due to the bulk
\begin{equation}
    \rho_{\mathrm{dos}}^{\mathrm{bulk}}(E) = \left| \frac{\rho_{\ket{0}^{\mathrm{bulk}}}(\mu)}{E'(\mu)} \right| = \frac{\frac{1}{2} \bar N \text{sech}(\pi \mu)}{2 \pi^2 J \tanh (\pi \mu) \text{sech}(\pi \mu)} = \frac{\bar N \coth (\pi \mu)}{4 \pi^2 J} = \frac{\bar N}{2 \pi \sqrt{4 \pi^2 J^2 - E^2}}
\end{equation}
where we used the spinon energy to write $\mu(E)=\frac{1}{\pi }\cosh ^{-1}\left(\frac{2 \pi  J}{E}\right)$.
The impurity part in the Kondo phase gives
\begin{align}
  { \rho_{\mathrm{dos}}}^{\mathrm{imp}}(E) &=\frac{4 \pi  J^2 \cos (\pi  (\beta\pm i \gamma))}{\sqrt{4 \pi ^2 J^2-E^2} \left(E^2 \cos (2 \pi  (\beta\pm i \gamma))-E^2+8 \pi ^2 J^2\right)}
  \label{impdos}
\end{align}

Let us now consider the ratio
\begin{align}
   R(E)_\pm= \frac{\bar N}{2}\frac{{ \rho_{\mathrm{dos}}}^{\mathrm{imp}}(E)}{{ \rho_{\mathrm{dos}}}^{\mathrm{bulk}}(E)}&=\frac{4 \pi ^2 J^2 \cos (\pi  (\beta\pm i \gamma))}{E^2 \cos (2 \pi  (\beta\pm i \gamma))-E^2+8 \pi ^2 J^2}
   \label{rdens}
\end{align}

 We compute the Kondo temperature as the energy scale  at which the integrated density of state is half of the total number of state
 \begin{equation}
    \int_0^{T_K} d E \rho_{i m p}(E)=\frac{1}{2} \int_0^{T_0} d E \rho_{i m p}(E)
\end{equation}
We obtain, the complex Kondo temperature for the left impurity
\begin{equation}
   {T_K}_{\pm}=\left(\frac{T_0}{\sqrt{1+\cos ^2(\pi (\beta\pm i\gamma)}}\right).
\end{equation}
where the negative sign is for the left impurity and the positive sign is for right impurity and $T_0=2\pi J$ is the maximum energy of  spinon. 

Notice that even though the ground state energy is real, the spectral weight $R(E)_\pm$ and the dynamically generated Kondo scale (${T_K}_{\pm}$) are complex for each impurity.

\subsection{Bound Mode Phase I (\texorpdfstring{$\frac{1}{2} < \beta < 1$}{1/2 < beta < 1}) and \texorpdfstring{$\gamma \neq 0$}{gamma ≠ 0}}

There are two unique complex solutions in this phase. The first one is

\begin{equation}
    \mu^{b}_+=\pm i\left(\frac12-b\right)=\pm i \left(-\beta -i \gamma +\frac{1}{2}\right)=\pm \gamma\pm i \left(\frac{1}{2}-\beta\right)
\end{equation}
whose energy can be computed using Eq.\eqref{cengeng}
\begin{align}
    E^b_+=E_b&=-2\pi\csc(\pi(\beta+i\gamma))
\end{align}

and
\begin{equation}
    \mu^{b}_-=\pm i\left(\frac12-b^*\right)=\pm i \left(-\beta +i \gamma +\frac{1}{2}\right)=\mp \gamma\pm i\left(\frac{1}{2}-\beta\right) 
\end{equation}
Using Eq.\eqref{cengeng}, the energy of this solutions is
\begin{equation}
      E^{b}_-=E_{b^*}=-2\pi\csc(\pi(\beta-i\gamma))
\end{equation}

The real part of the energies of both solutions $\mu_b$ and $\mu_{b^*}$ is negative in the region $\frac{1}{2}<\beta<1$. Thus, we need to add both of these solutions to construct the ground-state solution. Adding both of these solutions, the integral equation for the root density becomes
\begin{multline}
  \rho_{\ket{0}_{b,b^*}} = (2\bar N+1) a_{\frac{1}{2}}(\mu) - a_{\frac{3}{2} - \beta}(\mu - \gamma) - a_{\frac{3}{2} - \beta}(\mu + \gamma) - a_{-\frac{1}{2} + \beta}(\mu - \gamma) - a_{-\frac{1}{2} + \beta}(\mu + \gamma)\\-\sum_{\upsilon=\pm}\int_{-\infty}^\infty\mathrm{d}\mu~a_1(\mu+\upsilon\mu')\rho_{\left|0\right\rangle}(\mu')-\delta(\mu)
\end{multline}
The solution of above equation in the Fourier space is
\begin{equation}
    \tilde \rho_{\ket{0}_{b,b^*}}=\frac{1}{4} \text{sech}\left(\frac{| \omega | }{2}\right) \left(-e^{(\beta -1) | \omega | -i \gamma  \omega }-e^{\beta  | \omega | +i \gamma  \omega }-2 e^{-((\beta -1) | \omega | )} \cos (\gamma  \omega )-e^{\frac{| \omega | }{2}}+2\bar N+1\right)
    \label{rhobbs}
\end{equation}
Thus the total number of Bethe roots including the two complex roots $\mu_b$ and $\mu_{b^*}$ is 
\begin{equation}
    M_{\ket{0}_{b,b^*}}=2+\int\rho_{\ket{0}_{b,b^*}}(\mu)\mathrm{d}\mu =\frac{\bar N+2}{2}
\end{equation}
Hence, the spin in the ground state is
\begin{equation}
    S^z_{\ket{0}_{b,b^*}}=\frac{\bar N+2}{2}-M_{\ket{0}_{b,b^*}}=0
\end{equation}
Both impurities are screened in the ground state by two dissipative bound modes formed at the sites of the two impurities.

The energy of this state is 
\begin{equation}
    E_{\ket{0}_{b,b^*}}=E_{ar}+E_b+E_{b^*}
    \label{GSBM1Eng-eqn}
\end{equation}
where $E_{ar}$ is the energy of all real roots which upon using Eq.\eqref{cengeng} evaluates to
\begin{align}
     E_{ar} &= J (\bar N-1) - J (2\bar N+1) \log (4) + \pi  J - \frac{4 \beta  J}{\beta ^2+\gamma ^2} - \frac{2 J \left(\beta ^2-\gamma ^2-1\right)}{\beta ^4+2 \beta ^2 \gamma ^2-2 \beta ^2+\gamma ^4+2 \gamma ^2+1} \nonumber \\
     &+ 4 \operatorname{Re}\left(J \psi ^{(0)}\left(\frac{1}{2} (\beta +i \gamma )+\frac{1}{2}\right)\right) - 4 \operatorname{Re}\left(J \psi ^{(0)}\left(\frac{1}{2} (\beta +i \gamma )\right)\right)
     \label{Engallreal}
\end{align}

The ground state energy obtain from Bethe Ansatz given in Eq.\eqref{GSBM1Eng-eqn} and the corresponding result obtained from DMRG is shown in Fig.\ref{risbm1} for various parameters in the bound mode phase I. 

Notice that 
\begin{equation}
    E_{\ket{0}_{b,b^*}}=E_{\ket{0}}
    \label{gsengbm1phase}
\end{equation}
where $E_{\ket{0}}$ given by Eq.\eqref{gseng}.  Notice that for some values of $\gamma$, the energy density becomes independent of system size. For example for $\beta=0.7$ at $\gamma\approx 0.279$, the energy density becomes independent of the system size \footnote{The energy density becomes independent of the system size for some values of $\gamma$ for $\beta$ approximately from 0.35 to 1.17. However, since the energy function is written in terms of digamma functions which are difficult to invert, the analytic expression for the relation between $\beta$ and $\gamma$ where the energy density becomes independent of system size is difficult to obtain.}  which happens because the energy contribution from the impurity independent free boundary exactly cancels the impurity contribution from the two impurities. 

\begin{figure}[H]
\begin{minipage}[h]{0.47\linewidth}
\begin{center}
\includegraphics[width=1\linewidth]{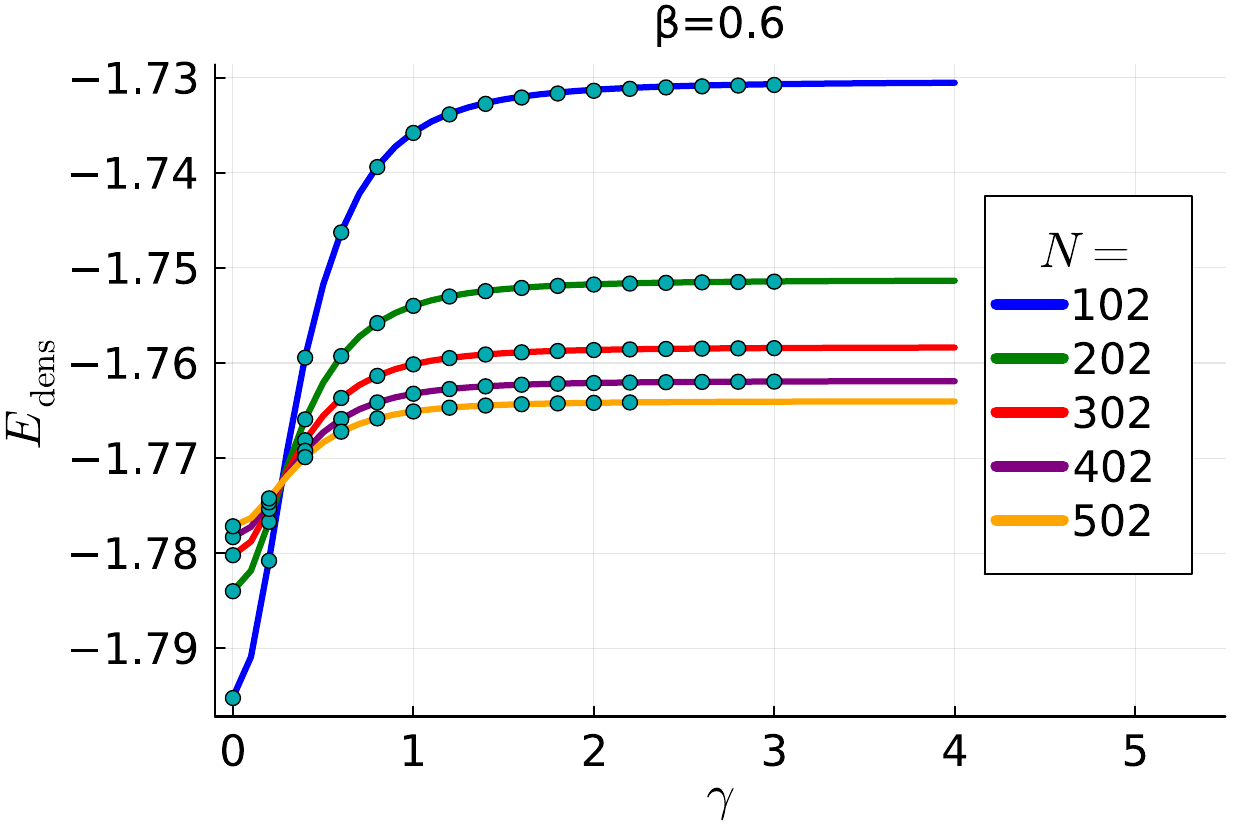} 
\label{qwe1bm1}
\end{center} 
\end{minipage}
\hfill
\vspace{0.2 cm}
\begin{minipage}[h]{0.47\linewidth}
\begin{center}
\includegraphics[width=1\linewidth]{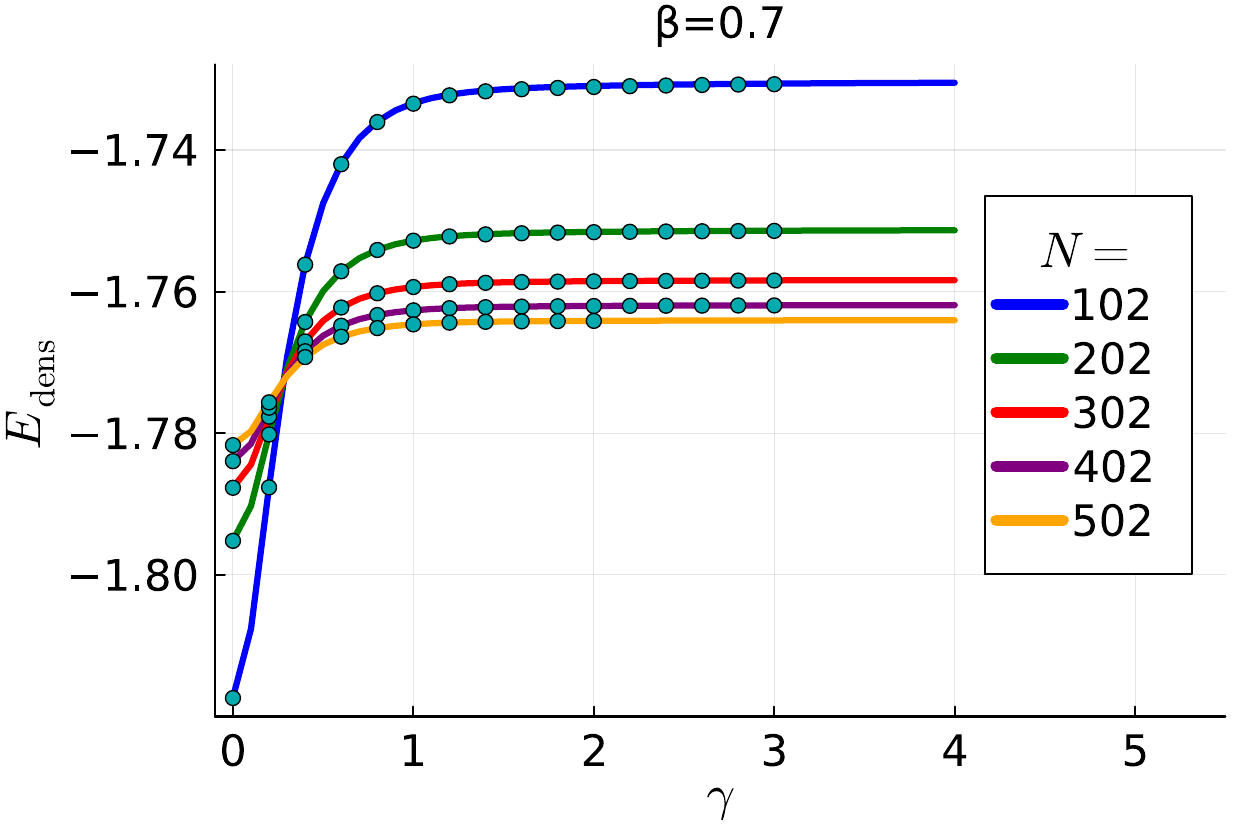} 
\label{qwe2bm1}
\end{center}
\end{minipage}
\vfill
\vspace{0.2 cm}
\begin{minipage}[h]{0.47\linewidth}
\begin{center}
\includegraphics[width=1\linewidth]{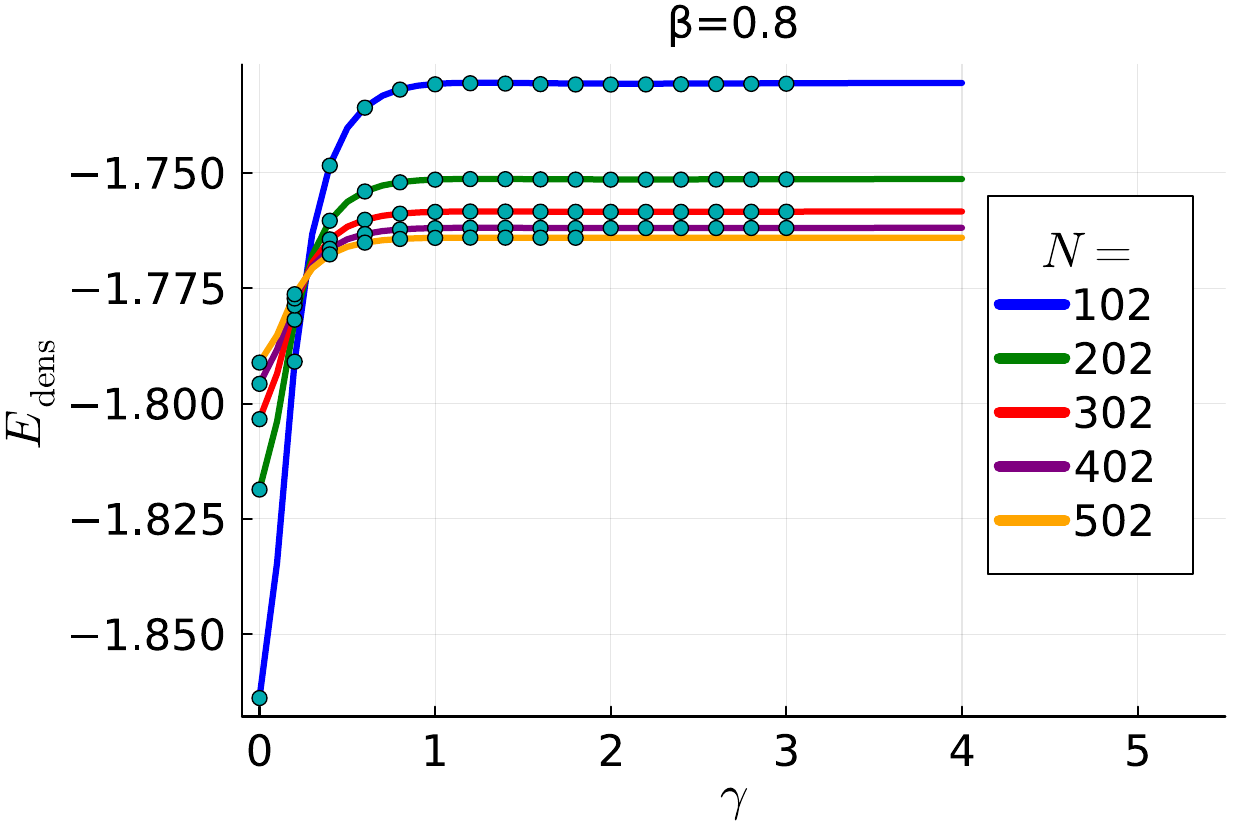} 
\label{qwe3bm1}
\end{center}
\end{minipage}
\hfill
\begin{minipage}[h]{0.47\linewidth}
\begin{center}
\includegraphics[width=1\linewidth]{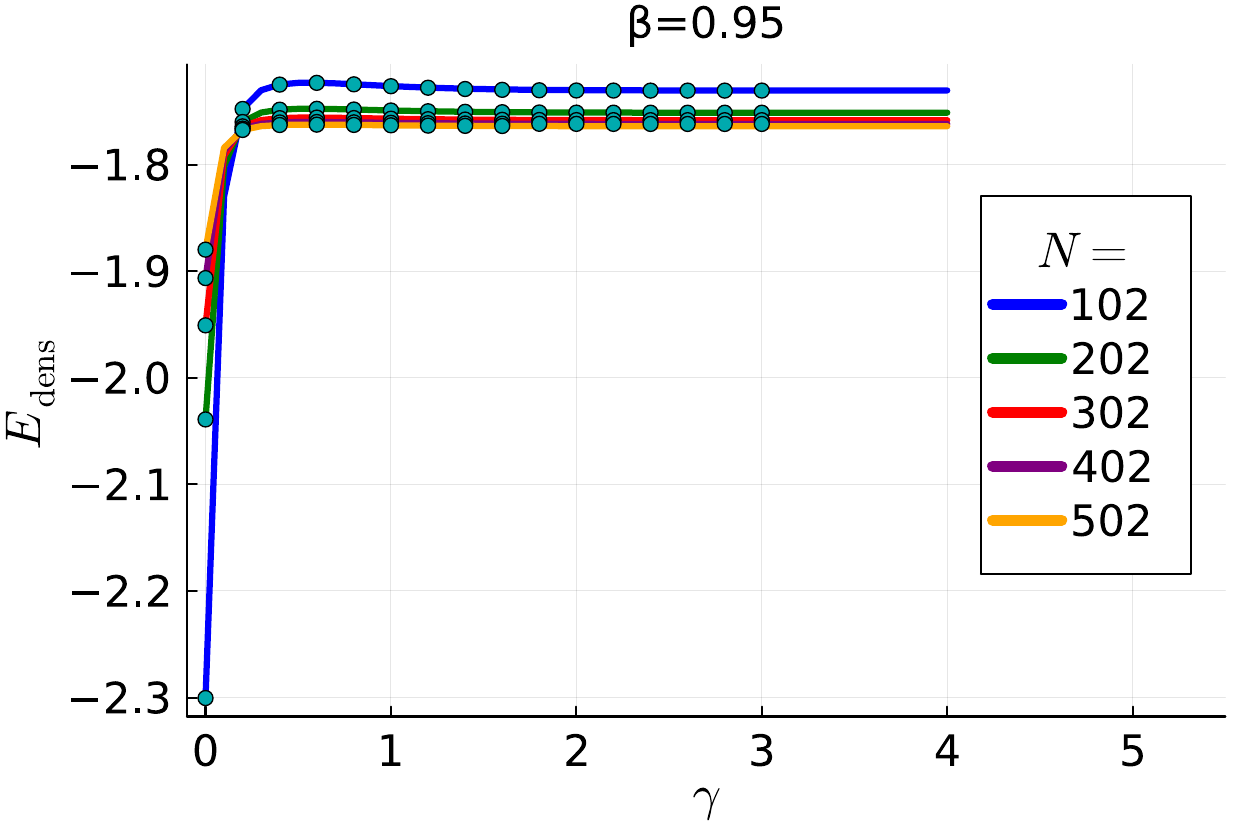} 
\label{qwe4bm1}
\end{center}
\end{minipage}
\caption{Bethe Ansatz and DMRG calculation of ground state energies for various values of complex impurity couplings. Note that the energy density obtained from Bethe Ansatz in the thermodynamic limit and the ones obtained from DMRG using finite size algorithm have a relative difference of about $10^{-4}\%$.}
\label{risbm1}
\end{figure}

There are unique boundary excitations in this phase. We could remove the boundary string solution $\mu_{b^*}$ and add a hole to construct a state whose solution density is given by an integral equation
  \begin{align}
2&\rho_{\ket{1}_{b,\theta}}(\mu)=(2\bar N+1)a_{\frac12}(\mu)-a_{\beta-\frac{1}{2}}(\mu-\gamma)-a_{\beta-\frac{1}{2}}(\mu+\gamma) +a_{\frac{1}{2}+\beta}(\mu-\gamma)+a_{\frac{1}{2}+\beta}(\mu+\gamma)\nonumber\\&-a_{\beta +i \gamma +\frac{1}{2}}(\mu)-a_{\frac{3}{2}-\beta -i \gamma }(\mu)-\sum_{\upsilon=\pm}\int_{-\infty}^\infty\mathrm{d}\mu~ a_1(\mu+\upsilon\mu')\rho_{\left|0\right\rangle}(\mu')-\delta(\mu)-\delta(\mu-\theta)-\delta(\mu+\theta)
\end{align}
The solution in the Fourier space reads
\begin{align}
\tilde \rho_{\ket{1}_{b,\theta}}& = -\frac{1}{4} e^{-\beta  | \omega | } \left(\tanh \left(\frac{| \omega | }{2}\right)+1\right) \left(e^{\beta  | \omega | } (2 \cos (\theta  \omega )+1)+4 \sinh \left(\frac{| \omega | }{2}\right) \cos (\gamma  \omega )\right)\nonumber\\
&+\frac{1}{4} \text{sech}\left(\frac{| \omega | }{2}\right) \left(-e^{| \omega |  (\beta +i \gamma -1)}-e^{-(| \omega |  (\beta +i \gamma ))}+2\bar N+1\right)
\end{align}

The total number of roots including one complex root $\mu_b$ is
\begin{equation}
    M_{\ket{1}_{b,\theta}}=1+\int \rho_{\ket{1}_{b,\theta}}(\mu)\mathrm{d}\mu =\frac{\bar N}{2}
\end{equation}
Thus, the spin of this state is
\begin{equation}
    S^z_{\ket{1}_{b,\theta}}=\frac{\bar N+2}{2}-M_{\ket{1}_{b,\theta}}=1
\end{equation}
The energy of this state is

\begin{equation}
    E_{\ket{1}_{b,\theta}}=E_{\ket{0}}+E_\theta-E_{b^*}=E_{ar}+E_\theta+E_b
\end{equation}
This state has unscreened impurity at right edge and one propagating spinon. Because of the SU(2) symmetry, the state is three fold degenerate where the two free spins make triplet pairing. There also exist a state where the unscreened impurity and the hole form singlet pairing. Such a state is constructed by adding both boundary strings $\mu_b$ and $\mu_{b^*}$, a hole with rapidity $\theta$, and the higher order boundary string 
\begin{equation}
    \mu_{hb^*}=\pm i \left(\frac{3}{2}-(\beta-i\gamma)\right)
\end{equation}
The root desnity of the resultant equation is
\begin{align}
    \tilde\rho_{\ket{0}_{b,b^*, hb^*,\theta}}=\frac{1}{4} \left(\tanh \left(\frac{| \omega | }{2}\right)+1\right) &\left(-e^{-\frac{1}{2} | \omega |  (2 \beta +2 i \gamma -1)}-2 e^{\frac{1}{2} | \omega |  (-2 \beta +2 i \gamma +1)}-e^{\frac{1}{2} | \omega |  (2 \beta -2 i \gamma -5)}\right.\nonumber\\
    &\left.-2 e^{\beta  | \omega | -\frac{3 | \omega | }{2}} \cos (\gamma  | \omega | )+2\bar N e^{-\frac{| \omega | }{2}}+e^{-\frac{| \omega | }{2}}-2 \cos (\theta  \omega )-1\right)
\end{align}
The total number of roots include three complex roots is
\begin{equation}
    M_{\ket{0}_{b,b^*, hb^*,\theta}}=3+\int\rho_{\ket{0}_{b,b^*, hb^*,\theta}}(\mu)\mathrm{d}\mu =\frac{\bar N+2}{2}
\end{equation}
such that the spin of this state is
\begin{equation}
    S^z_{\ket{0}_{b,b^*, hb^*,\theta}}=\frac{\bar N+2}{2}-M_{\ket{0}_{b,b^*, hb^*,\theta}}
\end{equation}
The energy of this state is
\begin{equation}
    E_{\ket{0}_{b,b^*,hb^*, \theta}}=E_{\ket{0}}+E_\theta-E_{b^*}=E_{\ket{1}_{b,\theta}}=E_{ar}+E_\theta+E_b
\end{equation}

Thus the fundamental boundary excitation involves removal of the bound model screening the impurity and introduction of a hole such that the two free spins form $2\otimes 2=3\oplus 1=4$ degenerate states.  These states, that have screened left impurity and unsceened impurity, have complex energy as $E_b^*$ is complex. 

One can also remove the $\mu_b$ solution and add a hole or add $h\mu_b$ and a hole to construct the four-fold degenerate state where the left impurity is unscreened and right impurity is screened. The energy of the resultant four states is
\begin{equation}
    E_{\ket{1}_{b^*,\theta}}=E_{\ket{0}}+E_\theta-E_b=E_{\ket{0}_{b,b^*,hb, \theta}}=E_{\ket{1}_{b,\theta}}=E_{ar}+E_\theta+E_{b^*}
\end{equation}
The energy of these four-fold degenerate state is complex as $E_b$ is complex.

We can also construct a state described by all real roots which have solution density
\begin{equation}
    \tilde\rho_{\ket{1}}=\frac{1}{4} e^{\frac{| \omega | }{2}} \text{sech}\left(\frac{| \omega | }{2}\right) \left(-\sinh \left(\frac{| \omega | }{2}\right) \left(4 e^{-\beta  | \omega | } \cos (\gamma  \omega )+2\bar N+1\right)+(2\bar N+1) \cosh \left(\frac{| \omega | }{2}\right)-1\right)
        \label{allreal}
\end{equation}
The total number of roots is
\begin{equation}
    M_{\ket{1}}=\int\rho_{\ket{1}}(\mu)\mathrm{d}\mu=\frac{\bar N}{2}
\end{equation}

The spin of this state is
\begin{equation}
    S^z_{\ket{1}}=\frac{\bar N+2}{2}-  M_{\ket{1}}=1
\end{equation}

This state contains two unscreened impurities that form triplet pairing. Because of $SU(2)$ symmetry, this state is three-fold degenerate and which has energy
\begin{equation}
    E_{\ket{1}}=E_{ar}=E_{\ket{0}}-E_b-E_{b^*}
    \label{engbothbm}
\end{equation}
Because $E_{b^*}=(E_b)^*$, these state have real energies.

By adding the boundary strings solutions $\mu_b$ and the higher order string solution $\mu_{hb}$, we obtain Bethe equation
\begin{align}
2 \rho_{\left|0\right\rangle_{b, hb}}(\mu) & =(2\bar N+1)a_{\frac{1}{2}}(\mu)-a_{\frac{3}{2}-\beta+i\gamma}(\mu)-a_{\frac{5}{2}-\beta+i\gamma}(\mu)-2a_{\beta-\frac{1}{2}+i\gamma}(\mu)\nonumber\\
&-a_{\beta-\frac{1}{2}-i\gamma}(\mu)+a_{\frac{1}{2}+\beta-i\gamma}(\mu) -\sum_{v=\pm}\int_{-\infty}^{\infty}d\mu a_1\left(\mu+v\mu^{\prime}\right)\rho_{\left|-\frac{1}{2}\right\rangle}\left(\mu^{\prime}\right)-\delta(\mu)
\end{align}
such that the solution density becomes
\begin{align}
    \tilde{\rho}_{\left|0\right\rangle_{b, hb}}(\omega) = \frac{1}{4} \text{sech}\left(\frac{| \omega | }{2}\right) \Bigg(&-e^{| \omega |  (-\beta +i \gamma +1)} - e^{| \omega |  (\beta +i \gamma -2)}- e^{| \omega |  (\beta +i \gamma -1)} - 2 e^{-(| \omega |  (\beta +i \gamma -1))} \nonumber\\
    &+ e^{-(| \omega |  (\beta -i \gamma ))} - e^{\frac{| \omega | }{2}} + 2\bar N + 1\Bigg)
    \label{allrealbshbs}
\end{align}
The total number of Bethe roots including the two complex roots is
\begin{equation}
    M_{\left|0\right\rangle_{b, hb}}=2+\int\rho_{\left|0\right\rangle_{b, hb}}(\mu)\mathrm{d}\mu=\frac{\bar N+2}{2}
\end{equation}
Thus, the total spin is
\begin{equation}
    S^z_{\left|0\right\rangle_{b, hb}}=\frac{\bar N+2}{2}-M_{\left|0\right\rangle_{b, hb}}=0
\end{equation}
Using Eq.\eqref{cengeng}, we obtain the energy of this state
\begin{equation}
    E_{_{\left|0\right\rangle_{b, hb}}}=E_{\ket{0}}-E_{b}-E_{b^*}=E_{ar}
\end{equation}
Thus, this state contains two free impurity spins that forms singlet pairing.

\subsection{\texorpdfstring{Bound Mode Phase II $\left(1<\beta<\frac{3}{2}\right)$ and $\gamma\neq 0$}{Bound Mode Phase II (1<beta<3/2) and gamma not equal to 0}}

The two complex solutions $\mu_b$ and $\mu_b^*$ are still valid solutions in this regime, but the real part of the energies of these solutions are positive. Therefore, the ground state is now made up of all roots. 

Thus, the four-fold degenerate state described by the root densities of Eq.\eqref{allreal} and Eq.\eqref{allrealbshbs} where both impurities are not screened is the ground state in this phase. These states have real energies  
\begin{align}
    E_{\ket{1}}&=J (\bar N-1) - J (2\bar N+1) \log (4) + \pi  J - \frac{4 \beta  J}{\beta ^2+\gamma ^2} - \frac{2 J \left(\beta ^2-\gamma ^2-1\right)}{\beta ^4+2 \beta ^2 \gamma ^2-2 \beta ^2+\gamma ^4+2 \gamma ^2+1} \nonumber \\
     &+ 4 \operatorname{Re}\left(J \psi ^{(0)}\left(\frac{1}{2} (\beta +i \gamma )+\frac{1}{2}\right)\right) - 4 \operatorname{Re}\left(J \psi ^{(0)}\left(\frac{1}{2} (\beta +i \gamma )\right)\right)
     \label{GSengBM2phase}
\end{align}

The plots of ground state equation obtained via Bethe Ansatz in Eq.\eqref{GSengBM2phase} and the corresponding result obtained by DMRG is shown in Fig.\ref{risbm2} for various values of $\beta$ in the bound mode phase II. 

\begin{figure}[H]
\begin{minipage}[h]{0.47\linewidth}
\begin{center}
\includegraphics[width=1\linewidth]{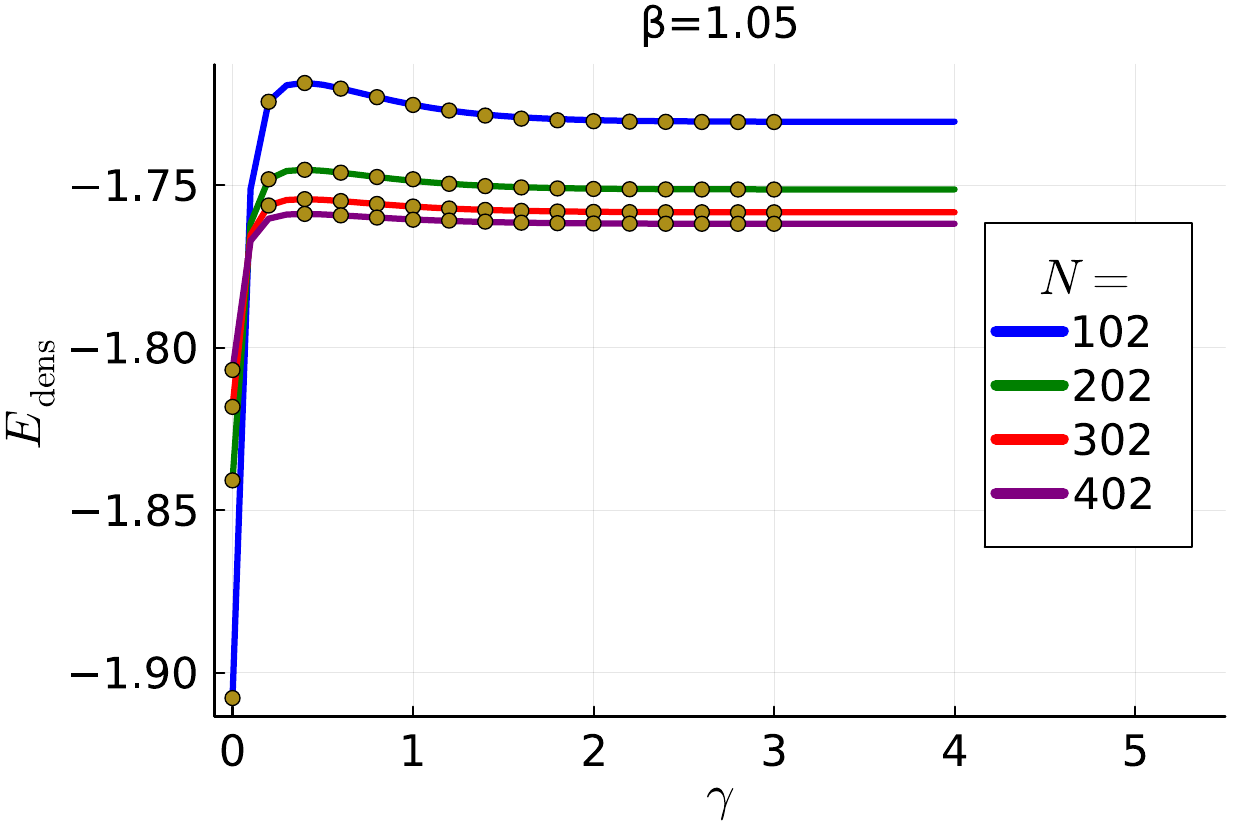} 
\label{qwe1bm}
\end{center} 
\end{minipage}
\hfill
\vspace{0.2 cm}
\begin{minipage}[h]{0.47\linewidth}
\begin{center}
\includegraphics[width=1\linewidth]{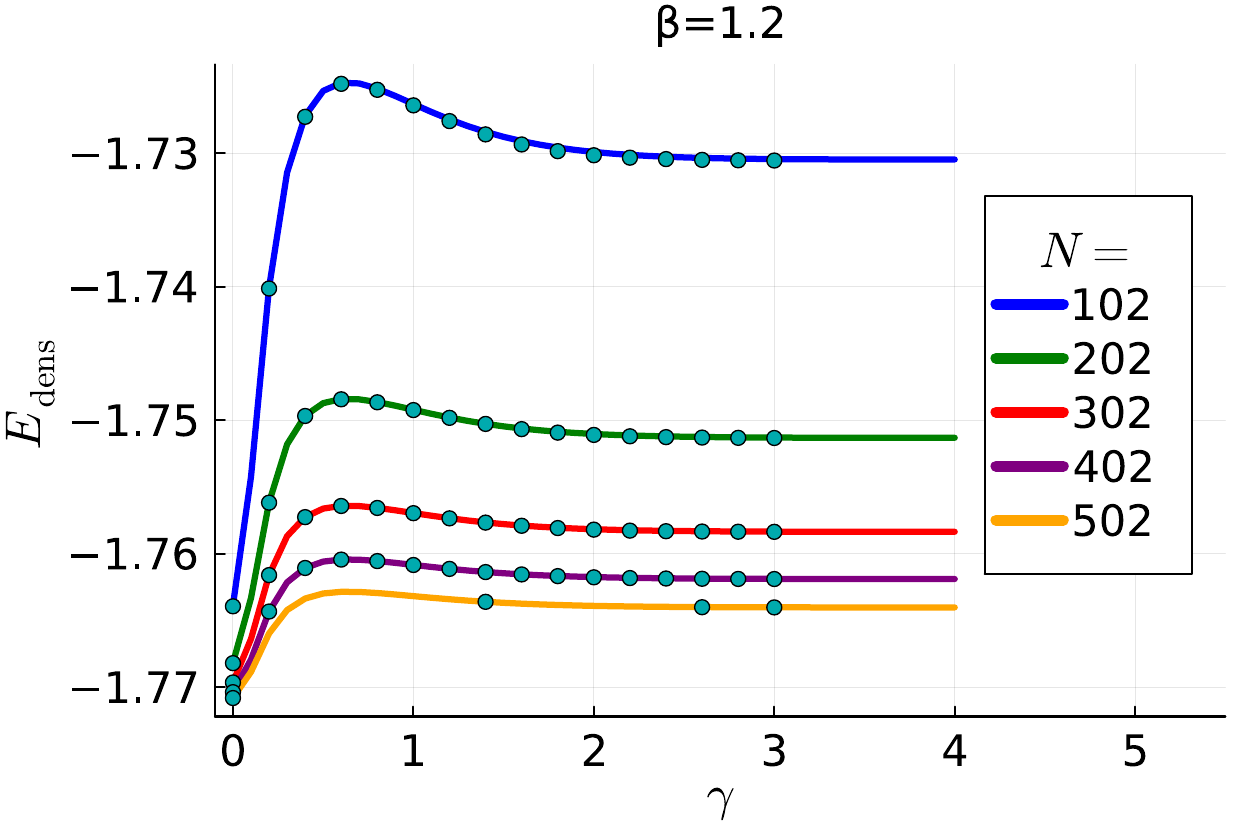} 
\label{qwe2bm}
\end{center}
\end{minipage}
\vfill
\vspace{0.2 cm}
\begin{minipage}[h]{0.47\linewidth}
\begin{center}
\includegraphics[width=1\linewidth]{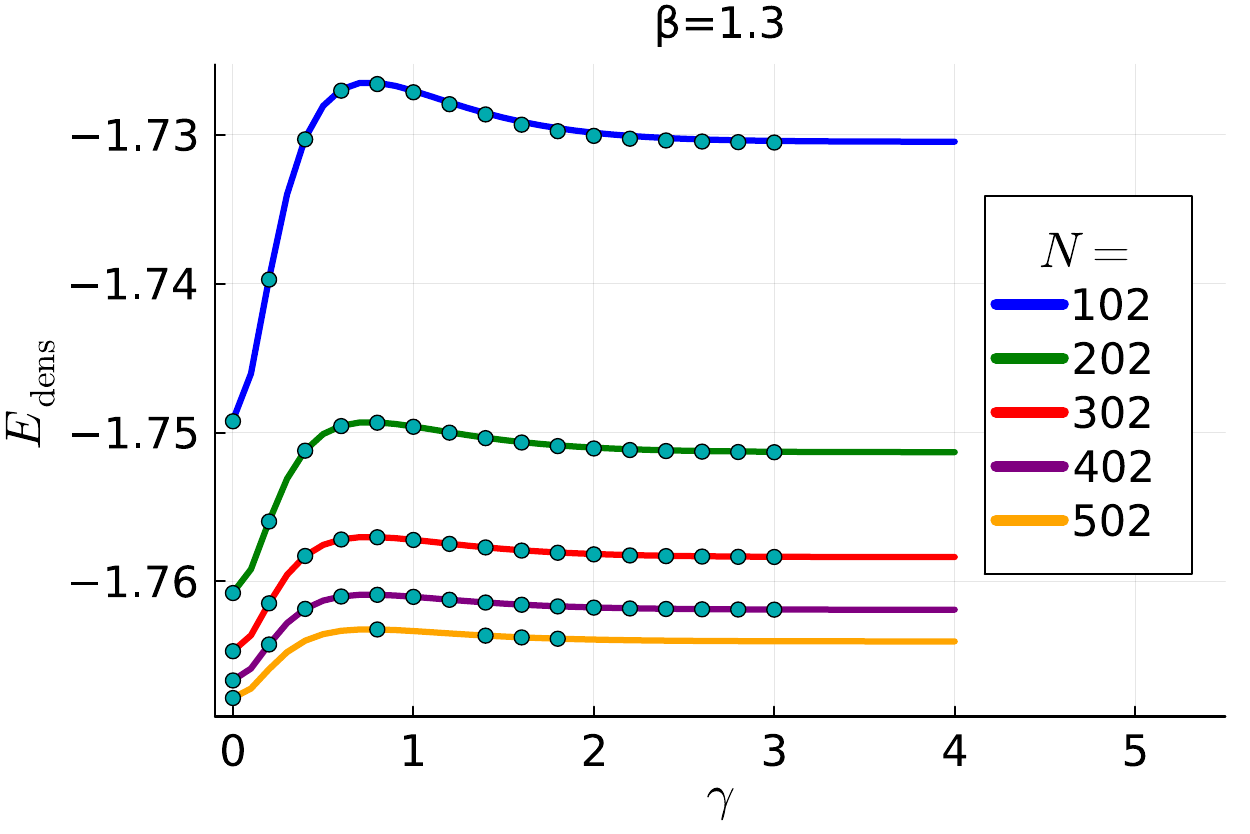} 
\label{qwe3bm}
\end{center}
\end{minipage}
\hfill
\begin{minipage}[h]{0.47\linewidth}
\begin{center}
\includegraphics[width=1\linewidth]{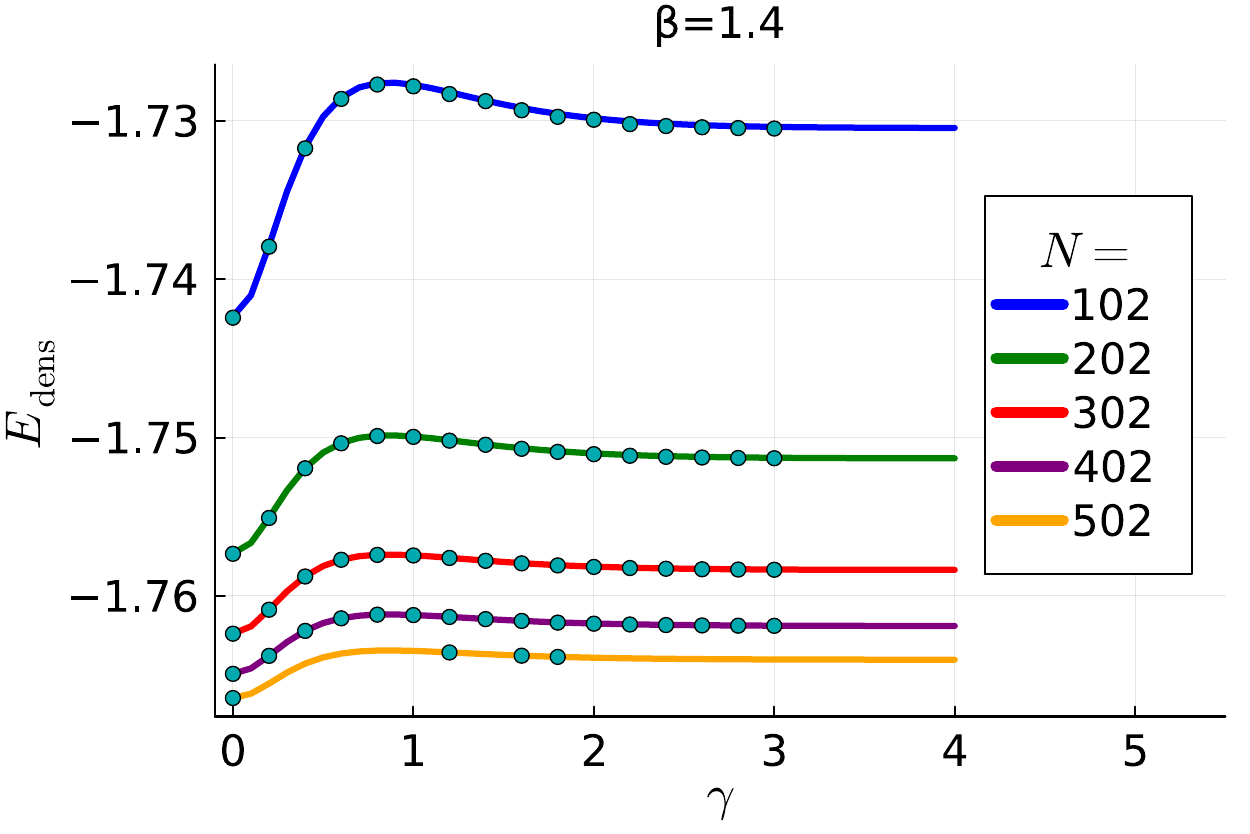} 
\label{qwe4bm}
\end{center}
\end{minipage}
\caption{Bethe Ansatz and DMRG calculation of ground state energies for various values of complex impurity couplings. Note that the energy density obtained from Bethe Ansatz in the thermodynamic limit and the ones obtained from DMRG using finite size algorithm have a relative difference of about $10^{-3}\%$.}
\label{risbm2}
\end{figure}
The excited states can be constructed by adding one (at each end) or both bound modes to the all real roots. For example, adding the bound modes to both left and right edges, creates a unique excited state where both impurities are screened. This state is described by root density Eq.\eqref{rhobbs}. This state has real energy $E_{\ket{0}_{b,b^*}}=E_{\ket{1}}+E_{b}+E_{b^*}=E_{\ket{0}}$ explicitly given by Eq.\eqref{gseng}.

We could also have a state with bound mode only at one end and a hole propagating at the bulk. This state is four-fold degenerate and has complex energy. The energy of the state containing bound mode only at the left end is
\begin{equation}
    E_{\ket{1}}+E_b+E_\theta
\end{equation}
and the four-fold degenerate state with bound mode only at the right edge have energy
\begin{equation}
    E_{\ket{1}}+E_{b^*}+E_\theta
\end{equation}
Thus, we have the same four unique states described in \ref{bothunscreened}-\ref{bothscreened}. By adding an even numbers of spinons, bulk strings, quartets, and higher order boundary strings to these four states, we can sort all the states in this phase into four distinct sets of excited states which are characterized by the number of bound modes. The sets of excited states containing an odd numbers of bound modes have complex energies, and the sets of excited states containing an even numbers of bound modes have real energies. This shows that the $\mathscr{PT}-$symmetry is spontaneously broken in this phase such that the states have either real energies or they appear in complex conjugate pairs.

\subsection{\texorpdfstring{Unscreened phase ($\beta>\frac{3}{2}$) and $\gamma\neq 0$}{Unscreened phase (beta > 3/2) and gamma not equal to 0}}

When $\beta>\frac{3}{2}$, the boundary string solution $\mu_b$ and $\mu_{b^*}$ becomes a wide string \cite{destri1982analysis}, and hence it has zero energy. 

Thus, the ground state is described by all real roots with the solution density given by Eq.\eqref{allreal}. This state has $S^z=1$ and real energy \footnote{In the Hermitian limit $\gamma\to 0$, notice that the expression in Equation \eqref{engallreal} exhibits superfluous divergence due to $E_b=\csc(\pi (\beta\pm i\gamma))$ blowing up for $\beta\in \mathbb{Z^+}$. However, there is also a divergent factor in $E_{\ket{0}}$ in the digamma functions that cancels out the divergences in the energy of the boundary string. Thus, one can explicitly write $E_{ar}$ as shown in Eq.\eqref{engallreal} which is divergence free.}
\begin{equation}
   E_{ar}= E_{\ket{0}}+E_b+E_{b^*}
    \label{engallreal}
\end{equation}

The plots of ground state equation obtained via Bethe Ansatz in Eq.\eqref{engallreal} and the corresponding result obtained by DMRG is shown in Fig.\ref{risus} for various values of $\beta$ in the unscreened phase. 

\begin{figure}[H]
\begin{minipage}[h]{0.47\linewidth}
\begin{center}
\includegraphics[width=1\linewidth]{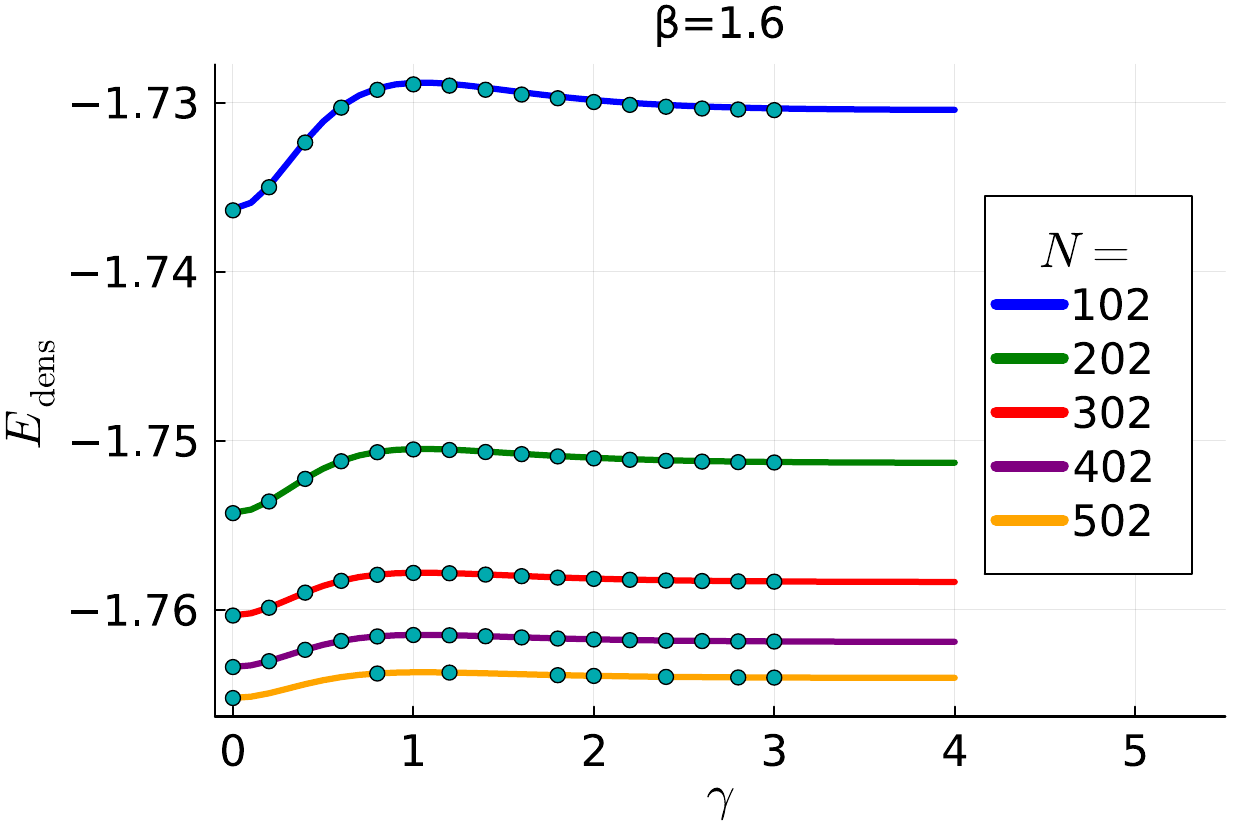} 
\label{qwe1us}
\end{center} 
\end{minipage}
\hfill
\vspace{0.2 cm}
\begin{minipage}[h]{0.47\linewidth}
\begin{center}
\includegraphics[width=1\linewidth]{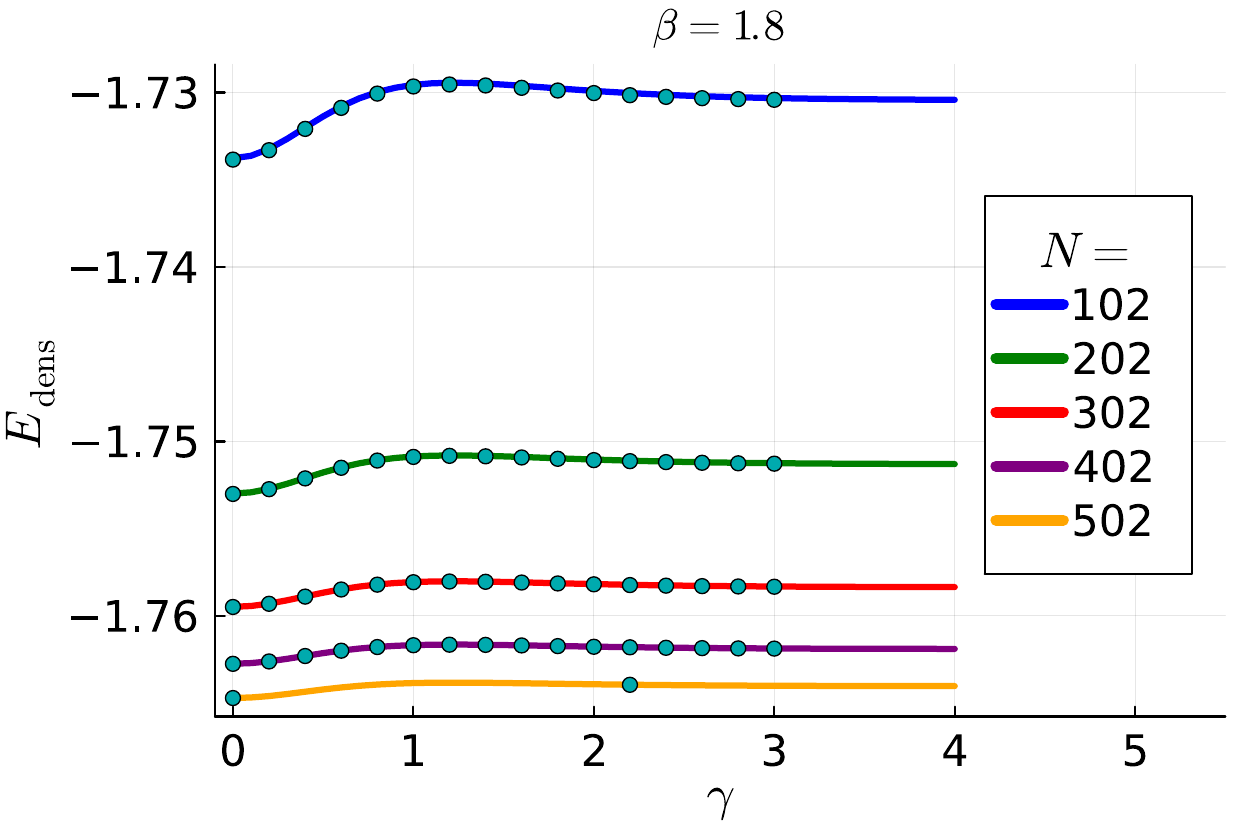} 
\label{qwe2us}
\end{center}
\end{minipage}
\vfill
\vspace{0.2 cm}
\begin{minipage}[h]{0.47\linewidth}
\begin{center}
\includegraphics[width=1\linewidth]{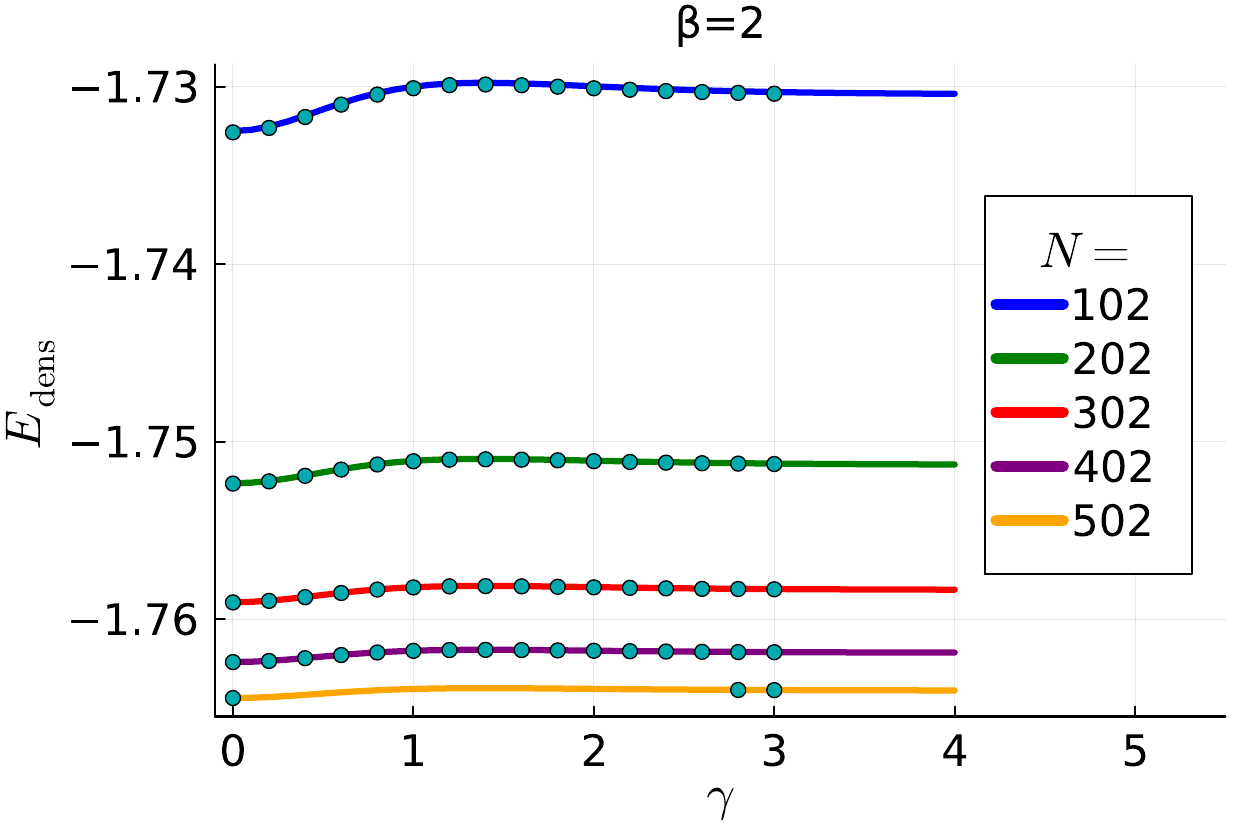} 
\label{qwe3us}
\end{center}
\end{minipage}
\hfill
\begin{minipage}[h]{0.47\linewidth}
\begin{center}
\includegraphics[width=1\linewidth]{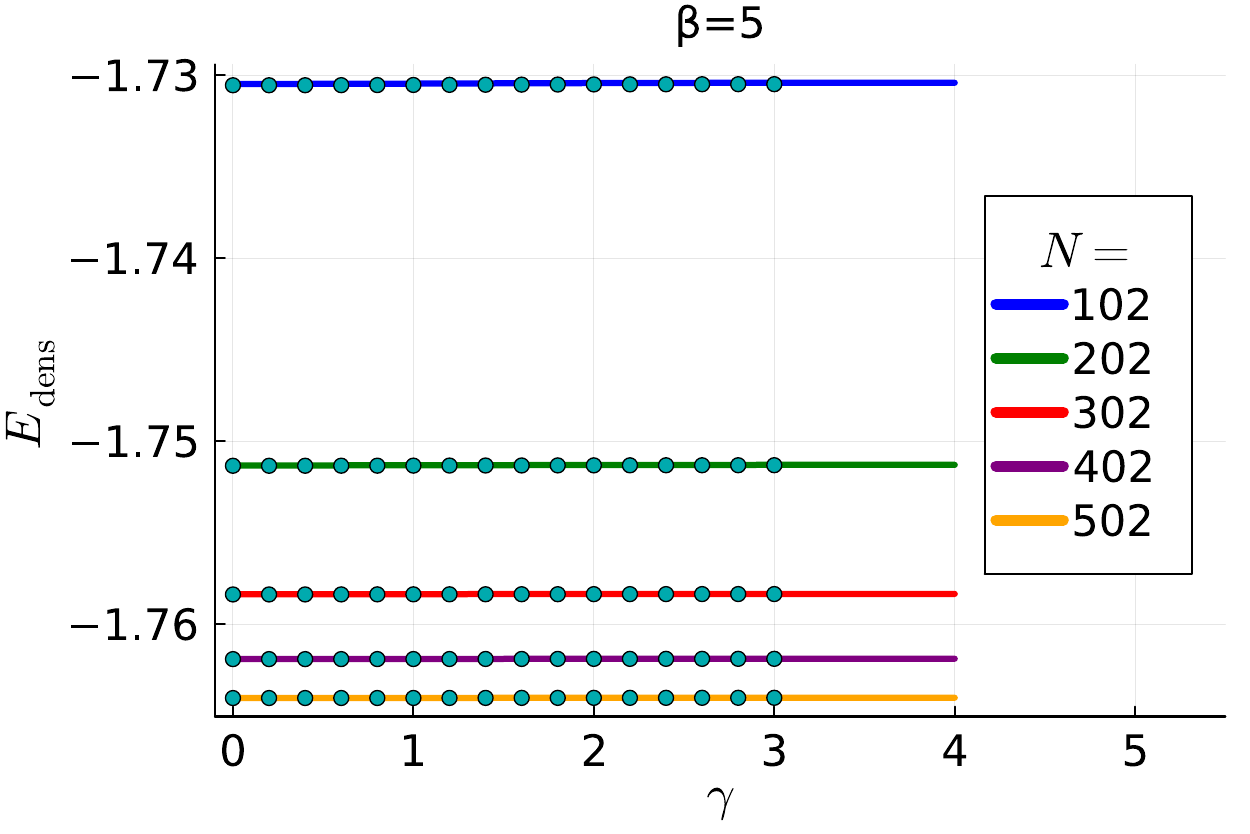} 
\label{qwe4us}
\end{center}
\end{minipage}
\caption{Bethe Ansatz and DMRG calculation of ground state energies for various values of complex impurity couplings. Note that the energy density obtained from Bethe Ansatz in the thermodynamic limit and the ones obtained from DMRG using finite size algorithm have a relative difference of about $10^{-3}\%$.}
\label{risus}
\end{figure}
One can add either of the boundary strings $\mu_b$ (or $\mu_{b^*}$) on top of all real roots and obtain a state described by continuous root distribution
\begin{align}
    \rho_{\ket{0}_b}&=\left(e^{-(| \omega |  (\beta +i \gamma -2))}-e^{-(| \omega |  (\beta +i \gamma ))}+2 e^{-\beta  | \omega | } \cos (\gamma  \omega )-2 e^{-((\beta -1) | \omega | )} \cos (\gamma  \omega )-e^{\frac{| \omega | }{2}}+2 \bar N+1\right)\nonumber\\
    &\times \frac{1}{4} \text{sech}\left(\frac{| \omega | }{2}\right) 
\end{align}
The total number of roots including the complex root $\mu_b$ is 
\begin{equation}
    M_{\ket{0}_b}=1+\int\rho_{\ket{0}_b}(\mu)\mathrm{d}\mu=\frac{\bar N+2}{2}
\end{equation}
Thus, the total spin is
\begin{equation}
    S^z=\frac{\bar N+2}{2}- M_{\ket{0}_b}=0
\end{equation}
This state contains the two free boundary impurity spins that form singlet pair. Since, the energy of the boundary string is zero, this state is degenerate to the triplets \textit{i.e.} the energy of this state is $E_{ar}$, which is real. 

Notice that there are no boundary exitations in this phase. Thus, all other excited states in this phase are constructed by adding an even number of spinons, bulk strings, quartets, higher order boundary strings all of which have real energies. Thus, all the states in this phase have real energies which shows that the $\mathscr{PT}-$symmetry remains unbroken in this phase. 

\section{DMRG results}\label{DMRG}
As mentioned in the main text, we extract the energy density in thermodynamic limit by fitting the energy density for various values of $N$ to an expression of the form
\begin{equation}
    E_{\mathrm{dens}}=E_{\mathrm{GS}}/N=\frac{E_{\partial B}}{N}+E_B
    \label{edensfiteqn}
\end{equation}

As a representative case, we consider $\beta=0.3$ and $\gamma=1$, we plot the energy density for $N=102,202,302,402,502,602,702,802,902$ and $1002$ as shown in Fig.\ref{fig:b0p3g1}. The $y-$intercept gives $E_B$ which is the bulk part of the energy density and the slope $E_{\partial B}$ gives the combined boundary and impurity contribution to the energy density.
\begin{figure}[H]
    \centering
    \includegraphics[scale=0.5]{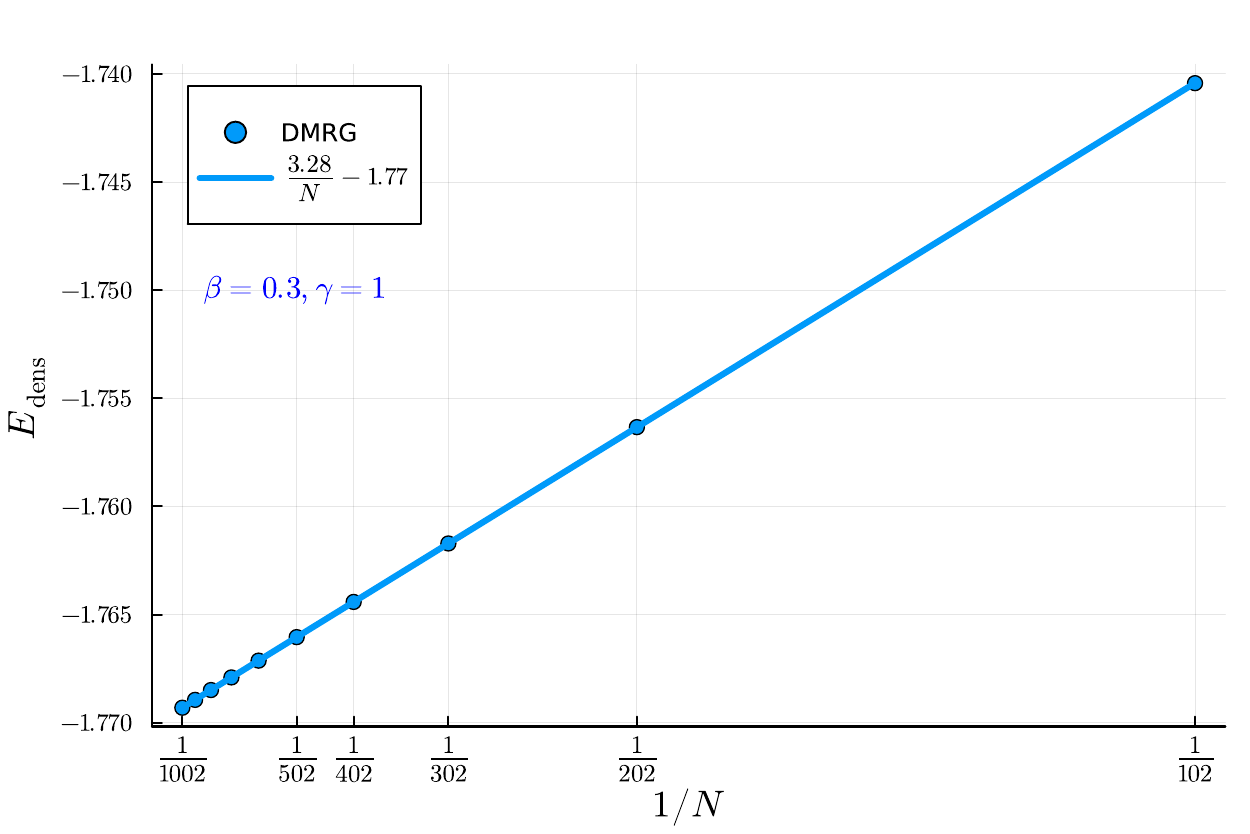}
    \caption{Energy density for various values of $N$ for fixed model parameters $\beta=0.3$ and $\gamma=1$ as a function of $\frac{1}{N}$.}
    \label{fig:b0p3g1}
\end{figure}
The relative percentage difference between the bulk part of the energy density obtained from Bethe Ansatz $E_{{B}_{BA}}=1-2\ln(4)$ and $E_{{B}_{DMRG}}=-1.7725786775$ is $5.67\times 10^{-4}\%$ and the relative percentage difference between the boundary and impurity part of the energy density $a$ is $0.21\%$.

We show for more representative cases of model parameter $\beta$ and $\gamma$ the extraction of boundary and bulk part of the energy density by plotting the energy density for various values of $N$ in Fig.\ref{extrapolation}.

\begin{figure}[H]
\begin{minipage}[h]{0.32\linewidth}
\begin{center}
\includegraphics[width=1\linewidth]{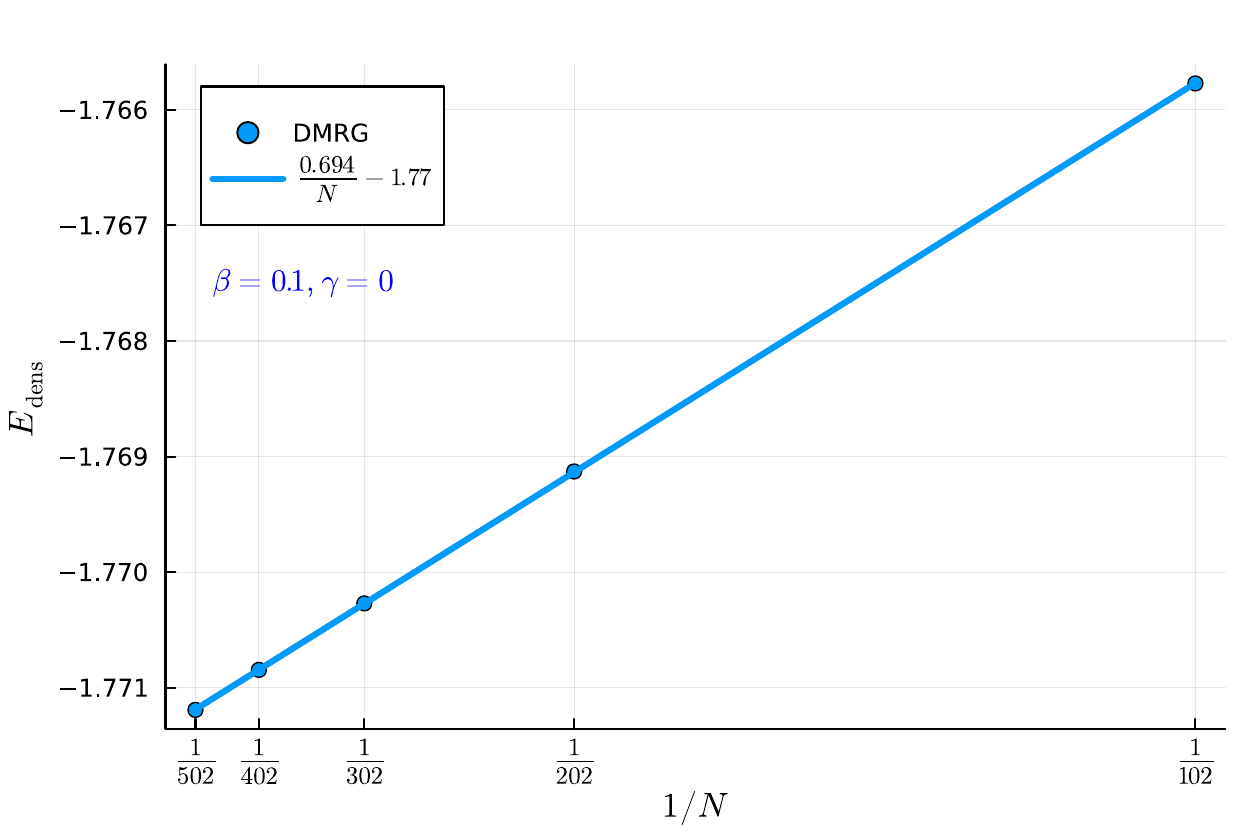} 
\end{center} 
\end{minipage}
\begin{minipage}[h]{0.32\linewidth}
\begin{center}
\includegraphics[width=1\linewidth]{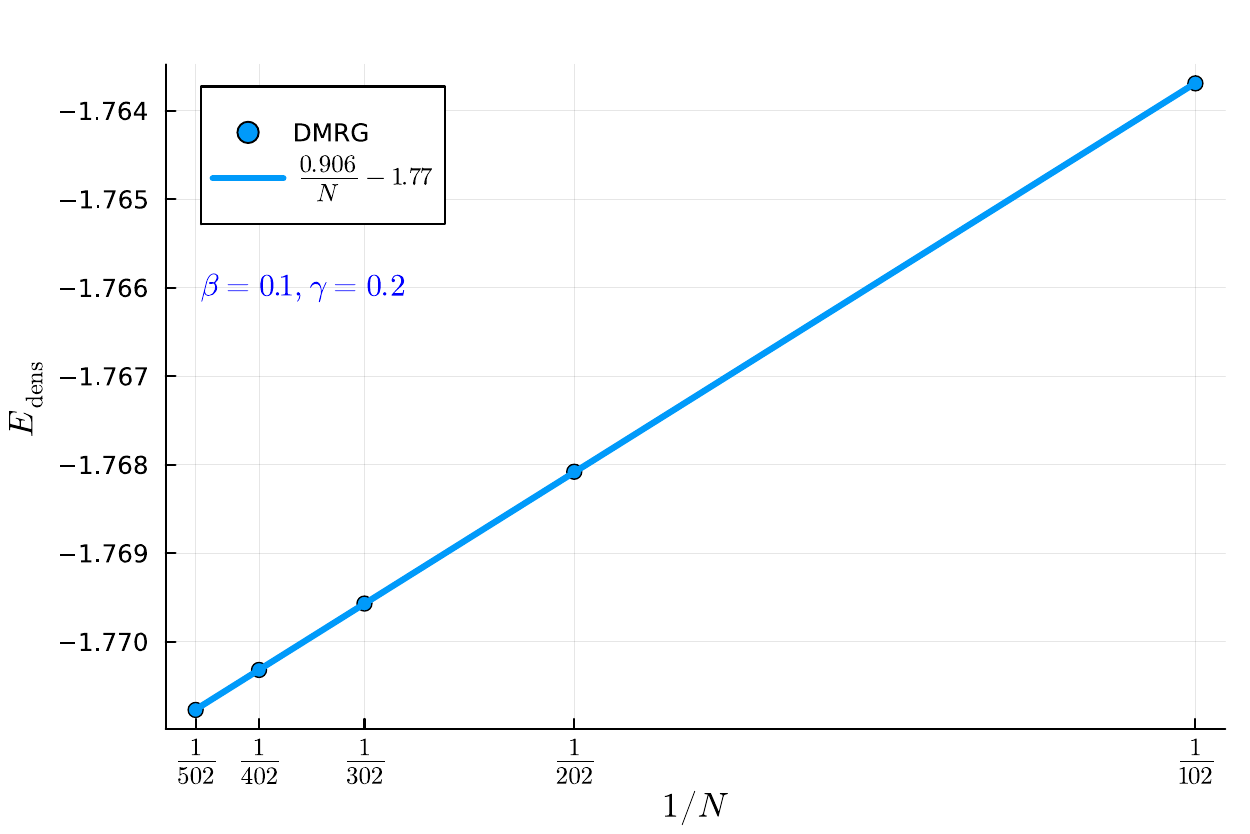} 
\end{center}
\end{minipage}
\begin{minipage}[h]{0.32\linewidth}
\begin{center}
\includegraphics[width=1\linewidth]{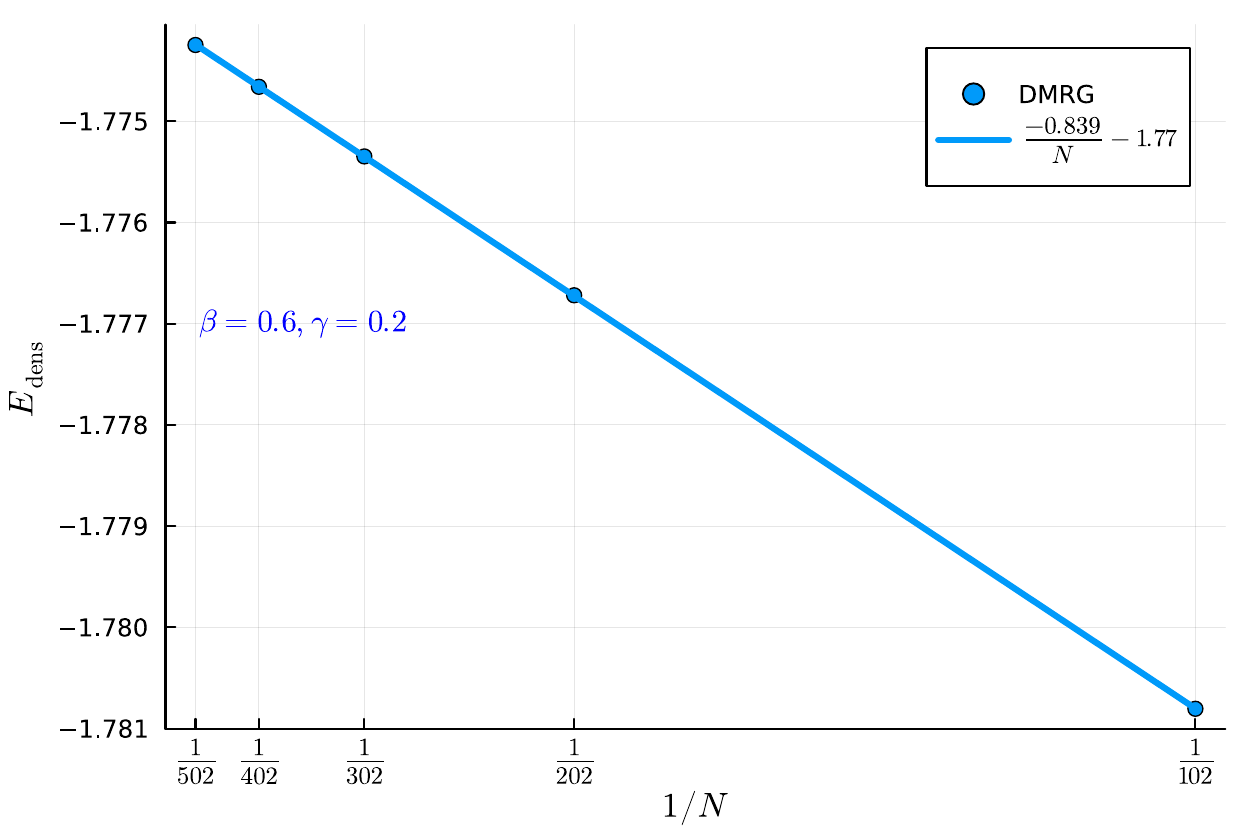} 
\end{center}
\end{minipage}
\begin{minipage}[h]{0.32\linewidth}
\begin{center}
\includegraphics[width=1\linewidth]{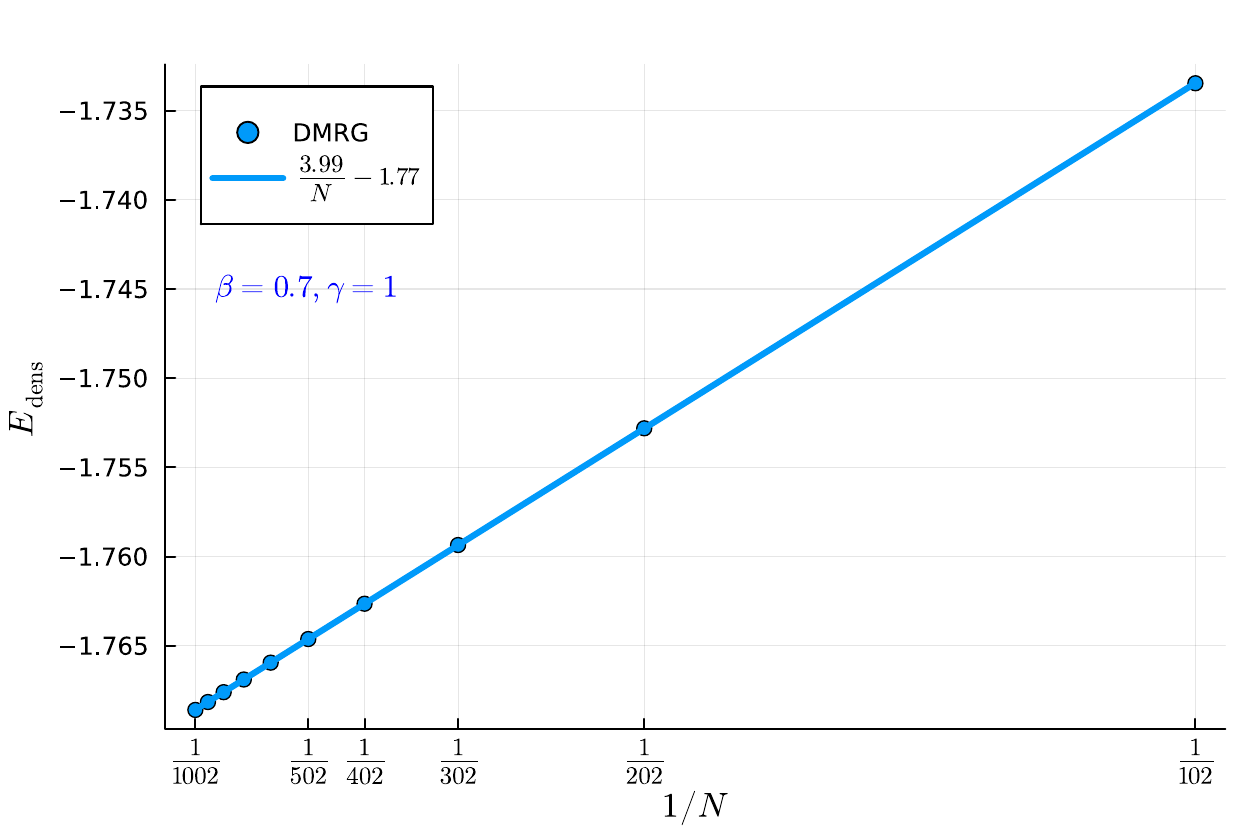} 
\end{center} 
\end{minipage}
\begin{minipage}[h]{0.32\linewidth}
\begin{center}
\includegraphics[width=1\linewidth]{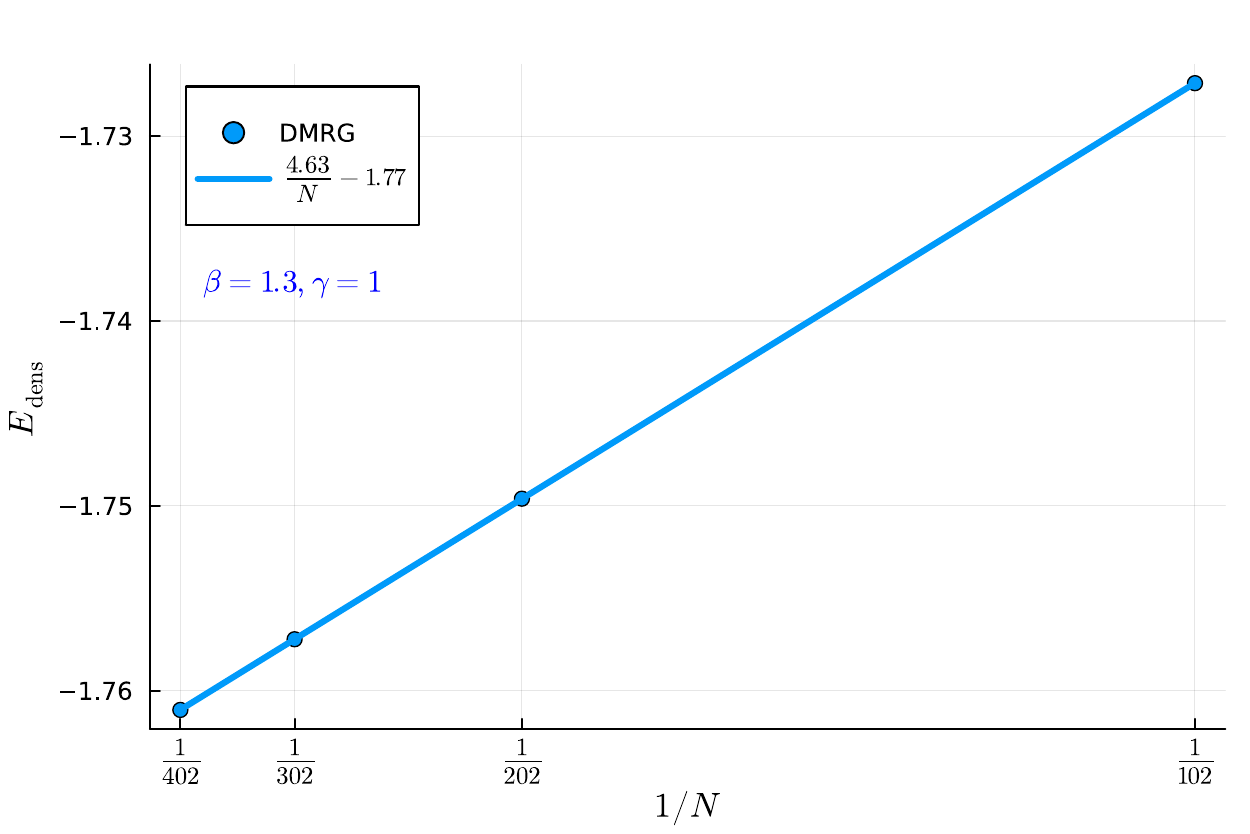} 
\end{center}
\end{minipage}
\begin{minipage}[h]{0.32\linewidth}
\begin{center}
\includegraphics[width=1\linewidth]{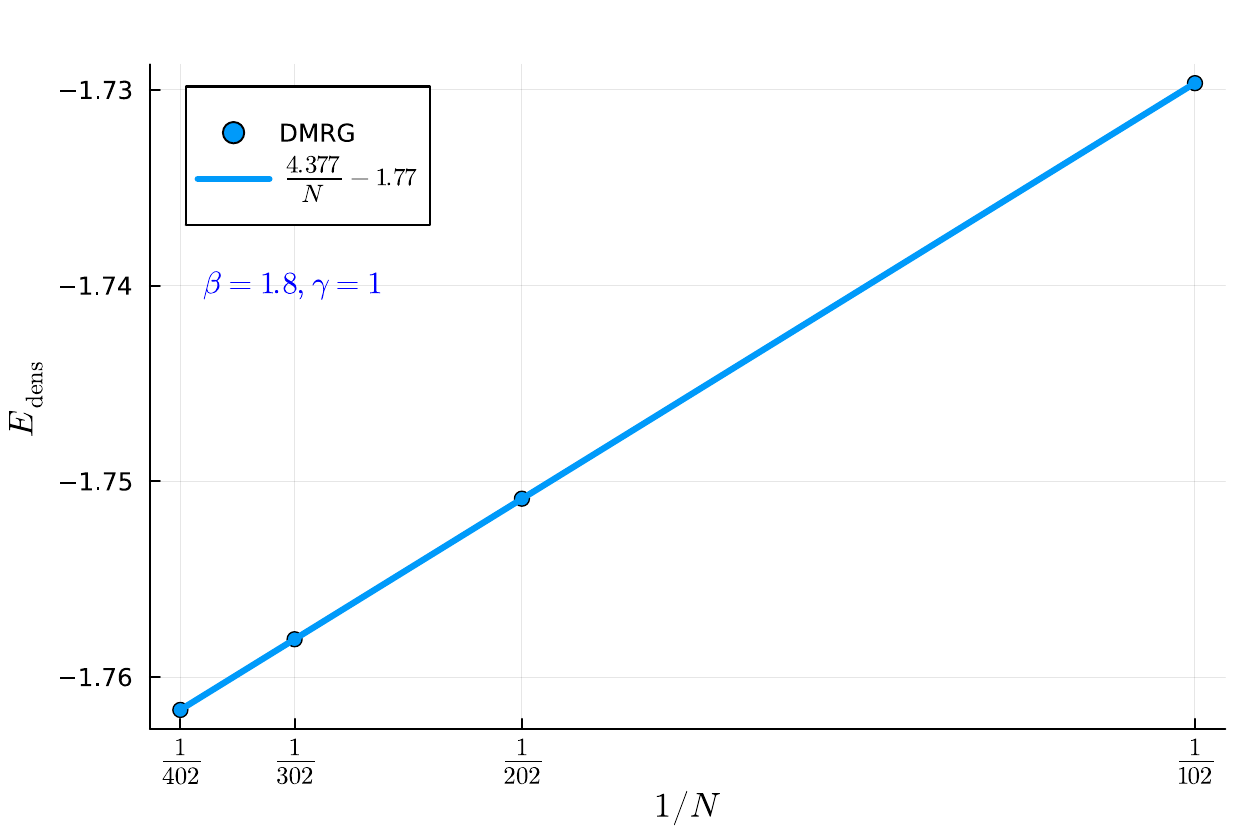} 
\end{center}
\end{minipage}
\caption{We extract the bulk and boundary part of the energy density for various values of model parameters $\beta$ and $\gamma$ by fitting the energy density obtained from DMRG for different values of $N$ to Eq.\eqref{edensfiteqn}.}
\label{extrapolation}
\end{figure}

\subsection{Hermitian limit}\label{herm-limit}
When either $\beta$ or $\gamma$ is set to zero, the model considered here reduces to the Hermitian case studied in \cite{kattel2023kondo} with equal strength of boundary couplings at the two ends of the spin chain. Here, we consider the case when $\gamma$ is set to zero and analyze the accuracy of DMRG calculation using ITensor library \cite{fishman2022itensor}.

We compute the ground state energy for various values of $\beta$ and extract the boundary and bulk part of the energy density. We notice that the bulk part of the energy density obtained from DMRG and Bethe Ansatz in thermodynamic limit have relative difference in order of $10^{-5}\%$ whereas the boundary part which get contribution from both the open boundary condition and the presence of impurities suffer from finite size effect and hence the difference between the result obtained from two method is relatively larger as shown in Fig.\ref{herm-boundary}.

\begin{figure}[H]
    \centering
    \includegraphics[scale=0.4]{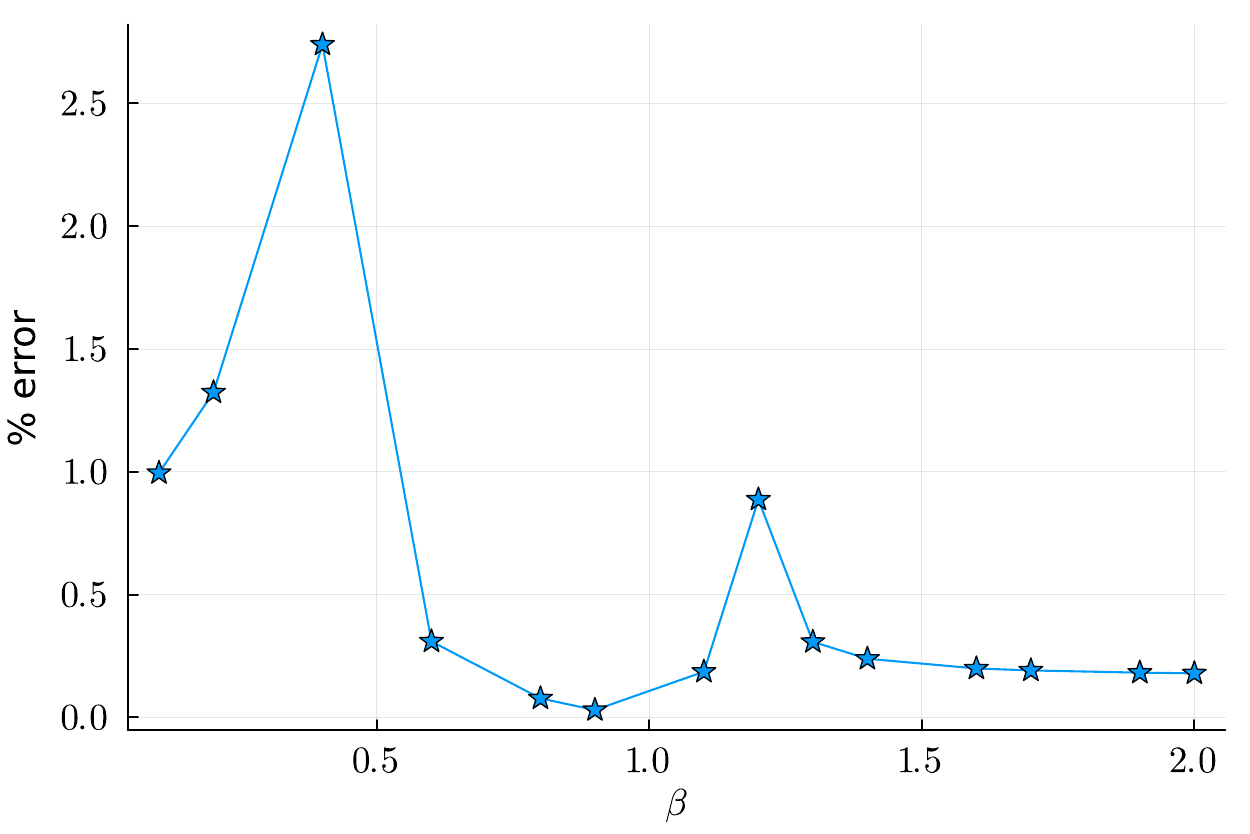}
    \caption{Relative difference between the boundary part of the grounds state energy density obtained from Bethe Ansatz in thermodynamic limit and Hermitian DMRG calculation by fitting the energy density for $N=102,202,302,402$ and $502$.}
    \label{herm-boundary}
\end{figure}

As shown in the first figure of Fig.\eqref{extrapolation}, for the Hermitian case of $\beta=0.1$ and $\gamma=0$, the energy density in thermodynamic limit is extracted by fitting energy density for various values of $N$. We find that the relative error between the Bethe Ansatz and DMRG result for bulk part of the energy density is $1.07\times 10^{-3}\%$ and the boundary part of the energy density is $1.27\%$. This shows that the relative errors in boundary and bulk part of the energy density for Hermitian case is similar to the ones for non-Hermitian case. This shows that the non-Hermitian DMRG implemented in ITensor library \cite{fishman2022itensor} used by turning on the \textit{ishermitian=false} flag  works equally well compared to the Hermitian DMRG implemented in ITensor library.

\end{widetext}
\end{appendix}

\end{document}